%% file: draft1.tex
\documentclass[prd,aps,a4paper,superscriptaddress,twocolumn,nofootinbib]{revtex4}
\usepackage{graphicx}
\usepackage{color}
\usepackage{dcolumn}
\usepackage{bm}
\usepackage{slashed}
\usepackage{amsmath}
\usepackage{latexsym}
\usepackage{amssymb}
\usepackage{mathrsfs}
\usepackage{amsfonts}
\usepackage{longtable}
\usepackage{physics}

\usepackage{url}
\allowdisplaybreaks
\begin{document}
\title{A higher-multipole gravitational waveform model for an eccentric binary black holes based on the effective-one-body-numerical-relativity formalism}

\author{Xiaolin Liu}
\affiliation{Department of Astronomy, Beijing Normal University,
Beijing 100875, China}
\author{Zhoujian Cao
\footnote{corresponding author}} \email[Zhoujian Cao: ]{zjcao@amt.ac.cn}
\affiliation{Department of Astronomy, Beijing Normal University,
Beijing 100875, China}
\affiliation{School of Fundamental Physics and Mathematical Sciences, Hangzhou Institute for Advanced Study, UCAS, Hangzhou 310024, China}
\author{Zong-Hong Zhu}
\affiliation{Department of Astronomy, Beijing Normal University,
Beijing 100875, China}

\begin{abstract}
We construct a new factorized waveform including $(l,|m|)=(2,2),(2,1),(3,3),(4,4)$ modes based on effective-one-body (EOB) formalism, which is valid for spinning binary black holes (BBH) in general equatorial orbit. When combined with the dynamics of $\texttt{SEOBNRv4}$, the $(l,|m|)=(2,2)$ mode waveform generated by this new waveform can fit the original $\texttt{SEOBNRv4}$ waveform very well in the case of a quasi-circular orbit. We have calibrated our new waveform model to the Simulating eXtreme Spacetimes (SXS) catalog. The comparison is done for BBH with total mass in $(20,200)M_\odot$ using Advanced LIGO designed sensitivity. For the quasi-circular cases we have compared our $(2,2)$ mode waveforms to the 281 numerical relativity (NR) simulations of BBH along quasi-circular orbits. All of the matching factors are bigger than 98\%. For the elliptical cases, 24 numerical relativity simulations of BBH along an elliptic orbit are used. For each elliptical BBH system, we compare our modeled gravitational polarizations against the NR results for different combinations of the inclination angle, the initial orbit phase and the source localization in the sky. We use the the minimal matching factor respect to the inclination angle, the initial orbit phase and the source localization to quantify the performance of the higher modes waveform. We found that after introducing the high modes, the minimum of the minimal matching factor among the 24 tested elliptical BBHs increases from 90\% to 98\%. Following our previous $\texttt{SEOBNRE}$ waveform model, we call our new waveform model $\texttt{SEOBNREHM}$. Our $\texttt{SEOBNREHM}$ waveform model can match all tested 305 SXS waveforms better than 98\% including highly spinning ($\chi=0.99$) BBH, highly eccentric ($e\approx0.15$) BBH and large mass ratio ($q=10$) BBH.
\end{abstract}

\maketitle

\section{Introduction}
LIGO \cite{advLIGO} and VIRGO \cite{advVirgo} have achieved the detection of gravitational waves (GW) \cite{PhysRevLett.116.061102_detectGW01,PhysRevX.9.031040_detecGW}. The gravitational waves from compact binary systems are the only detected GW sources, and they will be also the most likely gravitational wave events to be detected in the near future. For the reported more than 50 events in the O1/O2/O3a data, the accurate waveform model of the gravitational wave signal has played an important role. Based on the most accurate numerical relativistic (NR) simulation of BBHs, several inspiral-merger-ringdown (IMR) waveform models have been constructed. Effective-one-body (EOB) approach \cite{PhysRevD.59.084006_EOB01} is a widely used method. The $\texttt{EOBNR}$ model is based on EOB method and resums the results of post Newtonian (PN) approximation. Through the calibration to NR waveforms EOBNR obtains an accurate model of the time domain BBH waveform. Currently, the $\texttt{EOBNR}$ model \cite{PhysRevD.76.104049_EOBNR} has gradually developed to $\texttt{SEOBNRv4}$ model \cite{PhysRevD.86.024011_SEOBNRv1,PhysRevD.89.061502_SEOBNRv2,PhysRevD.95.024010_SEOBNRv3,PhysRevD.95.044028_SEOBNRv4}, which can accurately describe the BBH gravitational wave waveform of general spin-aligned BBH moving along an equatorial quasi-circular orbit.

In the source frame the gravitational wave can be decomposed by a set of spin weighted -2 spherical harmonics respect to different directions. For a two-body system, source frame with $z$ direction pointing to the orbital angular momentum is the most preferred frame choice. In such a frame the quadrupole components contribute most of the gravitational wave energy. In another word, the $(l,|m|)=(2,2)$ spin weighted -2 spherical harmonic components are the dominant modes. Consequently other modes than $(l,|m|)=(2,2)$ are called higher modes. When the difference between the two components masses, the spin precession and/or the orbit eccentricity becomes larger, the impacts of higher-order modes of gravitational waves appears.

For the quasi-circular orbit case, there are already many models describing the waveforms of higher-order modes, which include $\texttt{IMRPhenomHM}$, $\texttt{SEOBNRv4HM}$ \cite{PhysRevLett.120.161102_IMRPhenomHM,PhysRevD.98.084028_SEOBNRv4HM} and others \cite{PhysRevD.102.024077}. So far, most works about the detection of gravitational wave signal and parameter estimation do not consider the influence of the orbit eccentricity (but see \cite{Abbott_2019_searchEcc01,stz2996_searchEcc02,staa1176_searchEcc03,2020arXiv200106492R}). This is because people suspect that the orbit of a BBH system has become circular before it enter the LIGO frequency band. Recently the event GW190521 \cite{PhysRevLett.125.101102_GW190521} makes people rethink the orbit eccentricity problem for ground based detector \cite{Romero_Shaw_2020}.

There are many works investigating the binary system moving along an eccentric orbit. The properties of energy and angular momentum diffusion of eccentric binary system was seminally studied by Peters \cite{PhysRev.136.B1224_Peters}. The post-Newton (PN) waveforms for gravitational radiation have also been widely studied \cite{PhysRevD.91.084040_3PNWFInst,PhysRevD.100.044018_3PNWFtail,PhysRevD.100.084043_3PNWFmem}. In Ref.~\cite{PhysRevD.80.084001_PCmode}, the authors assumed a small eccentricity condition and got a post-circular (PC) waveform model based on the low PN order waveform in frequency domain. Later the PC model was improved by a phenomenological method and developed into an enhanced post-circular (EPC) model \cite{PhysRevD.90.084016_EPCmodel}, which recovers the TaylorF2 model in quasi-circular cases. The x-model \cite{PhysRevD.82.024033_xmodel} is a PN waveform with variable $x\equiv(M\omega)^{3/2}$ inspired by numerical relativity. In \cite{PhysRevD.97.024031_ENIGMA} Eccentric, Nonspinning, Inspiral-Gaussian-process Merger Approximant $(\texttt{ENIGMA})$ model was proposed, which combines analytical and NR results using machine learning algorithms. There are also some studies of elliptical BBH systems based on EOB formalism. In Ref.~\cite{PhysRevD.96.104048_eobecc}, an EOB eccentric waveform model was constructed based on an adiabatic approximation rather than NR calibration. We proposed $\texttt{SEOBNRE}$ \cite{PhysRevD.96.044028_SEOBNRE} waveform model based on $\texttt{SEOBNRv1}$ through adding elliptical correction terms to the EOB factorized waveform. $\texttt{SEOBNRE}$ model has been validated to NR catalog in \cite{PhysRevD.101.044049_validSEOBNRE}. The authors of \cite{PhysRevD.101.101501_TEOBeccc} modified $\texttt{TEOBiResumS\_SM}$, a highly NR-faithful EOB multipolar waveform model for quasi-circular orbits, to incorporate orbit eccentricity. They found that such a waveform model can also work for hyperbolic encounter BBHs \cite{Nagar:2020xsk,Nagar:2021gss}.

The existing $\texttt{SEOBNRE}$ waveform model admits only $(2,2)$ mode. In this paper, we resum the higher modes of gravitational waveforms for an equatorial orbit and combine them with the existing higher PN order EOB factorized waveforms to obtain a waveform model of higher modes for $\texttt{SEOBNRE}$. At the mean time some improvements on $\texttt{SEOBNRE}$ model have been implemented. For convenient reference we call the new waveform model proposed in the current work $\texttt{SEOBNREHM}$ to distinguish the previous $\texttt{SEOBNRE}$ model. In Sec.~\ref{secII}, we will show the construction details of our new waveform. Then in Sec.~\ref{secIII} we combine this new waveform with the dynamical system of $\texttt{SEOBNRv4}$ to get $\texttt{SEOBNREHM}$ model. Then we verify that the $(l,m)=(2,2)$ mode of our new model waveform can match the waveform of $\texttt{SEOBNRv4}$ well in the case of quasi-circular orbits. Meanwhile when comparing $(l,m)=(2,2)$ against NR catalogs, for both quasi-circular orbits cases and elliptical orbits cases, the fitting factors are bigger than 98\%. In Sec.~\ref{secIV} we test the performance of the higher modes in both quasi-circular and elliptical cases, we find that the matching factor between our model waveform and the NR waveform is bigger than 98\%. Based on our new waveform model we analyze the power fraction of higher modes in the Sec.~\ref{secou}. A summary is given in the last section. We adopt the geometric units $c=G=1$ throughout this paper.

\section{Effective-one-body multipolar waveform model for eccentric binary black hole}\label{secII}
\subsection{Effective-one-body dynamics}
For a binary system with component masses $m_1$ and $m_2$, the EOB conservative dynamics for the binary can be described by the Hamiltonian \cite{PhysRevD.59.084006_EOB01}
\begin{align}
H_{\text{EOB}} = M\sqrt{1+2\nu \left( \frac{H_{\text{eff}}}{\mu}-1 \right)}, \label{EOBE}
\end{align}
where $\mu=m_1m_2/M$ is the reduced mass, $\nu=\mu/M$ is the symmetric mass ratio and $M=m_1+m_2$ is the total mass. For a spin-aligned BBH we are concerned, we denote the spin angular momentums by $\boldsymbol{S}_i=\chi_i m_i^2\hat{\boldsymbol{L}}$, where $\chi_i$ is dimensionless spin parameters and $\hat{\boldsymbol{L}}=\boldsymbol{L}/|\boldsymbol{L}|$ is the direction of the orbital angular momentum. Here $i=1,2$ means the index of the binary components and we take the convention $m_1\geq m_2$. The explicit expression of the effective Hamiltonian $H_{\text{eff}}$ has been derived in \cite{PhysRevD.81.084024_EOBHam01,PhysRevD.84.104027_EOBHam02}. In the EOB framework, the spin-aligned orbital evolution is described by orbit phase $\phi$, radial distance $r$, and the corresponding canonical momentums $p_\phi$ and $p_r$. The other four parameters $(K, d_{\text{SO}}, d_{\text{SS}}, \delta^{22}_{\text{peak}})$ can be determined through calibration against numerical relativity simulations \cite{PhysRevD.95.044028_SEOBNRv4}. Here we adopt the conservative dynamics, i.e., the Hamiltonian from $\texttt{SEOBNRv4}$ model.

The gravitational waves would take away energy and angular momentum. Such effect can be described by the radiation-reaction force \cite{PhysRevD.95.044028_SEOBNRv4}. For elliptical orbits, we set the initial conditions of the orbital evolution as follows
\begin{align}
\pdv{H}{\hat{p}_\phi}&=\frac{\pi}{f_0},\label{IC1} \\
\pdv{H}{r} &= -\frac{e_0}{r^2},\label{IC2}
\end{align}
where $\hat{p}=p/\mu$, and $e_0\in(0,1)$ is the initial eccentricity. When $e_0=0$, the orbit is circular. When $e_0>0$ the orbit starts from the perihelion point. The procedure to solve the above equations for the initial condition is the same to the one described in the sec.~IV of \cite{PhysRevD.74.104005_IC}.
\subsection{Effective-one-body gravitational waveform modes}
The direction of the observer respect to the gravitational wave source frame corresponds to the angles $(\iota,\varphi_c)$ where the inclination angle $\iota$ represent the angle between the line of sight and the direction of the angular momentum of the binary system, and $\varphi_c$ denotes coalescence orbital phase. Consequently the gravitational wave can be decomposed by spin-weighted -2 spherical harmonics as
\begin{align}
h(\iota, \varphi_c; t) = & h_+(\iota, \varphi_c; t) - i h_\times(\iota, \varphi_c; t)\nonumber \\
= &\sum_{l=2}^{\infty} \sum_{m=-l}^{l} \,_{-2}Y_{lm}(\iota, \varphi_c)h_{lm}(t) \label{hPC}
\end{align}
Following \cite{PhysRevD.62.064015_EOBIMR}, we express the full inspiral-merger-ringdown EOB waveform modes as
\begin{equation}
h_{lm}^{\text{IMR}}(t) =
\begin{cases}
h_{lm}^{\text{insp}}(t), & t < t_{\text{match}}^{lm} \\
h_{lm}^{\text{RD}}(t), & t > t_{\text{match}}^{lm}
\end{cases}\label{WF}
\end{equation}
The definition of $t_{\text{match}}^{lm}$, and the ringdown waveform modes $h_{lm}^{\text{RD}}$ can be found in \cite{PhysRevD.98.084028_SEOBNRv4HM}.

The quasi-circular inspiral-plunge modes $h_{lm}^{\text{insp}}$ can be written in a factorized form \cite{PhysRevD.79.064004_FWF01,FWF02}
\begin{equation}
h_{lm}^{\text{insp}} = h^\text{F}_{lm} N_{lm} \label{FWF01},
\end{equation}
where $h^\text{F}_{lm}$ is a resummation of post-Newtonian waveform for quasi-circular orbit \cite{PNWF00,PhysRevD.79.104023_PNWF01, PhysRevD.87.044009_PNWF02}. The non-quasi-circular (NQC) term $N_{lm}$ is used to adjust EOB waveform modes in the late inspiral and merger phase. The Ref.~\cite{PhysRevD.95.044028_SEOBNRv4} has determined  $N_{22}$ as
\begin{align}
N_{22}=&\left[ 1+\left(\frac{\hat{p}_{r*}}{r\Omega}\right)^2\left(a_1^{h_{22}} + \frac{a_2^{h_{22}}}{r} + \frac{a_3^{h_{22}}}{r^{3/2}} \right) \right] \nonumber\\
&\times\exp{\frac{i\hat{p}_{r*}}{r\Omega} \left( b_1^{h_{22}} + b_2^{h_{22}}\hat{p}_{r*} \right)} \label{NQCV4}
\end{align}
where $\Omega=\dot{\phi}$ and $\hat{p}_{r*}$ is the tortoise radial angular momentum. The coefficients $a^{h_{22}}$ and $b^{h_{22}}$ will be determined by fitting amplitude, frequency and their derivatives of $h_{lm}^{\text{insp}}$ at $t_{\text{match}}^{lm}$ to the values predicted by NR simulations.

In a quasi-circular case, the prefactor $\hat{p}_{r*}/(r\Omega)$ is small during early inspiral. Consequently $N_{22}$ is about 1 and the NQC does not take effect. For an elliptical orbit $\hat{p}_{r*}/(r\Omega)$ may be quite large near the perihelion point, even for early inspiral. In order to make sure the NQC term takes effect only near merger, we introduce a window function to limit the scope of this term as
\begin{align}
N_{lm}(t)=&\left[ 1+\left(\mathcal{W}(t)\frac{\hat{p}_{r*}}{r\Omega}\right)^2\left(a_1^{h_{lm}} + \frac{a_2^{h_{lm}}}{r} + \frac{a_3^{h_{lm}}}{r^{3/2}} \right) \right] \nonumber\\
&\times\exp{\mathcal{W}(t)\frac{i\hat{p}_{r*}}{r\Omega} \left( b_1^{h_{lm}} + b_2^{h_{lm}}\hat{p}_{r*} \right)},\label{NQCNew}
\end{align}
where the window function $\mathcal{W}(t)$ reads
\begin{align}
\mathcal{W}(t) &= \frac{1}{1+\exp{\left[-\bar{\omega} (t-t_w)\right]}}.
\end{align}
The $t_w$ denotes the location of the window. Since $p_r=0$ at perihelion and aphelion, $p_r$ can be used as an indicator where the test particle locates respect to the orbit. We choose $t_w$ equals to the time after which $p_r<0$ until merge. In another word, we search backwards in time from $t_{\text{match}}^{lm}$ to the time where $p_r=0$ and set it as our $t_w$. When $e_0=0$ which corresponds to the quasi-circular case, the $p_r<0$ for the whole time. Consequently we have $t_w=-\infty$ under these circumstances, which means $\mathcal{W}(t) \equiv 1$, and our NQC recovers original NQC term in $\texttt{SEOBNRv4}$ model. The another parameter $\bar{\omega}$ controls the width of the window function. In principle this adjustable parameter should be determined by the calibration to the numerical relativity simulations. In the current work we did not do such extensive investigation. We instead choose $\bar{\omega}=\frac{\ln2}{3M}$ based on the comparison to the 24 elliptical BBH simulations of SXS. Different choices of $\bar{\omega}$ may result in a different matching factor. But the matching factor difference is less than 1\%.

\subsection{The factorized gravitational waveform modes for an eccentric orbit}

\begin{figure*}[t]
\centering
\begin{tabular}{cc}
\includegraphics[width=0.5\textwidth]{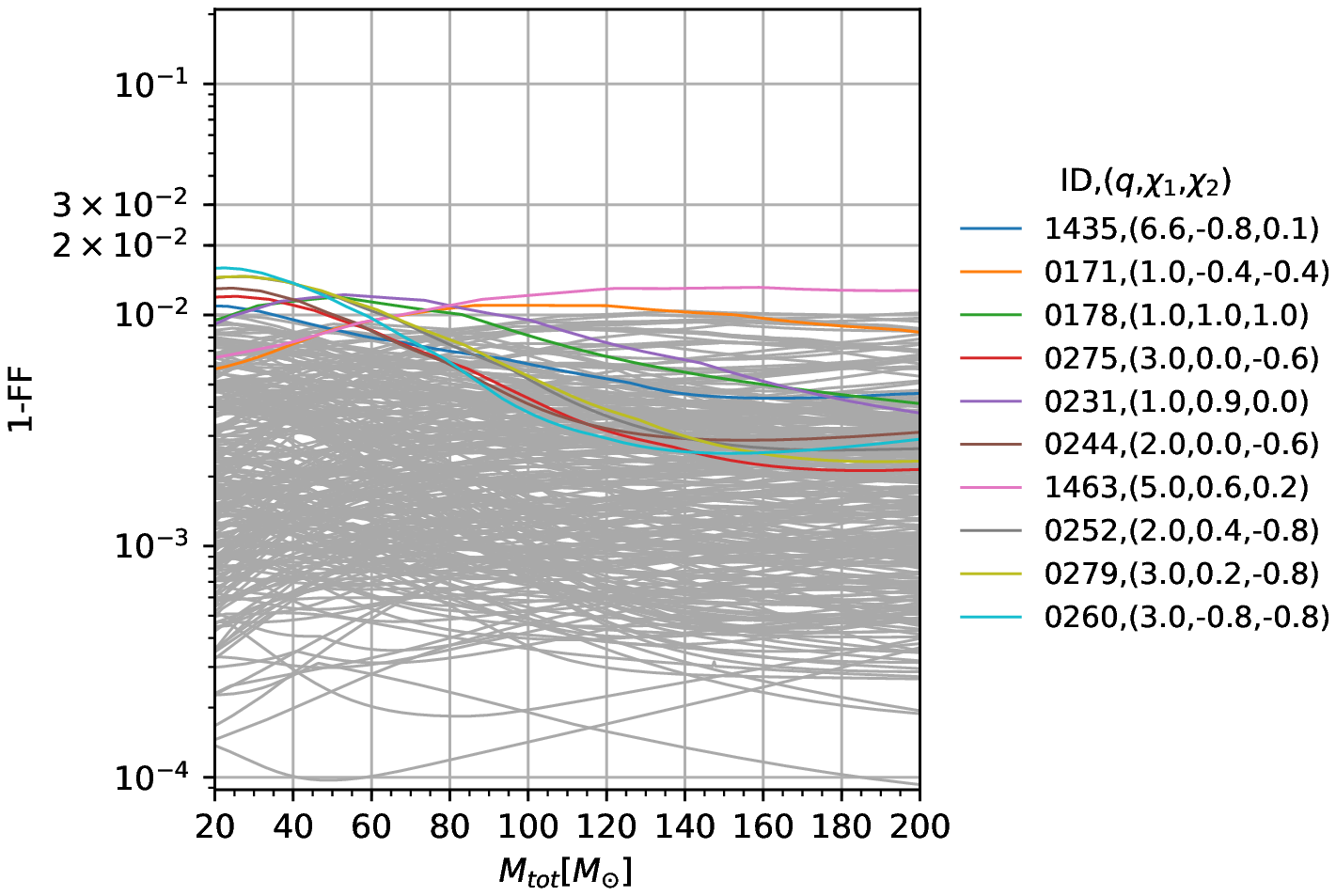}&
\includegraphics[width=0.5\textwidth]{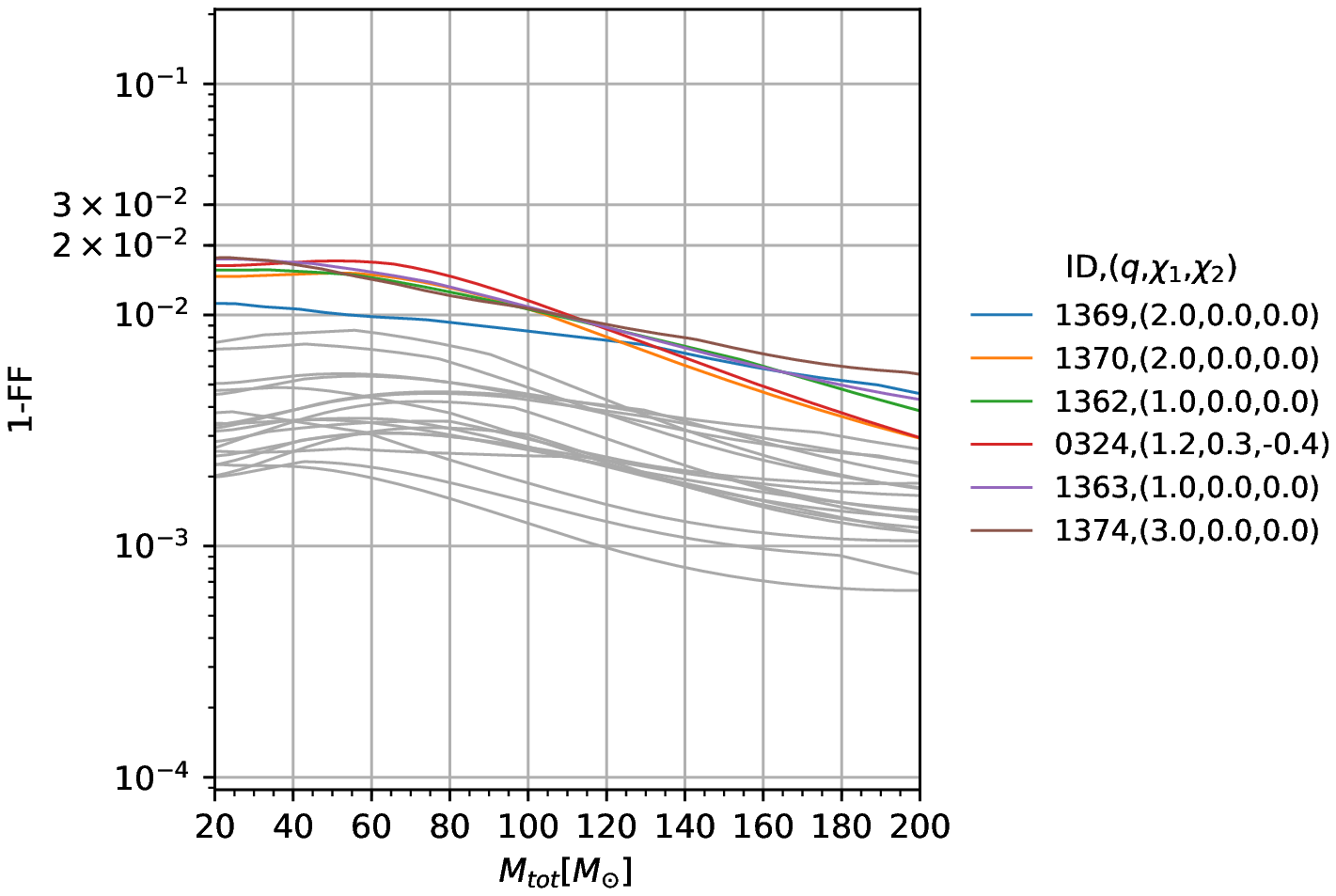}
\end{tabular}
\caption{The matching factor ${\rm FF}$ of the $(l,m)=(2,2)$ mode waveform for the spin-aligned BBHs between the new waveform model $\texttt{SEOBNREHM}$ proposed in the current work and the SXS catalog \cite{SXSBBH}. The Advanced LIGO designed PSD is used in this plot. We highlight the ones whose matching factor is less than 99\%. Left panel: 281 quasi-circular cases. Right panel: 24 eccentric cases.}\label{fig1}
\end{figure*}

\begin{figure*}[t]
\centering
\begin{tabular}{c}
\includegraphics[width=\textwidth]{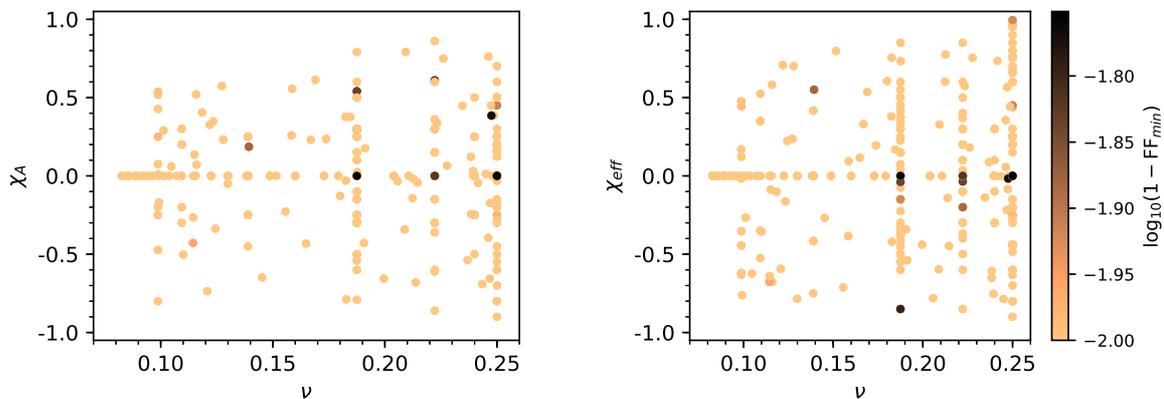}
\end{tabular}
\caption{Matching factors of the $(l,m)=(2,2)$ mode waveform for the eccentric spin-aligned BBH between the new waveform model $\texttt{SEOBNREHM}$ proposed in the current work and the SXS catalog \cite{SXSBBH}. The color means the minimal matching factor respect to the total mass in the range $(20,200)M_\odot$. This figure corresponds to the right panel of the Fig.~\ref{fig1}.}\label{fig2}
\end{figure*}
The factorized resummation of waveform modes proposed in \cite{PhysRevD.79.064004_FWF01,FWF02} are composed of $\nu\neq0$ circular waveform terms till 3PN and $\nu=0$ circular waveform terms till 5PN. In this work, we replace the 2PN order circular parts with the 2PN general equatorial form \cite{PhysRevD.96.044028_SEOBNRE}. Compare to the original $\texttt{SEOBNRE}$ waveform model where direct PN waveform for 2PN terms is used, this resumed waveform can hopefully improve the performance. And as we will show in the following part of the current paper, this resummation form improves quite much.

The factorized waveform modes read \cite{PhysRevD.79.064004_FWF01,FWF02}
\begin{align}
h^\text{F}_{lm} &= h_{lm}^{(N,\epsilon)}\hat{S}_{\text{eff}}^{(\epsilon)}T_{lm}e^{i\delta_{lm}}f_{lm}\label{FWF01}
\end{align}
where $\epsilon=0$ when $l+m$ is even, otherwise $\epsilon=1$. The amplitude correction factor $f_{lm}$ is resummed by $f_{lm}=(\rho_{lm})^l$ when $l$ is even and $f_{lm}=(\rho_{lm}^{\text{NS}})^l+f^{\text{S}}_{lm}$ when $l$ is odd. The explicit expression of these terms can be found in \cite{PhysRevD.79.064004_FWF01,FWF02}.

Quasi-circular Newtonian waveform modes $h_{lm}^{(N,\epsilon)}$ are controlled by the parameter $x\equiv (M\Omega)^{2/3}$. We generalize them to eccentric orbit cases \cite{PhysRevD.54.4813_2PNWF, PhysRevD.91.084040_3PNWFInst, PhysRevD.79.104023_PNWF01, PhysRevD.87.044009_PNWF02}
\begin{align}
h_{lm}^{N}=&\frac{M\nu}{R}n^{(\epsilon)}_{lm} c_{l+\epsilon}(\nu) Y^{l,-m}(\frac{\pi}{2},\phi_h)\hat{h}_{lm}\label{NewtWF02} \\
\hat{h}_{22}=&\frac{y_0^2}{2}-\frac{y_1^2}{2}+\frac{y_2^2}{2}+i y_1 y_2 \label{Newt22} \\
\hat{h}_{21}=&y_0^2y_2 \label{Newt21} \\
\hat{h}_{33}=&-\frac{2}{9} i y_1^3-\frac{2}{3} y_2 y_1^2+\frac{4}{9} i y_0^2 y_1\nonumber \\
&+\frac{2}{3} i y_2^2 y_1+\frac{2 y_2^3}{9}+\frac{7}{9} y_0^2 y_2 \label{Newt33} \\
\hat{h}_{44}=&\frac{7 y_0^4}{64}-\frac{9}{32} y_1^2 y_0^2+\frac{51}{64} y_2^2 y_0^2\nonumber \\
&+\frac{27}{32} i y_1 y_2 y_0^2+\frac{3 y_1^4}{32}\nonumber\\
&+\frac{3 y_2^4}{32}+\frac{3}{8} i y_1 y_2^3-\frac{9}{16} y_1^2 y_2^2-\frac{3}{8} i y_1^3 y_2 \label{Newt44}
\end{align}
The parameters $y_0$, $y_1$ and $y_2$ are defined as
\begin{equation}
y_0=\sqrt{\frac{1}{r_h}},\;
y_1=\frac{\boldsymbol{r}_h\cdot\boldsymbol{v}_h}{r_h},\;
y_2=r_h \Omega_h,\label{yDef}
\end{equation}
where the subscript $h$ means harmonic coordinates, used to distinguish it from the coordinates in EOB dynamics. Corresponding to the above harmonic coordinates, we can also use EOB coordinates to define three new variables
\begin{equation}
x_0=\sqrt{\frac{1}{r}},\;
x_1=\frac{\boldsymbol{r}\cdot\boldsymbol{v}}{r},\;
x_2=r\Omega.\label{xDef}
\end{equation}

Following \cite{PhysRevD.86.124012_GaugeTransform}, we apply a 2PN gauge transformation from the harmonic coordinates $(y_0,y_1,y_2,\phi_h)$ to the EOB coordinates $(x_0,x_1,x_2,\phi)$ (see the Appendix.~\ref{appendixA}). We replace the Newtonian waveform (\ref{NewtWF02}) with
\begin{align}
h_{lm}^{(N,\epsilon,e)} = \frac{M\nu}{R}n^{(\epsilon)}_{lm}c_{l+\epsilon}(\nu)Y^{l-\epsilon,-m}\left(\frac{\pi}{2}, \phi \right) x_0^{l+\epsilon}. \label{NewtWFNew}
\end{align}

We keep original factors $\hat{S}_{\text{eff}}^{(\epsilon)}$, $T_{lm}$ and $e^{i\delta_{lm}}$ unchanged, while change $f_{lm}$ in the following way. We introduce a Taylor expansion form
\begin{align}
\zeta_{lm}=c_0^{h_{lm}}+c_1^{h_{lm}} x_0 + c_2^{h_{lm}} x_0^2 + ...
\end{align}
We list the explicit expression of $c^{h_{lm}}_{0,1,2}$ in the Appendix.~\ref{appendixB}. The above Taylor expansion has poor convergence near merger. So we resum it by $(0,4)$ Pade approximation
\begin{equation}
f^{e}_{lm} =
\begin{cases}
P^0_4\left[\zeta_{lm}\right], & l\text{\ is\ even} \\
P^0_4\left[\zeta_{lm}^{\text{NS}}\right]+\zeta^{\text{S}}_{lm}, & l\text{\ is\ odd}
\end{cases}\label{f2PN}
\end{equation}
Here NS (Non-Spin) means dropping out the spin terms, while S (Spin) means solely spin terms. Explicitly $P^0_4$ means
\begin{align}
&P_4^0[A_0+A_1x+A_2x^2+A_3x^3+A_4x^4]\nonumber\\
&=\frac{A_0}{1+B_1x+B_2x^2+B_3x^3+B_4x^4}\\
B_1&=-\frac{A_1}{A_0} \\
B_2&=\frac{A_1^2-A_0A_2}{A_0^2}\\
B_3&=\frac{-A_1^3+2A_0A_1A_2-A_0^2A_3}{A_0^3}\\
B_4&=\frac{A_1^4-3A_0A_1^2A_2+A_0^2A_2^2+2A_0^2A_1A_3-A_0^3A_4}{A_0^4}.
\end{align}
We replace the 2PN resummed parts in the $\texttt{SEOBNRv4HM}$ waveform with the new 2PN form. The full factorized waveform modes are
\begin{align}
h^\text{F}_{lm} &= h^{(\text{F}, e)}_{lm} + h^{(\text{F}, c)}_{lm}\label{FullWF} \\
h^{(\text{F}, e)}_{lm} &= h_{lm}^{(N,\epsilon, e)}\hat{S}_{\text{eff}}^{(\epsilon)}T_{lm}e^{i\delta_{lm}}f_{lm}^e\label{FullWF2PN} \\
h^{(\text{F}, c)}_{lm}&=h_{lm}^{(N,\epsilon)}\hat{S}_{\text{eff}}^{(\epsilon)}T_{lm}e^{i\delta_{lm}} \left( f_{lm}-f_{lm}^{\text{2PN}} \right),
\end{align}
where $f^{\text{2PN}}_{lm}$ is
\begin{equation}
f^{\text{2PN}}_{lm} =
\begin{cases}
\left(\rho_{lm}^{\text{2PN}}\right)^l, & l\text{\ is\ even} \\
\left(\rho_{lm}^{\text{NS}_{\text{2PN}}}\right)^l+f^{\text{S}_{\text{2PN}}}_{lm}, & l\text{\ is\ odd}
\end{cases}\label{f2PN}
\end{equation}
The index 2PN represents 2PN parts.

\section{The (2,2) mode performance of $\texttt{SEOBNREHM}$}\label{secIII}
Given two waveforms $h_1(t)$ and $h_2(t)$, we define matching factor (FF) by inner product $\langle h_1|h_2\rangle$ with respect to a detector noise
\begin{align}
\langle h_1|h_2\rangle &=4\mathcal{R}\int_{f_{\text{min}}}^{f_{\text{max}}} \frac{\tilde{h}_1(f) \tilde{h}_2^*(f)}{S_n(f)} \dd f \nonumber \\
\text{FF} &\equiv \max\limits_{t_c,\varphi_c} \frac{\braket{ h_1 }{ h_2} }{\sqrt{\braket{ h_1}  \braket{ h_2} }} \label{FFeq}
\end{align}
where $S_n(f)$ is the one-sided power spectral density (PSD) of the detector noise. The detector frequency band $(f_{\text{min}}, f_{\text{max}}) = (10\text{Hz}, 8192\text{Hz})$ we used here is the designed sensitivity of the advanced LIGO. Specifically the detuned high power PSD is used \cite{Sho10}.

\subsection{Comparison with numerical relativity waveforms}
In this subsection we compare our new $(2,2)$ mode waveforms to the numerical relativity results catalog \cite{SXSBBH}. Here we use the BBH numerical waveform from Simulating eXtreme Spacetime (SXS) generated by Spectral Einstein code $(\texttt{SpEC})$ \cite{SXS01, SXS02}. Since the numerical relativity results can be scaled to any total mass, we plot the comparison results respect to different total masses in the left panel of the Fig.~\ref{fig1} for the 281 quasi-circular BBHs listed in the Appendix.~\ref{CIRCtable}. Each line corresponds to a parameter combination $\nu$, $\chi_{1z}$ and $\chi_{2z}$. Most matching factors are greater than 99\%. Only 10 parameter combinations among the 281 combinations in all admit matching factor in the range (98\%,99\%) for some total masses. These 10 cases have been highlighted out in the left panel of the Fig.~\ref{fig1}. We find out the minimal matching factor ${\rm FF}_{\rm 22min}$ respect to the total mass in the range $M\in[20, 200]M_\odot$ for each line and list them respect to the parameter combinations in the Appendix.~\ref{CIRCtable}.

For elliptical cases, we firstly estimate the initial eccentricity based on the first orbit of NR simulation through
\begin{equation}
e_{\text{NR}} = r_i^2 \ddot{r}|_{r=r_i} \label{NRecc}
\end{equation}
where subscript $i$ means the initial perihelion. Regarding to our $\texttt{SEOBNREHM}$ model we set the reference frequency to the frequency when the eccentricity is estimated by the Eq.~(\ref{NRecc}) and search the initial eccentricity which maximizes the matching factor between the our model (2,2) mode waveform and the numerical relativity waveform. We should notice that the eccentricity defined by the Eq.~(\ref{NRecc}) equivalents to the eccentricity defined in the Eq.~(\ref{IC2}) only at Newtonian order. Similar analysis has been done before \cite{PhysRevD.96.044028_SEOBNRE,PhysRevD.101.044049_validSEOBNRE}. This fact explains part of the deviations of the $\texttt{SEOBNREHM}$ estimated eccentricity $e_{\rm EOB}$ from the NR estimated one $e_{\rm NR}$. We have listed the results in the Table.~\ref{table1}. We plot the (2,2) mode comparison results in the right panel of the Fig.~\ref{fig1}. At the mean time we find out the minimal matching factor for each line and plot the results in the Fig.~\ref{fig2}. The minimal matching factors are also listed in the Table.~\ref{table1}.

\begin{table*}[t]
\caption{SXS Numerical relativity waveforms of eccentric BBH used for the waveform validation in the current work. ID corresponds to the id number in the SXS catalog. $f_o$ is the orbital frequency at the first perihelion from the NR simulation. $e_{\rm NR}$ is the estimated initial eccentricity by the Eq.~(\ref{NRecc}). $e_{\rm EOB}$ is the initial eccentricity at reference frequency $Mf_0=2Mf_o$ of our $\texttt{SEOBNREHM}$ model which maximizes the matching factor between the $\texttt{SEOBNREHM}$ waveform and the NR waveform for (2,2) mode. Matching factors between the $\texttt{SEOBNEHM}$ waveform model and the SXS Numerical relativity waveforms are also listed here. ${\rm FF}_{\rm 22min}$ means the minimal matching factor for the (2,2) mode respect to the total mass. $\overline{\rm FF}$ and ${\rm FF}_{\rm min}$ are respectively the averaged matching factor and the minimal matching factor respect to the three angles $(\kappa, \iota, \varphi)$ (check the Sec.~\ref{secIV} for detail explanation).}
\begin{center}
\begin{ruledtabular}
\begin{tabular}{cccccccccc}
ID & $q$ & $\chi_1$ & $\chi_2$ & $Mf_o$ & $e_{\rm NR}$ & $e_{\rm EOB}$ & FF${}_{\rm 22min}$ & $\overline{\rm FF}$ & ${\rm FF}_{\rm min}$\\
\hline \hline
1355 & 1.00 & 0.00 & 0.00 & $4.98\times10^{-3}$ & 0.047 & 0.065 & $99.76\%$ & $99.85\%$ & $99.74\%$ \\
1356 & 1.00 & 0.00 & 0.00 & $5.66\times10^{-3}$ & 0.083 & 0.114 & $99.51\%$ & $99.63\%$ & $99.56\%$ \\
1357 & 1.00 & 0.00 & 0.00 & $7.30\times10^{-3}$ & 0.085 & 0.124 & $99.46\%$ & $99.61\%$ & $99.53\%$ \\
1358 & 1.00 & 0.00 & 0.00 & $7.46\times10^{-3}$ & 0.084 & 0.122 & $99.58\%$ & $99.60\%$ & $99.53\%$ \\
1359 & 1.00 & 0.00 & 0.00 & $7.57\times10^{-3}$ & 0.083 & 0.120 & $99.53\%$ & $99.56\%$ & $99.47\%$ \\
1360 & 1.00 & 0.00 & 0.00 & $8.53\times10^{-3}$ & 0.114 & 0.167 & $99.25\%$ & $99.15\%$ & $99.10\%$ \\
1361 & 1.00 & 0.00 & 0.00 & $8.61\times10^{-3}$ & 0.115 & 0.168 & $99.14\%$ & $99.15\%$ & $99.05\%$ \\
1362 & 1.00 & 0.00 & 0.00 & $9.81\times10^{-3}$ & 0.146 & 0.219 & $98.43\%$ & $98.30\%$ & $98.25\%$ \\
1363 & 1.00 & 0.00 & 0.00 & $9.87\times10^{-3}$ & 0.147 & 0.219 & $98.24\%$ & $98.21\%$ & $98.15\%$ \\
\hline
1364 & 2.00 & 0.00 & 0.00 & $6.34\times10^{-3}$ & 0.037 & 0.039 & $99.65\%$ & $99.59\%$ & $99.46\%$ \\
1365 & 2.00 & 0.00 & 0.00 & $6.61\times10^{-3}$ & 0.050 & 0.054 & $99.64\%$ & $99.64\%$ & $99.50\%$ \\
1366 & 2.00 & 0.00 & 0.00 & $7.36\times10^{-3}$ & 0.080 & 0.118 & $99.67\%$ & $99.56\%$ & $99.42\%$ \\
1367 & 2.00 & 0.00 & 0.00 & $7.43\times10^{-3}$ & 0.080 & 0.117 & $99.44\%$ & $99.53\%$ & $99.39\%$ \\
1368 & 2.00 & 0.00 & 0.00 & $7.50\times10^{-3}$ & 0.079 & 0.116 & $99.54\%$ & $99.49\%$ & $99.33\%$ \\
1369 & 2.00 & 0.00 & 0.00 & $9.55\times10^{-3}$ & 0.143 & 0.217 & $98.87\%$ & $98.69\%$ & $98.47\%$ \\
1370 & 2.00 & 0.00 & 0.00 & $9.72\times10^{-3}$ & 0.139 & 0.239 & $98.48\%$ & $98.54\%$ & $98.40\%$ \\
\hline
1371 & 3.00 & 0.00 & 0.00 & $6.58\times10^{-3}$ & 0.047 & 0.052 & $99.74\%$ & $99.65\%$ & $99.41\%$ \\
1372 & 3.00 & 0.00 & 0.00 & $7.34\times10^{-3}$ & 0.076 & 0.112 & $99.69\%$ & $99.78\%$ & $99.73\%$ \\
1373 & 3.00 & 0.00 & 0.00 & $7.40\times10^{-3}$ & 0.076 & 0.112 & $99.65\%$ & $99.60\%$ & $99.52\%$ \\
1374 & 3.00 & 0.00 & 0.00 & $9.49\times10^{-3}$ & 0.135 & 0.230 & $98.23\%$ & $99.08\%$ & $98.98\%$ \\
\hline
0321 & 1.22 & +0.33 & -0.44 & $4.96\times10^{-3}$ & 0.040 & 0.054 & $99.77\%$ & $99.23\%$ & $98.55\%$ \\
0322 & 1.22 & +0.33 & -0.44 & $6.53\times10^{-3}$ & 0.048 & 0.051 & $99.62\%$ & $99.21\%$ & $98.51\%$ \\
0323 & 1.22 & +0.33 & -0.44 & $7.20\times10^{-3}$ & 0.078 & 0.112 & $99.69\%$ & $99.33\%$ & $98.88\%$\\
0324 & 1.22 & +0.33 & -0.44 & $9.35\times10^{-3}$ & 0.143 & 0.240 & $98.29\%$ & $98.44\%$ & $98.31\%$
 \label{table1}
 \end{tabular}
 \end{ruledtabular}
 \end{center}
\end{table*}

\begin{figure}
\begin{tabular}{c}
\includegraphics[width=0.5\textwidth]{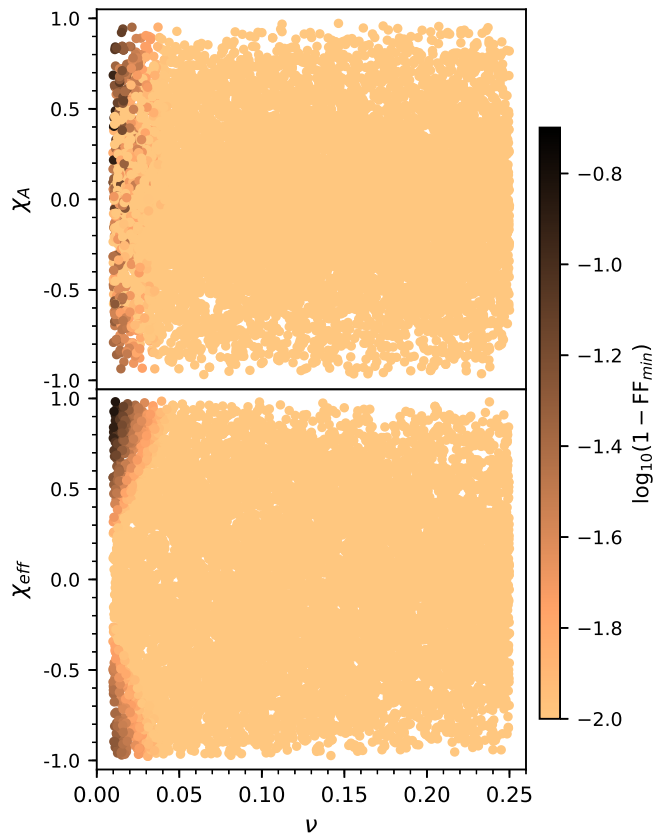}
\end{tabular}
\caption{Matching factors of the $(l,m)=(2,2)$ mode waveform for the quasi-circular spin-aligned BBH between the new waveform model proposed in the current work and $\texttt{SEOBNRv4}$. Uniformly distributed 10000 points in the parameter space $\nu$, $\chi_{1z}$ and $\chi_{2z}$ are chosen. The plot convention is the same to that of the Fig.~\ref{fig2}.}\label{fig3}
\end{figure}
\subsection{Comparison with $\texttt{SEOBNRv4}$}
$\texttt{SEOBNRv4}$ waveform model is one of the most widely used waveform model by LIGO data analysis. In this subsection, we compare our new $(2,2)$ mode waveforms to the corresponding ones given by $\texttt{SEOBNRv4}$ \cite{PhysRevD.95.044028_SEOBNRv4} for quasi-circular spin-aligned BBHs. We uniformly chose 10000 samples at random from the parameter space of symmetric mass ratio $\nu\in[0,0.25]$, black hole's dimensionless spin $\chi_{1z}\in[-0.99,0.99]$, $\chi_{2z}\in[-0.99,0.99]$. For each chosen $\nu$, $\chi_{1z}$ and $\chi_{2z}$ we scan the total mass in the range $M\in[20, 200]M_\odot$ to search the minimal matching factor ${\rm FF}_{\rm min}$ respect to the total mass. We plot the ${\rm FF}_{\rm min}$ respect to different mass ratio and black hole spin in the Fig.~\ref{fig3}. For convenient comparison to the figures in \cite{PhysRevD.95.044028_SEOBNRv4} we have used variables $\chi_A=(\chi_1-\chi_2)/2$ and $\chi_{\text{eff}}=(m_1\chi_1 + m_2 \chi_2)/M$ in the plot.

\begin{figure}
\begin{tabular}{c}
\includegraphics[width=0.5\textwidth]{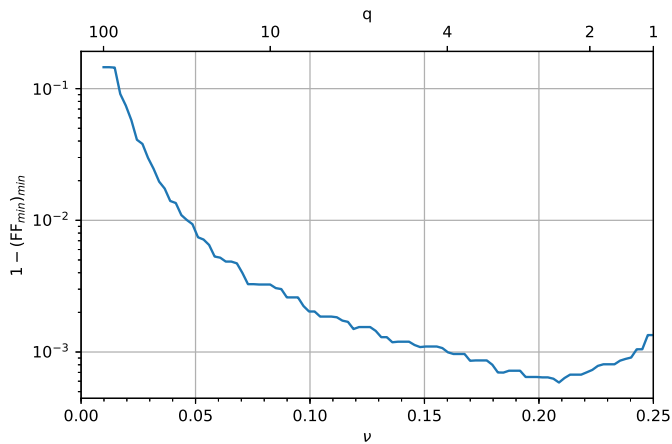}
\end{tabular}
\caption{The dependence of the minimal ${\rm FF}_{\rm min}$ respect to black holes' spin on the mass ratio, where the subscript min in ${\rm FF}_{\rm min}$ means the minimum respect to the total mass ranged in (20,200)M${}_\odot$. The minimal operation respect to black holes' spin is done based on the data shown in the Fig.~\ref{fig3}.}\label{fig4}
\end{figure}

The above comparison results indicate that our new waveform model can recover $\texttt{SEOBNRv4}$ very well in the case of quasi-circular spin-aligned BBH. Especially the fitting factor between our new waveform model $\texttt{SEOBNREHM}$ and the $\texttt{SEOBNRv4}$ waveform model is bigger than 99.9\% for spinless BBHs. At the mean time we find that the worst fitting factor respect to the black holes' spin depends on the mass ratio. In the Fig.~\ref{fig4} we plot such dependence between $\min({\rm FF}_{\rm min})\equiv\min_{\chi_{1z},\chi_{2z}}({\rm FF}_{\rm min})$ and the mass ratio $\nu$ ($q\equiv\frac{m_1}{m_2}\geq1$). When $\nu\gtrsim0.05$ ($q\lesssim20$) the fitting factor is bigger than 99\%; when $\nu\gtrsim0.01$ ($q\lesssim60$) the fitting factor is bigger than 90\%.

\subsection{Comparison with $\texttt{SEOBNRE}$}
\begin{figure*}[t]
\centering
\begin{tabular}{c}
\includegraphics[width=\textwidth]{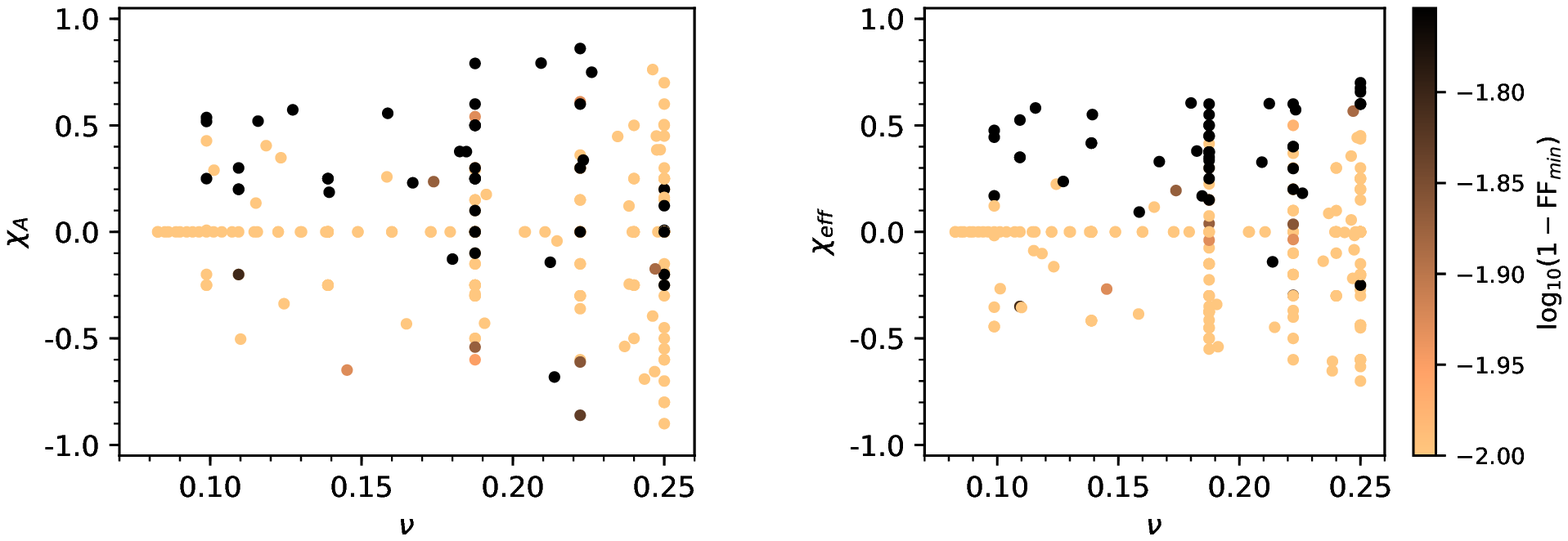}\\
\includegraphics[width=\textwidth]{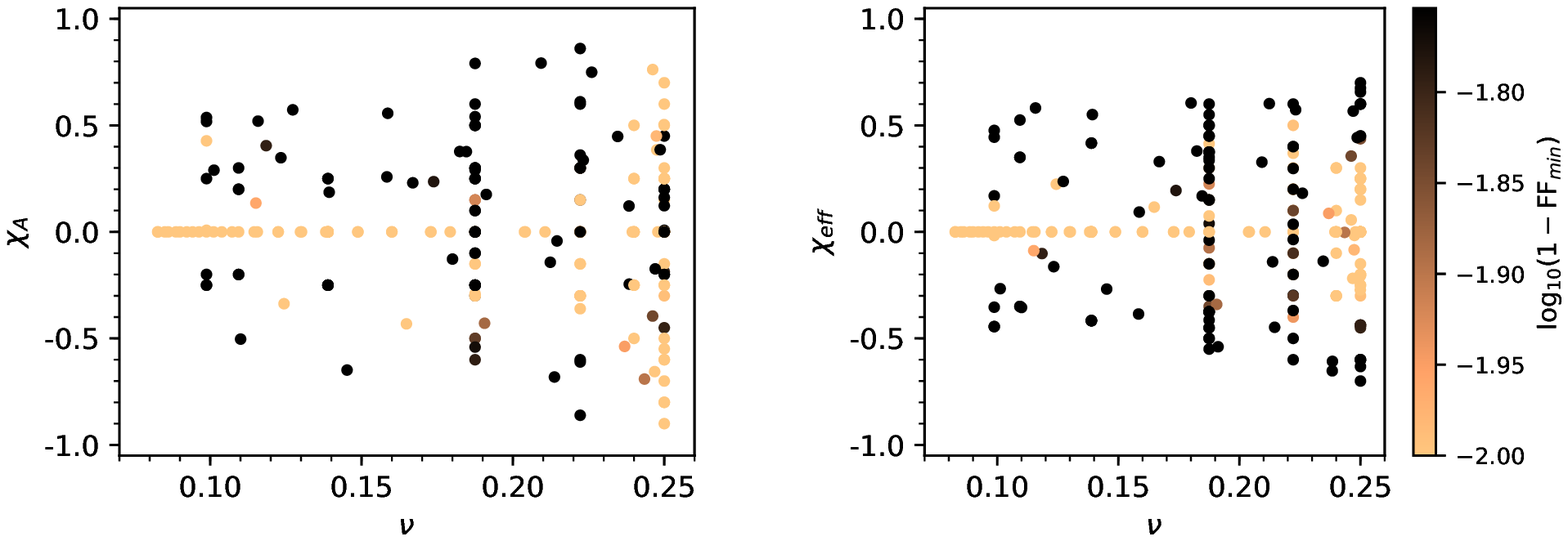}
\end{tabular}
\caption{Matching factors of the $(l,m)=(2,2)$ mode waveform for the eccentric spin-aligned BBH for the $\texttt{SEOBNRE}$ waveform model. Top row is for the matching factors between the $\texttt{SEOBNRE}$ waveform model and the SXS catalog \cite{SXSBBH}. The bottom row is for the matching factors between the $\texttt{SEOBNRE}$ waveform model and the $\texttt{SEOBNREHM}$ waveform model. The plot convention is the same to the Fig.~\ref{fig2}.}\label{fig5}
\end{figure*}
\begin{figure*}[t]
\begin{tabular}{c}
\includegraphics[width=\textwidth]{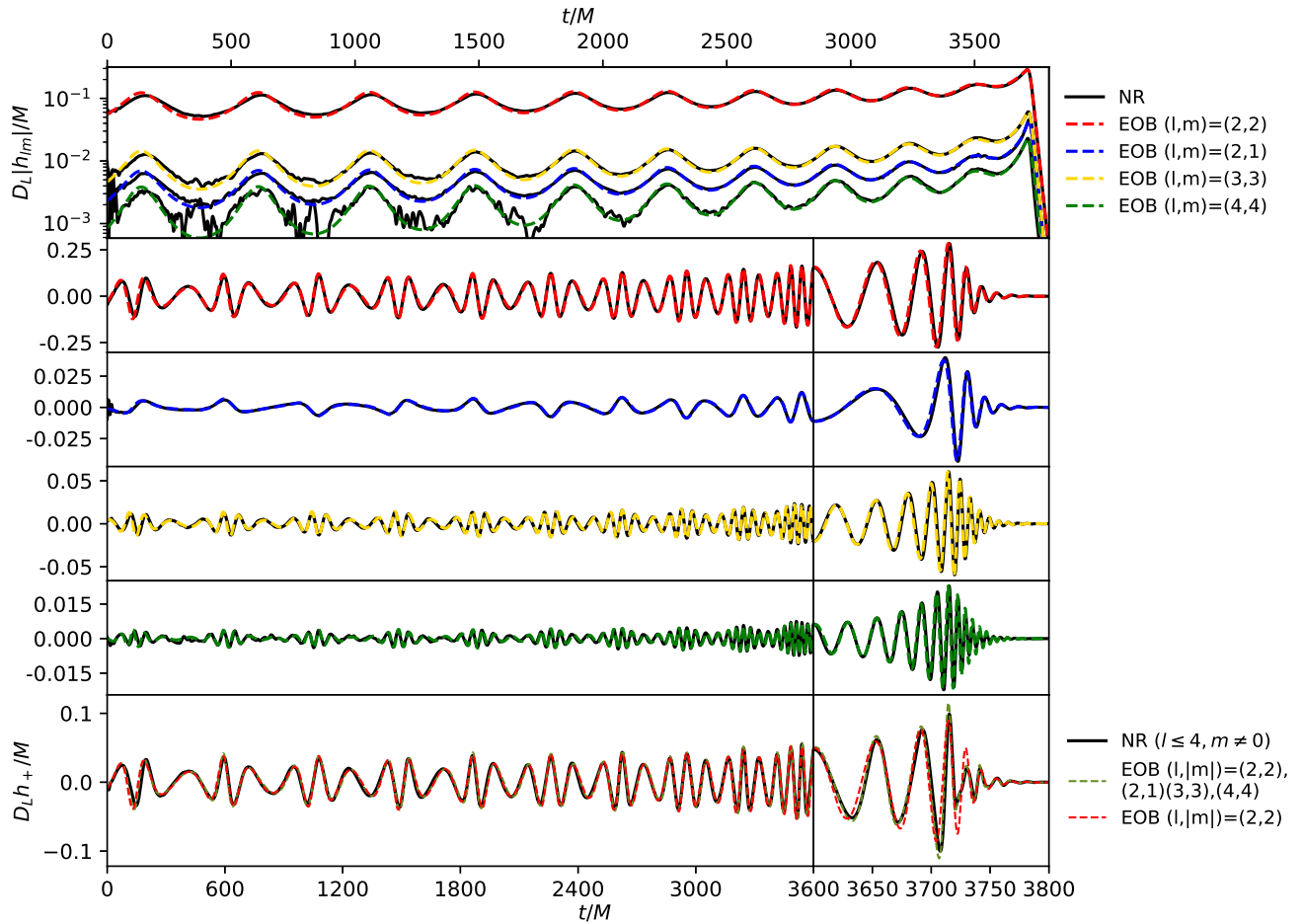}
\end{tabular}
\caption{Waveform comparison between our $\texttt{SEOBNREHM}$ waveform model and SXS:BBH:1374, corresponding to spinless BBH with mass ratio $q=3$. The top panel shows the comparison of the amplitudes for the $(l,m)=(2,2),(2,1),(3,3),(4,4)$ modes. The second to fifth row corresponds to the comparison of the real part of the $(l,m)=(2,2),(2,1),(3,3),(4,4)$ modes respectively between the $\texttt{SEOBNREHM}$ waveform and the NR waveform. The bottom row shows $h_+(\iota,\varphi,\kappa)$ with choice $\iota=\pi/2, \varphi=\pi, \kappa=0$. The red line means only (2,2) mode is used to combine the gravitational wave strain. The green line means the (2,2) mode and the higher modes (2,1), (3,3) and (4,4) are used to combine the gravitational wave strain.}\label{fig6}
\end{figure*}

\begin{figure*}
\begin{tabular}{c}
\includegraphics[width=\textwidth]{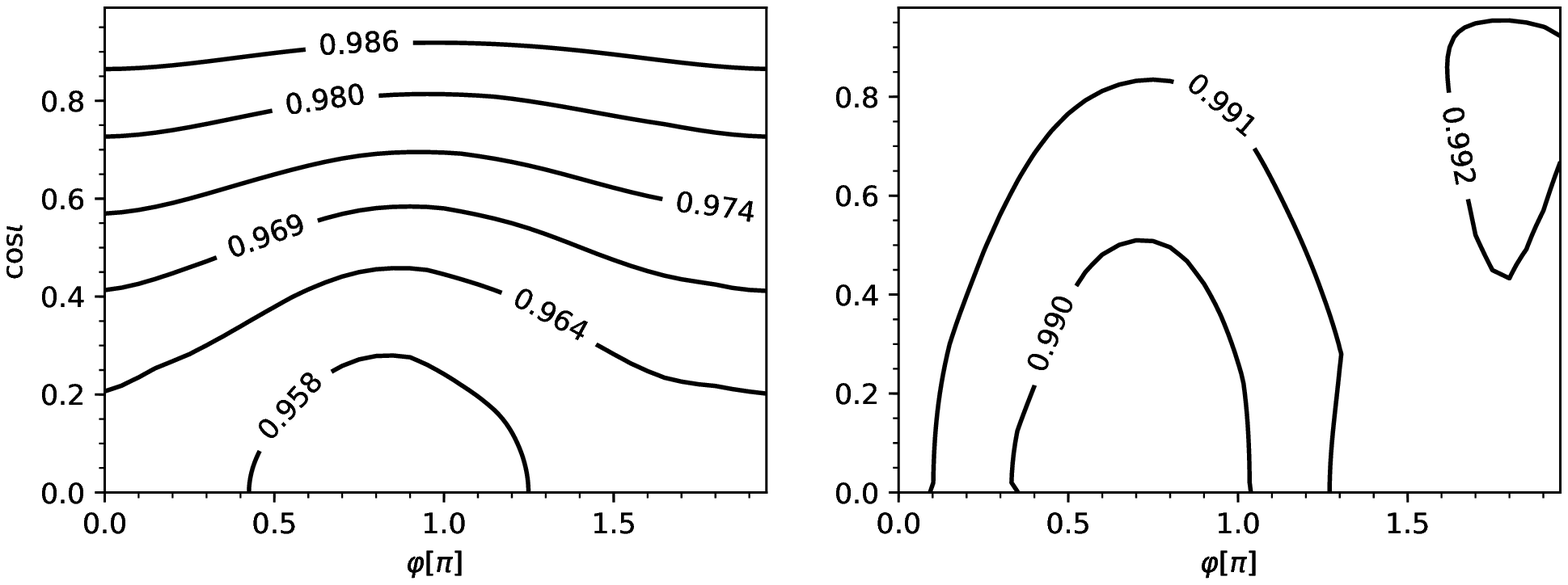}\\
\includegraphics[width=\textwidth]{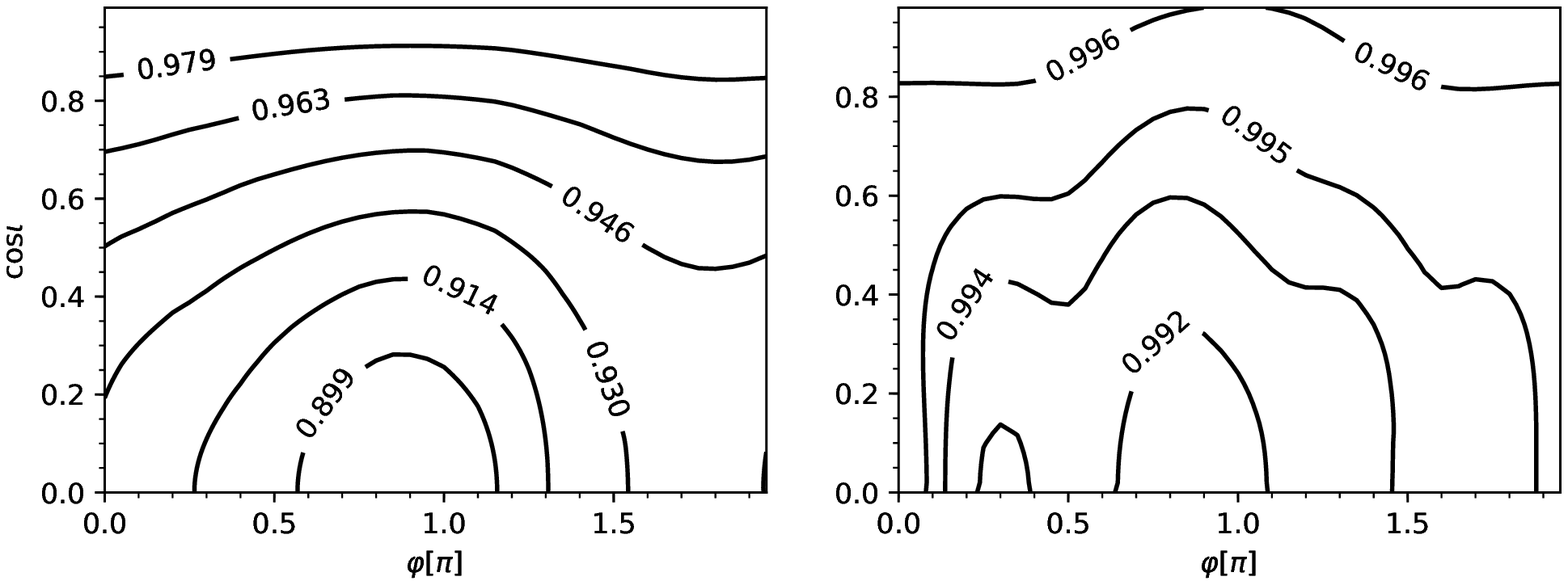}
\end{tabular}
\caption{The dependence on the angles $(\varphi,\iota)$ of the matching factor between the $\texttt{SEOBNREHM}$ model waveform and the SXS NR $(l\leq4,m\neq0)$ waveform. The contour lines together with the contour values of the matching factor FF defined in the Eq.~(\ref{eq1}) are shown in the figure. The intrinsic parameters except the total mass corresponds to the case SXS:BBH:1374 (corresponding to the waveform shown in the Fig.~\ref{fig6}). In addition $\kappa=0$ and the Advanced LIGO designed PSD are used in this figure. The top row corresponds to the total mass $M=40M_\odot$ and the bottom row corresponds to the total mass $M=180M_\odot$. In the left column, only $(l,|m|)=(2,2)$ mode of the $\texttt{SEOBNREHM}$ model is used to combine the gravitational wave strain. In the right column, modes $(l,|m|)=(2,2),(2,1),(3,3),(4,4)$ of the $\texttt{SEOBNREHM}$ model are used.}\label{fig7}
\end{figure*}
\begin{figure*}[htb]
\begin{tabular}{c}
\includegraphics[width=\textwidth]{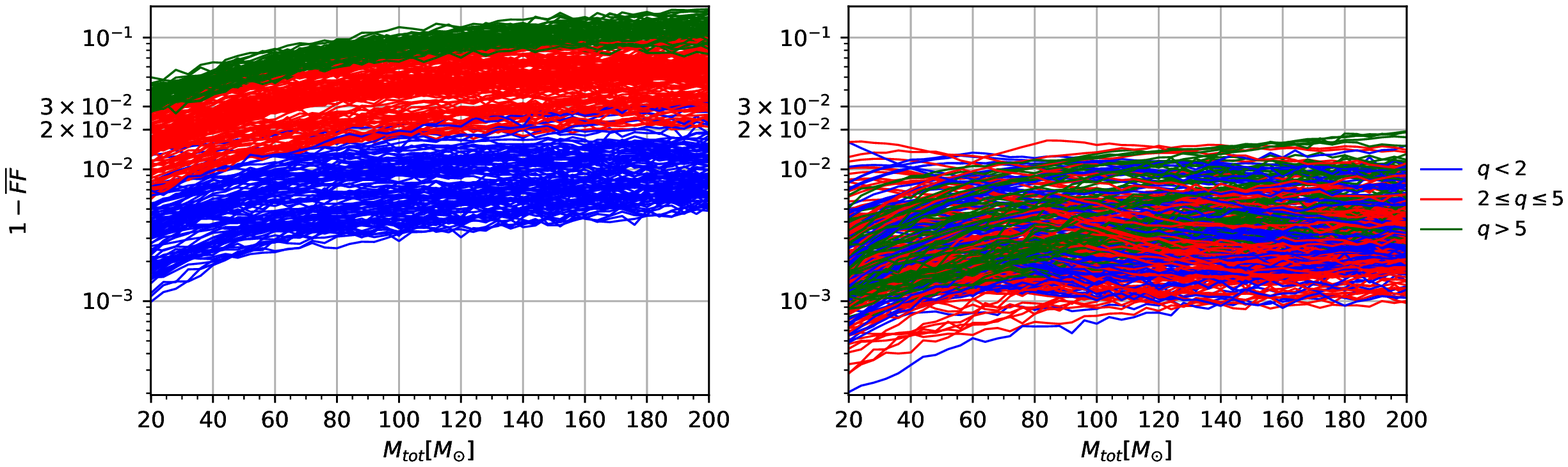}\\
\includegraphics[width=\textwidth]{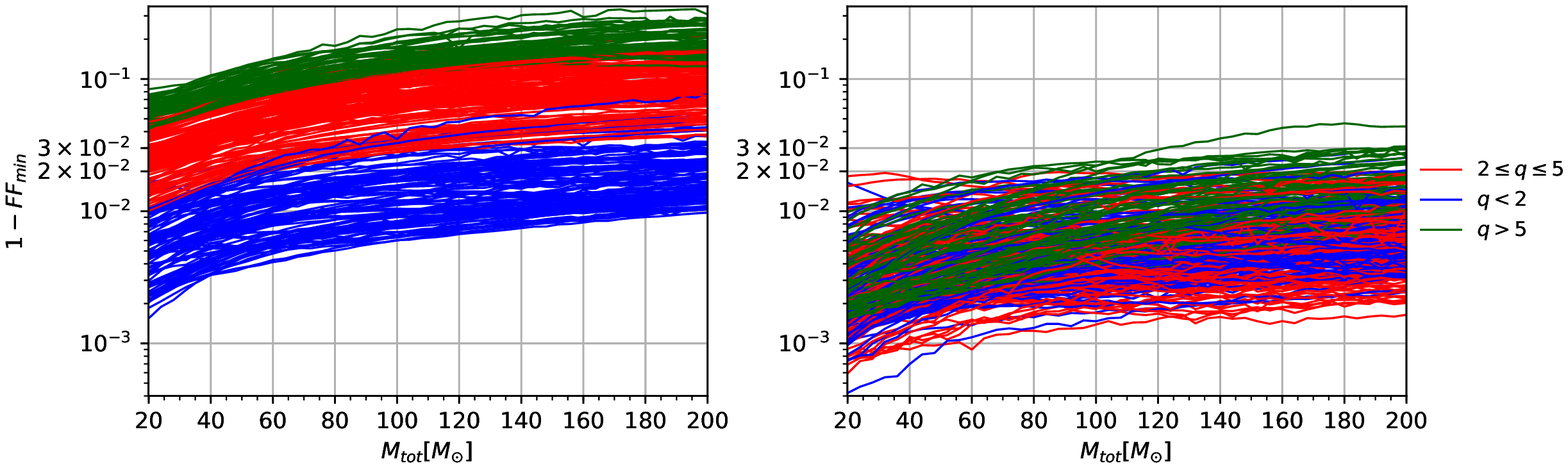}
\end{tabular}
\caption{The matching factors between the $\texttt{SEOBNREHM}$ waveform model and the SXS NR waveforms for the quasi-circular BBHs listed in the Appendix.~\ref{CIRCtable}. The NR waveform modes $(l\leq4,m\neq0)$ and the Advanced LIGO design zero-detuned high-power noise PSD are used in this figure. The top row is for the averaged matching factor over angles $\kappa,\varphi,\iota$ which is defined in the Eq.~(\ref{eq2}). The bottom row is for the minimal matching factor which is defined in the Eq.~(\ref{eq3}). In the left column only $(l,|m|) = (2,2)$ $\texttt{SEOBNREHM}$ waveform mode is used. In the right column the $\texttt{SEOBNREHM}$ waveform modes $(l,|m|)=(2,2),(2,1),(3,3),(4,4)$ are used.}\label{fig8}
\end{figure*}
\begin{table*}[htb]
\caption{Matching factors between the models waveform and the SXS numerical relativity waveform for the 5 mostly mismatched cases between $\texttt{SEOBNREHM}$ and SXS waveforms among the 281 quasi-circular simulations listed in the Appendix.~\ref{CIRCtable}. ID corresponds to the id number in the SXS catalog. $\overline{\rm FF}$ and ${\rm FF}_{\rm min}$ are respectively the averaged matching factor and the minimal matching factor respect to the three angles $(\kappa, \iota, \varphi)$. `E' means model $\texttt{SEOBNREHM}$ and `C' means model $\texttt{SEOBNRv4HM}$.}
\begin{center}
\begin{ruledtabular}
\begin{tabular}{cccccccc}
ID & $q$ & $\chi_1$ & $\chi_2$ & $\overline{\rm FF}(C)$ & ${\rm FF}_{\rm min}(C)$ & $\overline{\rm FF}(E)$ & ${\rm FF}_{\rm min}(E)$\\
\hline \hline
1422 & 7.95 & -0.80 & -0.46 & 98.87\% & 96.95\% & 98.75\% & 96.97\% \\
1424 & 6.46 & -0.66 & -0.80 & 99.03\% & 97.94\% & 98.87\% & 96.98\% \\
1425 & 6.12 & -0.80 & +0.67 & 99.22\% & 97.12\% & 98.98\% & 97.06\% \\
1427 & 7.41 & -0.61 & -0.73 & 99.39\% & 98.32\% & 99.27\% & 98.26\% \\
1428 & 5.52 & -0.80 & -0.70 & 98.95\% & 97.18\% & 98.69\% & 96.91\%
 \label{table2}
 \end{tabular}
 \end{ruledtabular}
 \end{center}
\end{table*}

In \cite{PhysRevD.96.044028_SEOBNRE,PhysRevD.101.044049_validSEOBNRE} we have constructed $\texttt{SEOBNRE}$ waveform model which extends the waveform model $\texttt{SEOBNRv1}$ for spin-aligned circular BBH to spin-aligned eccentric cases. $\texttt{SEOBNRE}$ waveform model works only for (2,2) mode waveform. Our current $\texttt{SEOBNREHM}$ waveform model not only extends $\texttt{SEOBNRE}$ to high modes but also improves the accuracy for (2,2) mode due to the resummation skill and the calibration to $\texttt{SEOBNRv4}$ waveforms. In this subsection we compare our new (2,2) mode waveform to our previous $\texttt{SEOBNRE}$ model.

In order to understand the difference between the waveform model $\texttt{SEOBNRE}$ and $\texttt{SEOBNREHM}$, we use the numerical relativity waveform as a reference. Accordingly we calculate the waveforms for the parameters combination shown in the Fig.~\ref{fig2}. We show the comparison results in the Fig.~\ref{fig5}. Firstly we note that some points shown in the Fig.~\ref{fig2} are missing in the Fig.~\ref{fig5}. This is because $\texttt{SEOBNRE}$ model breaks down for such high spin cases. In addition the matching behavior to the NR waveform of $\texttt{SEOBNREHM}$ improves in general than $\texttt{SEOBNRE}$. In most cases the improvement is less than 1\%. For several cases including SXS:BBH:0292, SXS:BBH:1439 and SXS:BBH:1432, the improvement is more than 10\%.

\section{The multipolar waveform performance of $\texttt{SEOBNREHM}$}\label{secIV}
$\texttt{SEOBNRv4HM}$ waveform model is one of the most widely used higher modes waveform models used in LIGO data analysis. Here we calibrate our new $\texttt{SEOBNREHM}$ waveform model to $\texttt{SEOBNRv4HM}$ for quasi-circular BBHs. The eccentric BBH waveform model $\texttt{TEOBiResumS\_SM}$ also involves high modes. But unfortunately we can not access such waveform data.

Since the waveform of higher modes is very weak, the numerical error of NR results is large for higher modes. In order to treat this issue, previous works on higher modes used combined waveform comparison to numerical relativity simulations. In the current work we adopt the same strategy.
\subsection{Comparison with numerical relativity waveforms}
The gravitational wave strain detected by a detector can be described as \cite{PhysRevD.98.084028_SEOBNRv4HM},
\begin{align}
&h(\iota, \varphi_c, t_c,\boldsymbol{\theta}, \alpha, \delta, \psi ; t) = \nonumber \\
&F_+(\alpha, \delta, \psi) h_+(\iota, \varphi_c, t_c, \boldsymbol{\theta}; t) + F_\times(\alpha, \delta, \psi)h_\times(\iota, \varphi_c, t_c, \boldsymbol{\theta}; t) \label{STRAIN}\\
&F^+(\alpha,\delta,\psi)\equiv\frac{1}{2}(1+\cos^2\alpha)\cos2\delta\cos2\psi\nonumber\\
&\,\,\,\,-\cos\alpha\sin2\delta\sin2\psi,\\
&F^\times(\alpha,\delta,\psi)\equiv+\frac{1}{2}(1+\cos^2\alpha)\cos2\delta\sin2\psi\nonumber\\
&\,\,\,\,+\cos\alpha\sin2\delta\cos2\psi,
\end{align}
where $\boldsymbol{\theta}=(m_1,m_2,\chi_1,\chi_2, e_0)$ is the intrinsic parameters of a BBH. $(\alpha, \delta)$ denotes the sky location of the GW source, and $\psi$ is the polarization angle. These three angles determine the antenna pattern functions $F_+, F_\times$~\cite{ATFUNC01,ATFUNC02}. $(\iota, \varphi_c)$ corresponds to the direction of the observation line respect to the gravitational source frame (check the Eq.~(\ref{hPC})) and $t_c$ is the coalescence time. Following \cite{PhysRevD.98.084028_SEOBNRv4HM}, we define the angle $\kappa\in\left[0\right.\left.,2\pi\right)$ (effective polarization \cite{PhysRevD.89.102003_HMEffects}) and the amplitude $\mathcal{A}$ as
\begin{align}
e^{i\kappa(\alpha, \delta, \psi)} &= \frac{F_+(\alpha, \delta, \psi) + iF_\times(\alpha, \delta, \psi)}{\mathcal{A}(\alpha, \delta)} \\
\mathcal{A}(\alpha, \delta) &= \sqrt{F_+^2(\alpha, \delta, \psi) + F_\times^2(\alpha, \delta, \psi)}.
\end{align}

Firstly we show one comparison example for eccentric BBH between our $\texttt{SEOBNREHM}$ waveform and the NR waveform. The Fig.~\ref{fig6} shows the waveform comparison for the most eccentric BBHs in the SXS catalog which corresponds to SXS:BBH:1374. We can see the consistence of the modes amplitude and of the real part of each mode between the $\texttt{SEOBNREHM}$ waveforms and the NR waveforms. Regarding to the combined gravitational wave strain, the higher modes than (2,2) improves the matching behavior near merger.

When comparing a waveform with higher modes, we define the matching factor as \cite{PhysRevD.98.084028_SEOBNRv4HM}
\begin{equation}
\text{FF}(\kappa, \varphi, \iota, \boldsymbol{\theta}) = \max\limits_{t_c,\kappa',\varphi'} \left[ \eval{\frac{ \braket{ h_1}{ h_2} }{\sqrt{\braket{ h_1} \braket{ h_2}}}
}_{\mbox{\tiny$\begin{array} {l}
\kappa_1 = \kappa' \\
\kappa_2 = \kappa \\
\varphi_{c1} = \varphi' \\
\varphi_{c2} = \varphi \\
\iota_1=\iota_2=\iota \\
\boldsymbol{\theta}_1 = \boldsymbol{\theta}_2 = \boldsymbol{\theta}
 \end{array}$}} \right].\label{eq1}
\end{equation}
In this work, the sky location of the source is set to be the same for EOB/NR waveforms. For a given intrinsic parameter $\boldsymbol{\theta}$ the matching factor is a function of angles $(\kappa,\varphi,\iota)$. As an example we show the matching factor behavior corresponding to SXS:BBH:1374 respect to the angles $(\varphi,\iota)$ in the Fig.~\ref{fig7}. Along with $\cos\iota$ decreases from 1 to 0, the orientation of the orbital plane is from face-on to edge-on with respect to the observer. For the spin-aligned BBHs, the $\cos\iota<0$ part is symmetric to the $\cos\iota>0$ part \cite{PhysRevD.96.044028_SEOBNRE}. Compared to the combination solely with (2,2) mode, the higher modes improve the matching factor about several percentages.

We can find an interesting feature in the Fig.~\ref{fig7}. $\iota=0$ always results in better matching than $\iota=\pi/2$ for $\texttt{SEOBNREHM}$ model to NR waveform. This feature is true for both with and without higher modes. We have also checked that this feature is common for other spin-aligned BBHs. This feature has also been reported in \cite{PhysRevD.98.084028_SEOBNRv4HM}. We understand this fact in the following way. Following (2,2) mode, the next strongest mode is (3,3). Noting that the spin weighted spherical harmonic functions read
\begin{align}
&{}_{-2}Y_{2\pm2}(\iota,\varphi_c)=\frac{1}{8}e^{-2i\varphi_c}\sqrt{\frac{5}{\pi}}(1\mp\cos\iota)^2,\\
&{}_{-2}Y_{3\pm3}(\iota,\varphi_c)=\pm\frac{1}{8}e^{-3i\varphi_c}\sqrt{\frac{21}{2\pi}}(1\mp\cos\iota)^2\sin\iota.
\end{align}
So we have $\left|\frac{{}_{-2}Y_{2\pm2}}{{}_{-2}Y_{3\pm3}}\right|\propto\frac{1}{\sin\iota}$ which means the contribution of (3,3) increases when $\iota$ increases from 0 to $\pi/2$. Unless the waveform model for (3,3) mode is as accurate as (2,2) mode, the matching factor may decrease when $\iota$ increases from 0 to $\pi/2$.

In order to quantify the matching behavior of our $\texttt{SEOBNREHM}$ model waveform to the NR waveform we have defined an averaged and a minimal matching factor over $(\kappa,\varphi,\iota)$ respectively as \cite{PhysRevD.98.084028_SEOBNRv4HM}
\begin{align}
&\overline{\rm FF}(\boldsymbol{\theta}) \equiv \frac{1}{8\pi^2}\int_{0}^{2\pi}\dd \kappa\int_0^{2\pi} \dd \varphi\int_{-1}^{1} \text{FF}(\kappa, \varphi, \iota, \boldsymbol{\theta}) \dd (\cos\iota),\label{eq2}\\
&{\rm FF}_{\rm min}(\boldsymbol{\theta}) \equiv\min_{\kappa, \varphi, \iota}\text{FF}(\kappa, \varphi, \iota, \boldsymbol{\theta}).\label{eq3}
\end{align}

Firstly we check the quasi-circular BBHs whose parameters have been listed in the Appendix.~\ref{CIRCtable}. In all we have checked 281 NR simulations for quasi-circular BBH. The NR multipolar waveform was constructed by using $l\leq4,m\neq0$ modes. Respect to different total mass, we plot the resulted matching factor in the Fig.~\ref{fig8}. Since the higher modes are stronger when the mass ratio ia larger, we have divided the eccentric BBH cases into three groups, including $q<2$, $2\leq q\leq 5$ and $q>5$. When the mass ratio increases the matching factor for solely (2,2) mode decreases because the higher modes become stronger. After we combine the (2,2), (2,1), (3,3) and (4,4) modes together, the averaged matching factors $\overline{\rm FF}$ are always bigger than 98\%. The minimal matching factors ${\rm FF}_{\rm min}$ are bigger than 98\% for all cases except SXS:BBH:1422, SXS:BBH:1424, SXS:BBH:1425 and SXS:BBH:1428. Together with SXS:BBH:1427 we list these 5 smallest ${\rm FF}_{\rm min}$ cases among the 281 quasi-circular BBHs in the Table.~\ref{table2}. These 5 simulations were not available when the paper \cite{PhysRevD.98.084028_SEOBNRv4HM} was published. In the Table.~\ref{table2} we also list the matching results for $\texttt{SEOBNRv4HM}$ waveform model. We find that the behavior of $\texttt{SEOBNRv4HM}$ is similar to that of $\texttt{SEOBNREHM}$ for these 5 cases.

\begin{figure*}[htb]
\begin{tabular}{c}
\includegraphics[width=\textwidth]{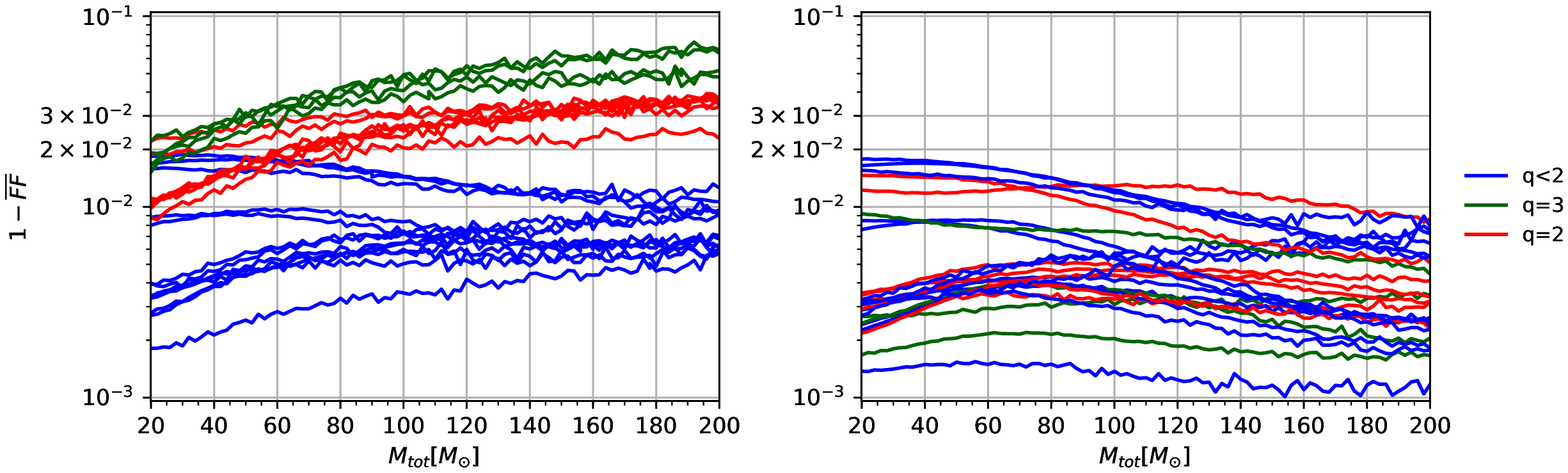}\\
\includegraphics[width=\textwidth]{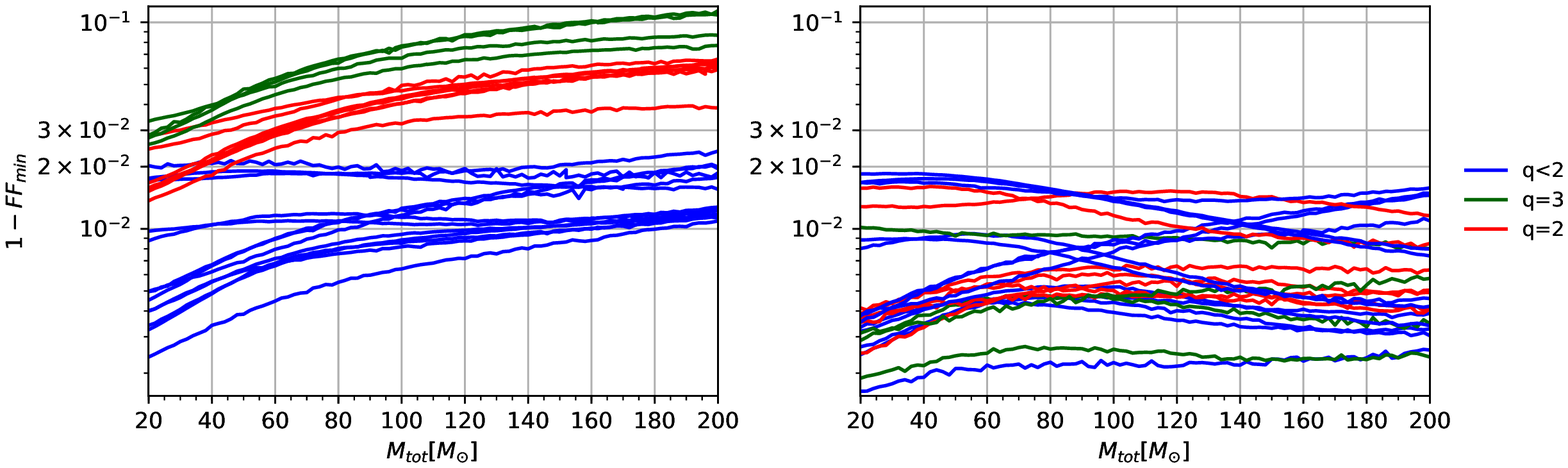}
\end{tabular}
\caption{The matching factors between the $\texttt{SEOBNREHM}$ waveform model and the SXS NR waveforms for the eccentric BBHs listed in the Table.~\ref{table1}. The NR waveform modes $(l\leq4,m\neq0)$ and the Advanced LIGO design zero-detuned high-power noise PSD are used in this figure. The top row is for the averaged matching factor over angles $\kappa,\varphi,\iota$ which is defined in the Eq.~(\ref{eq2}). The bottom row is for the minimal matching factor  which is defined in the Eq.~(\ref{eq3}). In the left column only $(l,|m|) = (2,2)$ $\texttt{SEOBNREHM}$ waveform mode is used. In the right column the $\texttt{SEOBNREHM}$ waveform modes $(l,|m|)=(2,2),(2,1),(3,3),(4,4)$ are used. When including higher modes, all of the matching factors are bigger than 98\%. Here we have used the initial eccentricity $e_{\rm EOB}$ listed in the Table.~\ref{table1} to generate the $\texttt{SEOBNREHM}$ waveform.}\label{fig9}
\end{figure*}

We are more interested in the performance of higher modes of the new waveform in elliptical cases. We also calculate the averaged matching factor $\overline{\rm FF}$ and the minimal matching factor ${\rm FF}_{\rm min}$ between the new higher modes waveform model $\texttt{SEOBNREHM}$ and the NR waveforms. The comparison results are plotted in the Fig.~\ref{fig9}. Here we again divide the BBHs into three groups based on mass ratio, including $q<2$, $q=2$ and $q=3$. As expected, when $q\geq2$ all cases with solely (2,2) mode admit matching factor less than 98\%. Even some cases with $q<2$ also admit matching factor less than 98\%. In the contrast, when higher modes including (2,1), (3,3) and (4,4) are used, all the matching factors increases to larger than 98\%. For the cases with initial eccentricity $e_{\rm EOB}<0.2$ (shown in the Table.~\ref{table1}), the matching factors are bigger than 99\%. We list the results in the Table.~\ref{table1}.

\subsection{Comparison with $\texttt{SEOBNRv4HM}$}
\begin{figure*}
\begin{tabular}{c}
\includegraphics[width=\textwidth]{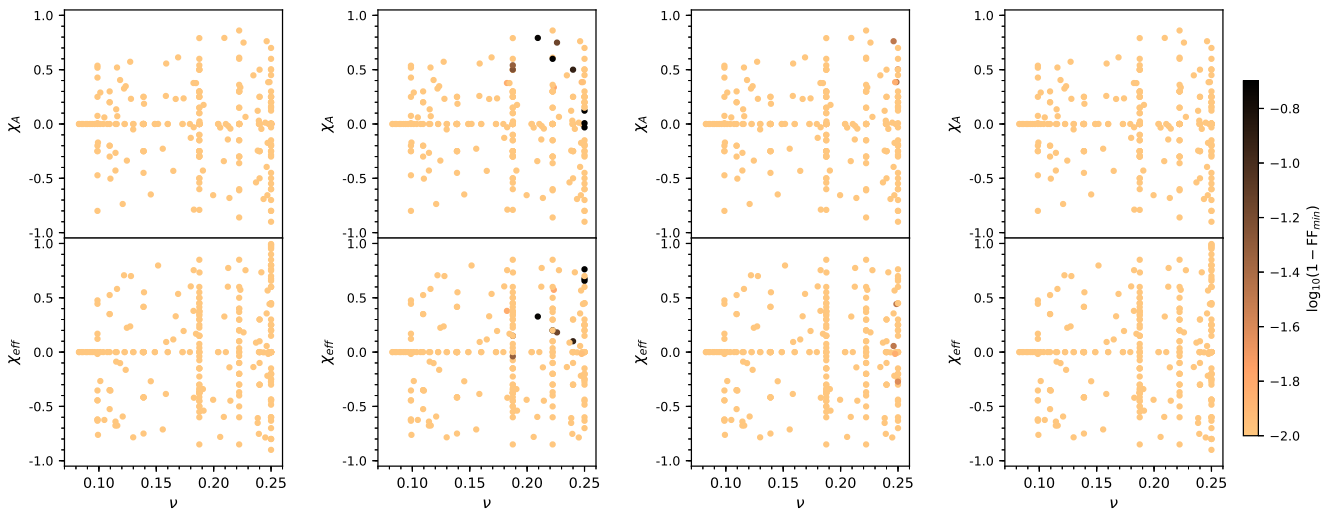}
\end{tabular}
\caption{Matching factors of the $(l,m)=(2,2),(2,1),(3,3),(4,4)$ modes waveform for the quasi-circular spin-aligned BBH between the new waveform model $\texttt{SEOBNREHM}$ proposed in the current work and $\texttt{SEOBNRv4HM}$ for the BBH parameters listed in the Appendix.~\ref{CIRCtable}. From left to right, the ${\rm FF}_{\rm min}$ respect to different mass ratio for the $(l,m)=(2,2),(2,1),(3,3),(4,4)$ modes are shown respectively. The plot convention is the same to the Fig.~\ref{fig2}.}\label{fig10}
\end{figure*}
\begin{table*}[htb]
\caption{Matching factors between the models waveform and the SXS numerical relativity waveform for the 10 mostly mismatched cases between $\texttt{SEOBNREHM}$ and $\texttt{SEOBNRv4HM}$. ID corresponds to the id number in the SXS catalog. $\overline{\rm FF}$ and ${\rm FF}_{\rm min}$ are respectively the averaged matching factor and the minimal matching factor respect to the three angles $(\kappa, \iota, \varphi)$. `E' means model $\texttt{SEOBNREHM}$ and `C' means model $\texttt{SEOBNRv4HM}$.}
\begin{center}
\begin{ruledtabular}
\begin{tabular}{cccccccc}
ID & $q$ & $\chi_1$ & $\chi_2$ & $\overline{\rm FF}(C)$ & ${\rm FF}_{\rm min}(C)$ & $\overline{\rm FF}(E)$ & ${\rm FF}_{\rm min}(E)$\\
\hline \hline
0025 & 1.50 & +0.50 & -0.50 & 99.77\% & 99.48\% & 99.53\% & 99.04\% \\
0254 & 2.00 & +0.60 & -0.60 & 99.82\% & 99.55\% & 99.69\% & 99.41\% \\
0305 & 1.22 & +0.33 & -0.44 & 99.58\% & 99.54\% & 99.35\% & 98.83\% \\
1453 & 2.35 & +0.80 & -0.78 & 99.74\% & 99.40\% & 99.71\% & 99.44\% \\
1474 & 1.28 & +0.72 & -0.80 & 99.04\% & 98.53\% & 98.10\% & 96.42\% \\
1481 & 1.00 & +0.73 & +0.79 & 99.66\% & 99.53\% & 99.63\% & 99.48\% \\
1495 & 1.00 & +0.78 & +0.53 & 99.52\% & 99.37\% & 99.53\% & 99.41\% \\
1496 & 1.16 & +0.80 & +0.03 & 98.95\% & 98.75\% & 98.78\% & 98.21\% \\
1497 & 1.00 & +0.68 & +0.67 & 99.68\% & 99.58\% & 99.63\% & 99.54\% \\
1498 & 1.03 & +0.22 & -0.78 & 99.78\% & 99.49\% & 99.40\% & 98.69\%
 \label{table3}
 \end{tabular}
 \end{ruledtabular}
 \end{center}
\end{table*}

In this subsection, we compare our new (2,2), (2,1), (3,3) and (4,4) modes waveforms to the corresponding ones given by $\texttt{SEOBNRv4HM}$ \cite{PhysRevD.98.084028_SEOBNRv4HM} for quasi-circular spin-aligned BBHs. Similar to the above (2,2) mode comparison to the $\texttt{SEOBNRv4}$ model, we use the matching factor ${\rm FF}$ defined in the Eq.~(\ref{FFeq}) for each mode to do the comparison. In order to use the NR simulations as a reference we check the BBH parameters corresponding to the SXS quasi-circular BBH simulations listed in the Appendix.~\ref{CIRCtable}. For each given combination of $\nu$, $\chi_{1z}$ and $\chi_{2z}$ we scan the total mass in the range $M\in[20, 200]M_\odot$ for each mode to search the minimal matching factor ${\rm FF}_{\rm min}$ respect to the total mass.

We plot the ${\rm FF}_{\rm min}$ respect to different mass ratio and black hole spin in the Fig.~\ref{fig10}. Following the Fig.~\ref{fig3} we have used variables $\chi_A=(\chi_1-\chi_2)/2$ and $\chi_{\text{eff}}=(m_1\chi_1 + m_2 \chi_2)/M$ in the plot.

For spinless cases, all matching factors for all (2,2), (2,1), (3,3) and (4,4) modes are bigger than 99.9\%. Regarding to (2,2) and (4,4) modes, all matching factors are bigger than 99\% even for highly spinning BBHs.

\begin{figure*}
\begin{tabular}{c}
\includegraphics[width=\textwidth]{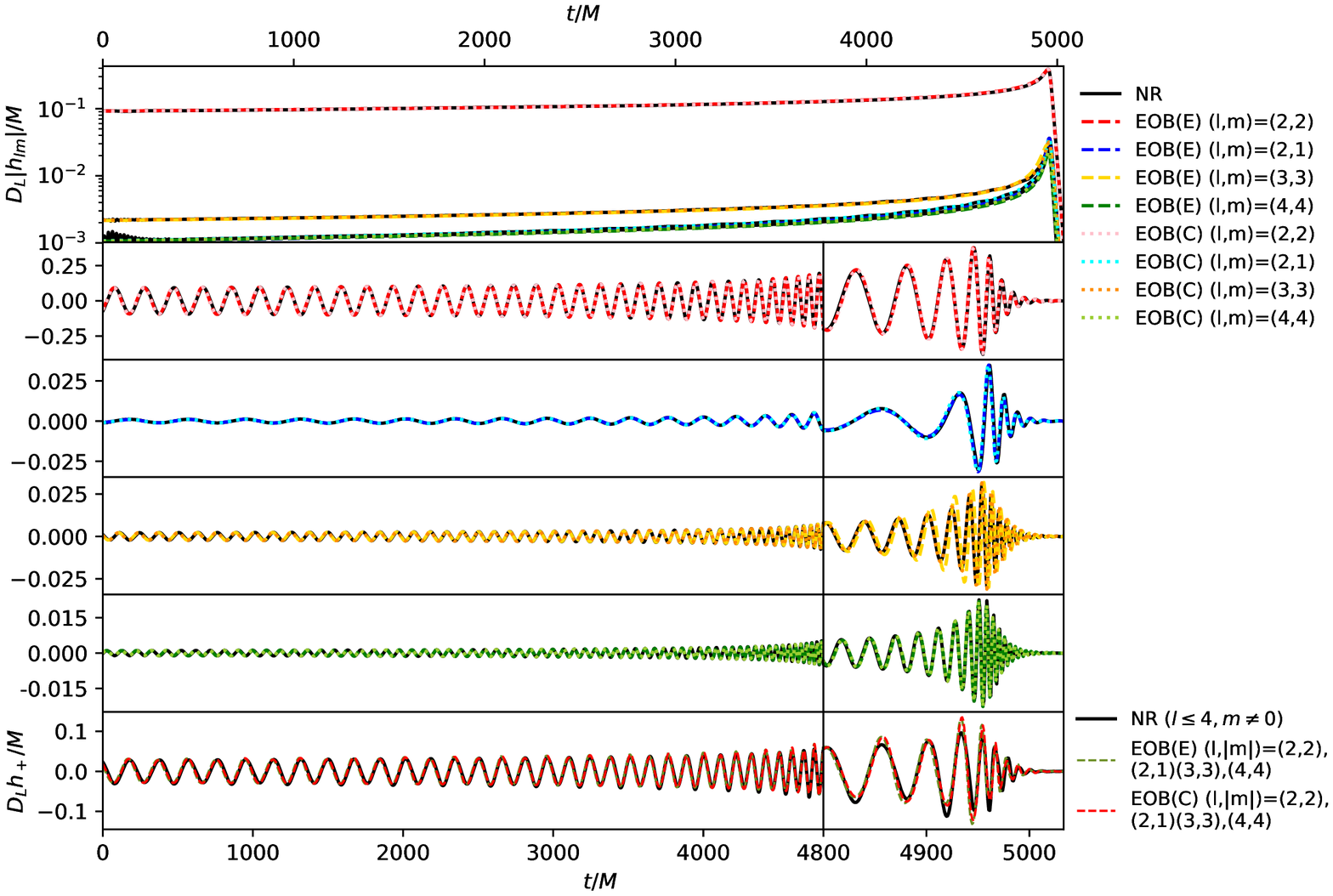}
\end{tabular}
\caption{Waveform comparison between model waveform and NR waveform for case SXS:BBH:1474. EOB(C) means $\texttt{SEOBNRv4HM}$ and EOB(E) means $\texttt{SEOBNREHM}$.}\label{fig11}
\end{figure*}

\begin{figure*}
\begin{tabular}{c}
\includegraphics[width=\textwidth]{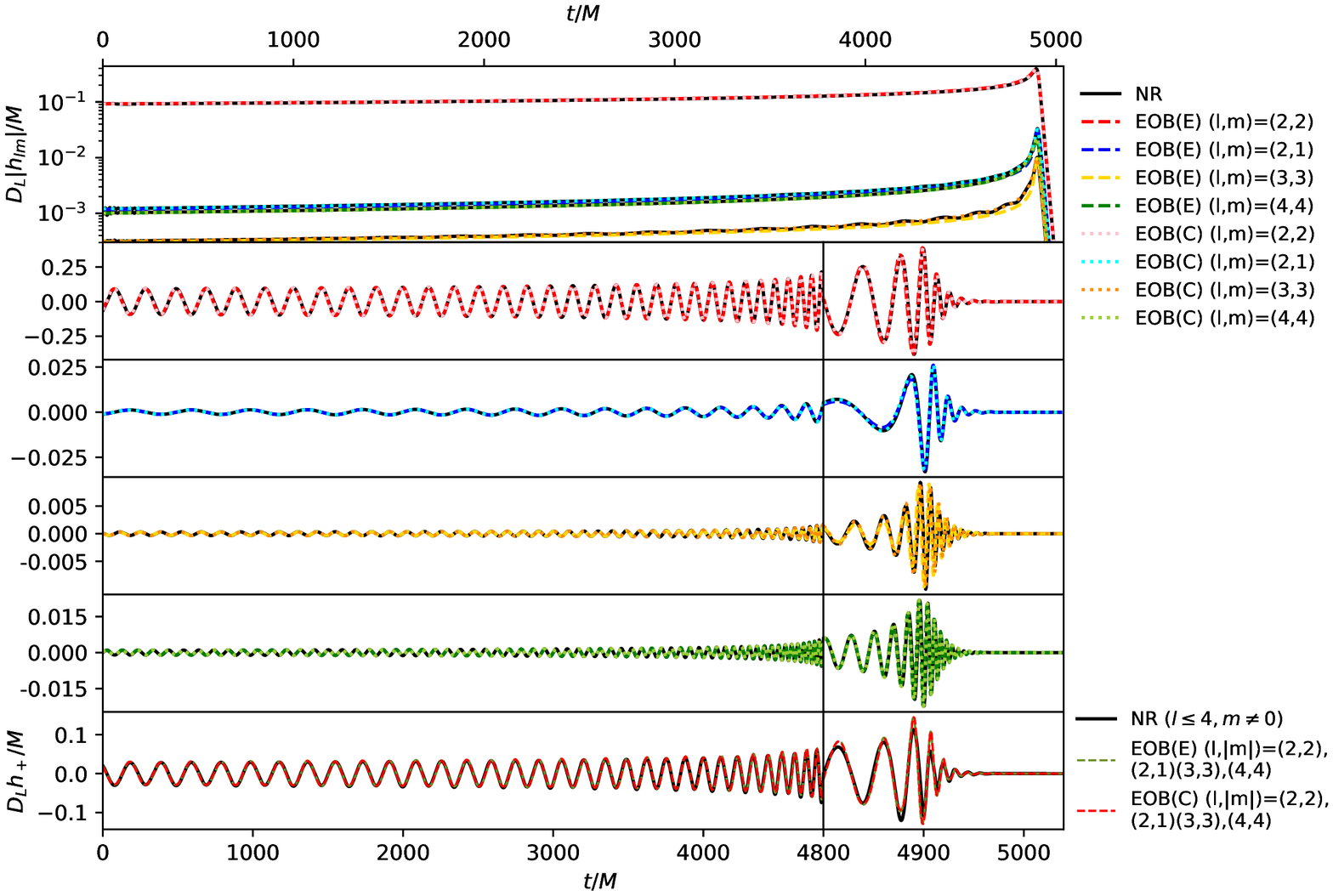}
\end{tabular}
\caption{Waveform comparison between model waveform and NR waveform for case SXS:BBH:1498. EOB(C) means $\texttt{SEOBNRv4HM}$ and EOB(E) means $\texttt{SEOBNREHM}$.}\label{fig12}
\end{figure*}

For (3,3) mode, there are 4 cases, including SXS:BBH:0305, SXS:BBH:1474, SXS:BBH:1496 and SXS:BBH:1498, admit minimal matching factor less than 99\%. All other minimal matching factors are bigger than 99\%. Among these 4 cases, the matching factor of SXS:BBH:1474 is smallest, which is 96.6\%. We plot the waveform comparison for SXS:BBH:1474 and SXS:BBH:1498 respectively in the Figs.~\ref{fig11} and \ref{fig12}. Regarding to the waveform alignment for different modes, we would like to mention one point. All modes waveforms share the unique shifted time which is determined by (2,2) mode. In another word, we did not maximize the matching factors through adjusting parameter $t_c$ except (2,2) mode.

\begin{figure*}
\begin{tabular}{c}
\includegraphics[width=\textwidth]{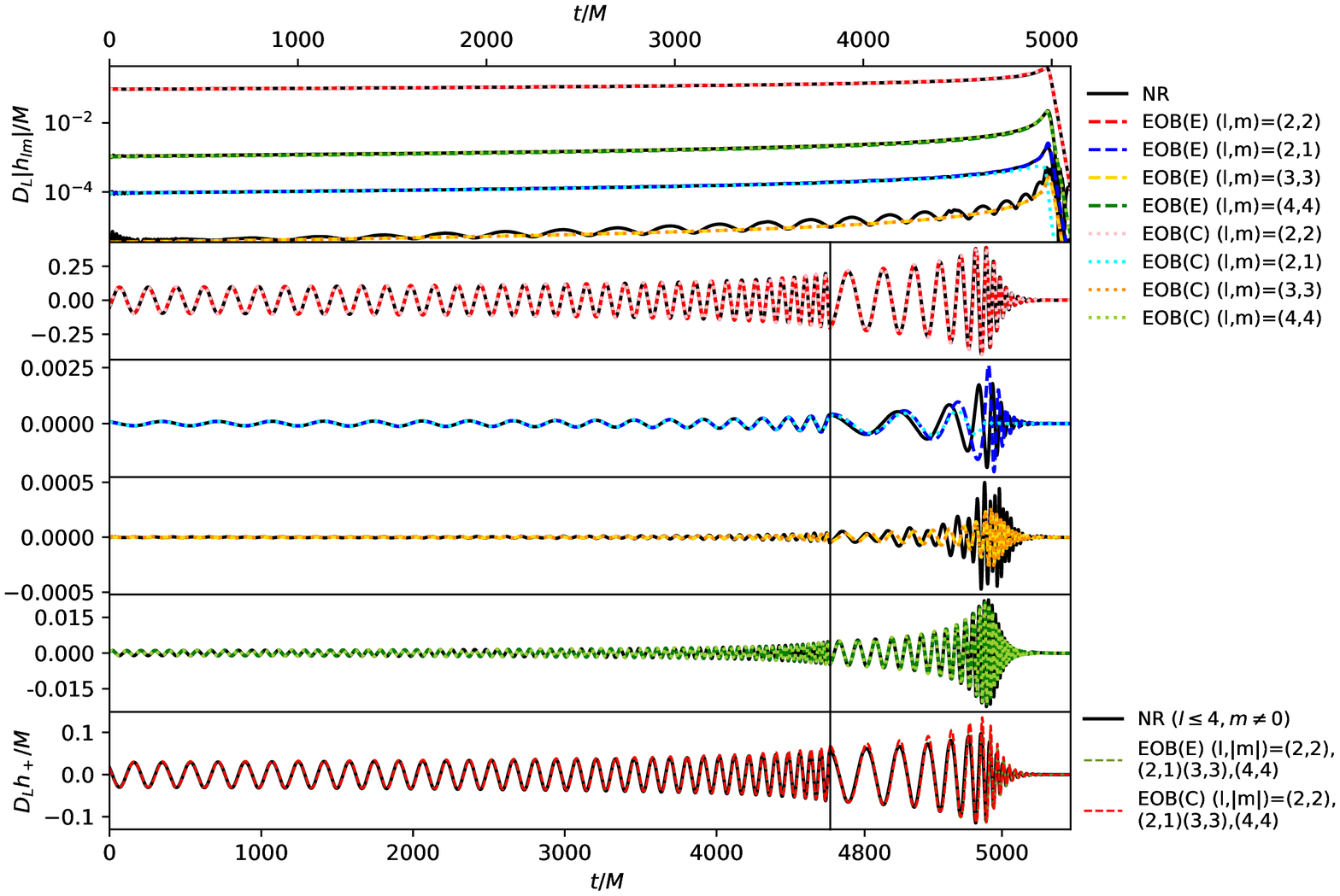}
\end{tabular}
\caption{Waveform comparison between model waveform and NR waveform for case SXS:BBH:1481. EOB(C) means $\texttt{SEOBNRv4HM}$ and EOB(E) means $\texttt{SEOBNREHM}$.}\label{fig13}
\end{figure*}

\begin{figure*}
\begin{tabular}{c}
\includegraphics[width=\textwidth]{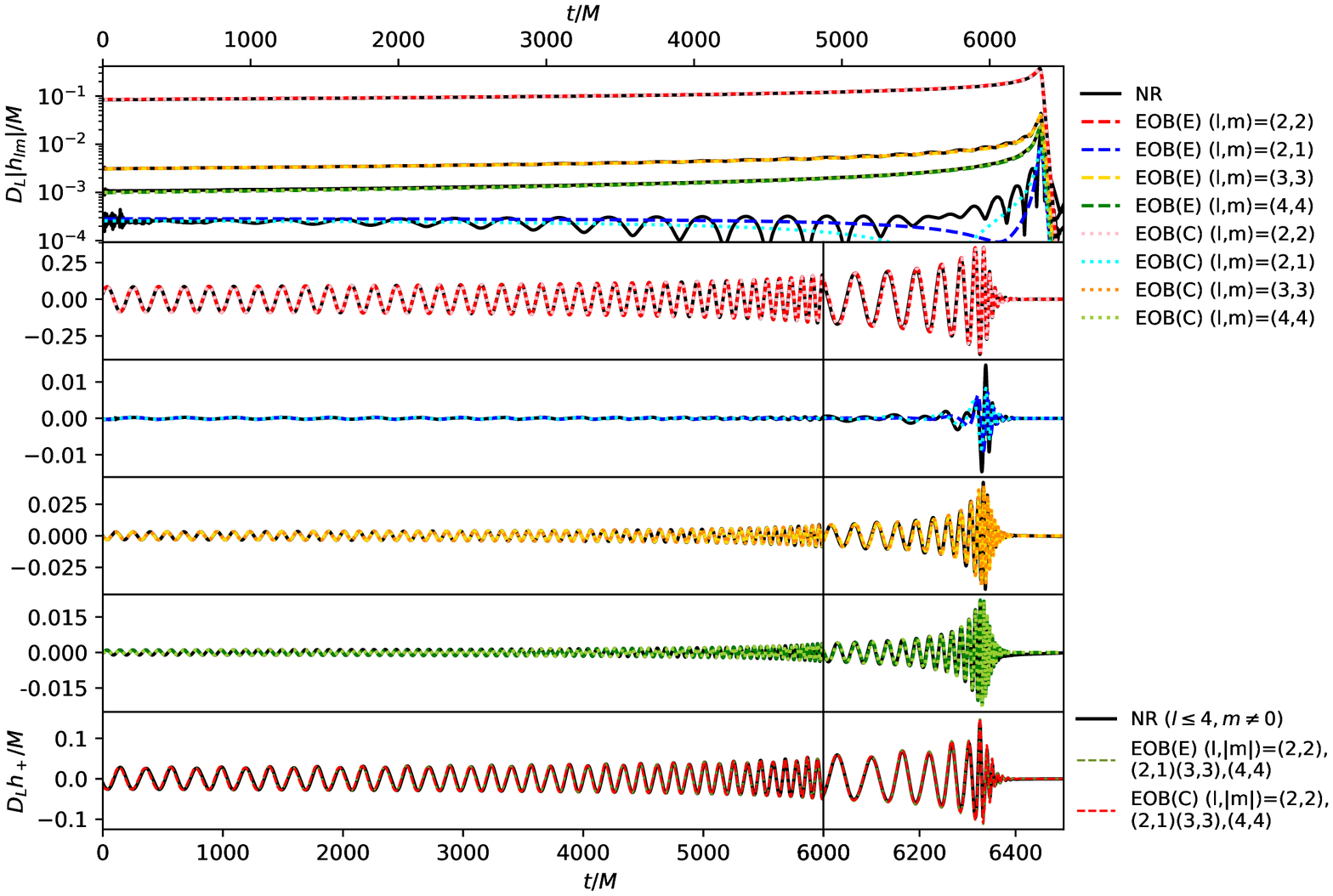}
\end{tabular}
\caption{Waveform comparison between model waveform and NR waveform for case SXS:BBH:0025. EOB(C) means $\texttt{SEOBNRv4HM}$ and EOB(E) means $\texttt{SEOBNREHM}$.}\label{fig14}
\end{figure*}

For (2,1) mode,  all minimal matching factors are bigger than 90\% except 6 cases including SXS:BBH:0025, SXS:BBH:0254, SXS:BBH:1453, SXS:BBH:1481, SXS:BBH:1495 and SXS:BBH:1497. Among them, the matching factor of SXS:BBH:1481 is 53.7\% which is the smallest one. We plot the waveform comparison for SXS:BBH:1481 and SXS:BBH:0025 in the Figs.~\ref{fig13} and \ref{fig14}.
At the mean time we note that the corresponding (2,1) mode and (3,3) mode for the above mentioned 10 cases are quite weak. This fact makes it hard to extract information from NR results and accordingly construct and tune model waveform. On the other hand this fact makes the mismatch mentioned above weakly affect the waveform matching to NR results directly. Specifically, the resulted averaged matching factor $\overline{\rm FF}$ and minimal matching factor ${\rm FF}_{\rm min}$ respect to the angles $\kappa,\varphi,\iota$ are quite similar between $\texttt{SEOBNREHM}$ and $\texttt{SEOBNRv4HM}$. For convenient reference we list these matching factors in the Table.~\ref{table3}.
\section{Power fraction of higher modes for spin-aligned BBHs}\label{secou}
\begin{figure*}
\begin{tabular}{c}
\includegraphics[width=\textwidth]{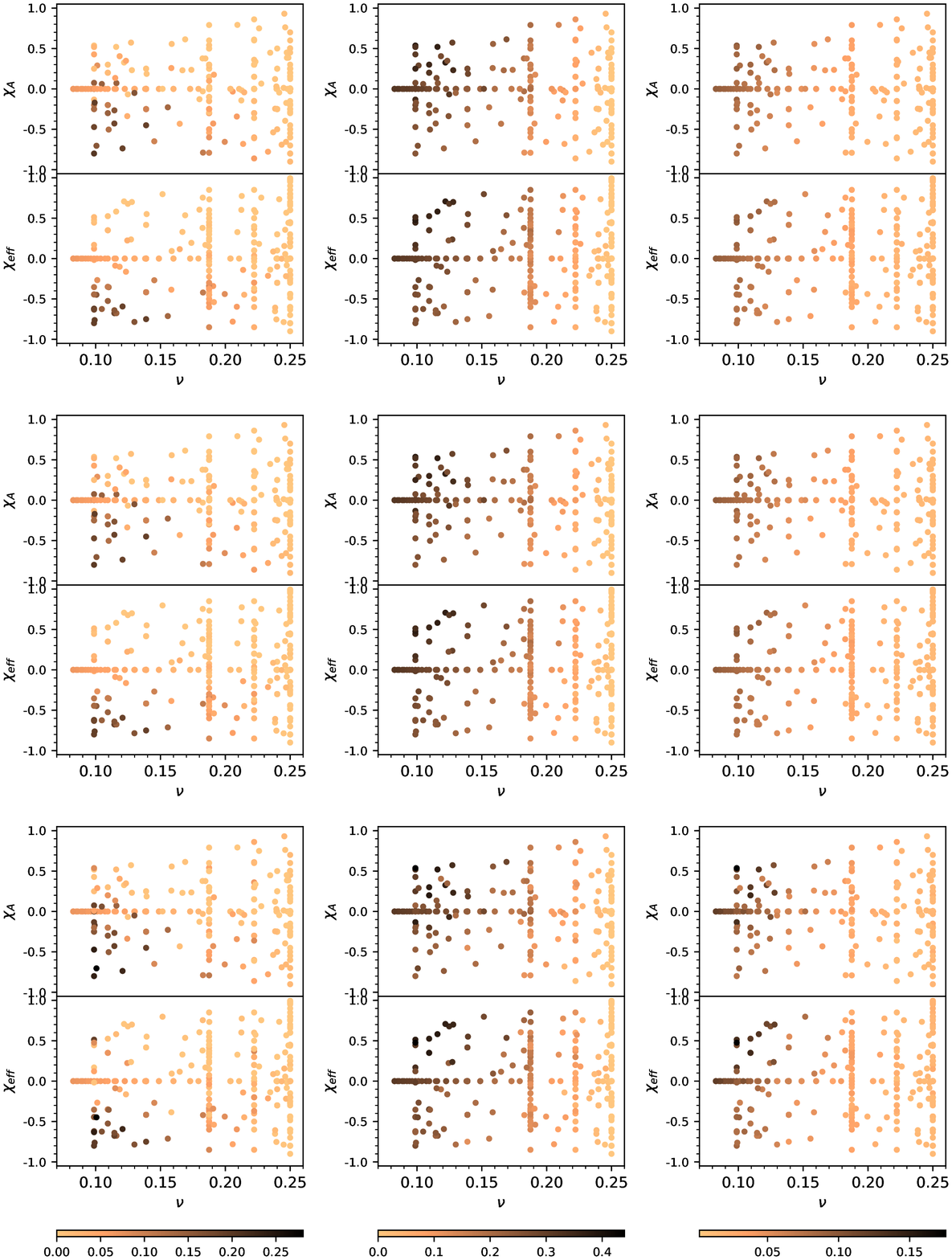}
\end{tabular}
\caption{Power ratio $\eta_{lm}$ for modes (2,1), (3,3) and (4,4) respectively from the left to the right. The top row corresponds to the quasi-circular SXS simulations listed in the Appendix.~\ref{CIRCtable}. The middle row corresponds to the $\texttt{SEOBNREHM}$ waveform model results with $e_0=0$ and BBH parameters corresponding to the top row. The bottom row corresponds to the $\texttt{SEOBNREHM}$ waveform model results with $e_{0}=0.5$ at reference frequency $Mf_0=0.002$ and other BBH parameters the same to the top row.}\label{fig15}
\end{figure*}
As we have seen in the previous sections that when the contribution of higher modes increases the higher modes become more and more important to gravitational waveform template for data analysis. Roughly we expect the contribution of higher modes increases along with the mass ratio $q$ and orbital eccentricity increase. In this section we would like to investigate the power fraction of higher modes.

We define the power of each mode as \cite{PhysRevD.96.044028_SEOBNRE}
\begin{align}
P_{lm}=\frac{1}{16\pi}\left|\dot{h}_{lm}\right|^2.
\end{align}
At about the merger time $P_{lm}$ admits a peak. But usually different mode approaches such peak at different time. As an estimation of power fraction of higher modes, we calculate the ratio between the peak $P_{lm}$ and the peak $P_{22}$
\begin{align}
\eta_{lm}\equiv\frac{\max_tP_{lm}}{\max_tP_{22}},
\end{align}
where the maximum is taken respect to time corresponding to the peak value.

For the quasi-circular BBHs, we plot $\eta_{21}$, $\eta_{33}$ and $\eta_{44}$ in the top row of the Fig.~\ref{fig15} which corresponds to the 281 SXS simulation results listed in the Appendix.~\ref{CIRCtable}. In the second row we plot the results of the $\texttt{SEOBNREHM}$ waveform model with $e_0=0$ and BBH parameters corresponding to the top row. We can see that the ratio gotten by the $\texttt{SEOBNREHM}$ waveform model is very close to the one gotten by the NR simulations. As expected the ratio for each mode increases with mass ratio $q$. Regarding to the effect of BH spin, the ratio for each mode increase with the magnitude of spin on the one hand. On the other hand we interestingly find that positive spin respect to the direction of orbital angular momentum results in much bigger ratio for (3,3) and (4,4) modes. Instead negative spin results in much bigger ratio for (2,1) mode.

In order to check the effect of eccentricity on the power ratio we have calculated different initial eccentricity but keep other BBH parameters unchanged. We found the ratio $\eta_{lm}$ roughly does not change. As an example we plot the result with $e_0=0.5$ at reference frequency $Mf_0=0.002$ in the bottom row of the Fig.~\ref{fig15}. We suspect this fact implies that the gravitational radiation strongly circularizes the orbit near the merger and the quasi-circular state is approached before the final BH forms.
\begin{figure*}
\begin{tabular}{c}
\includegraphics[width=\textwidth]{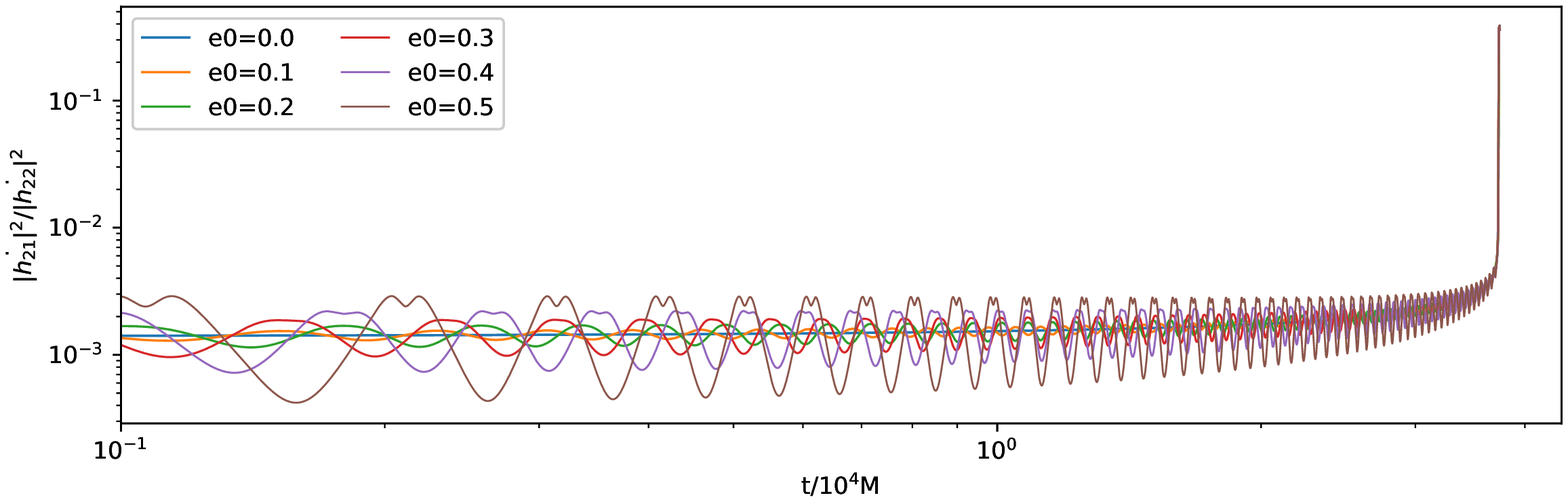}\\
\includegraphics[width=\textwidth]{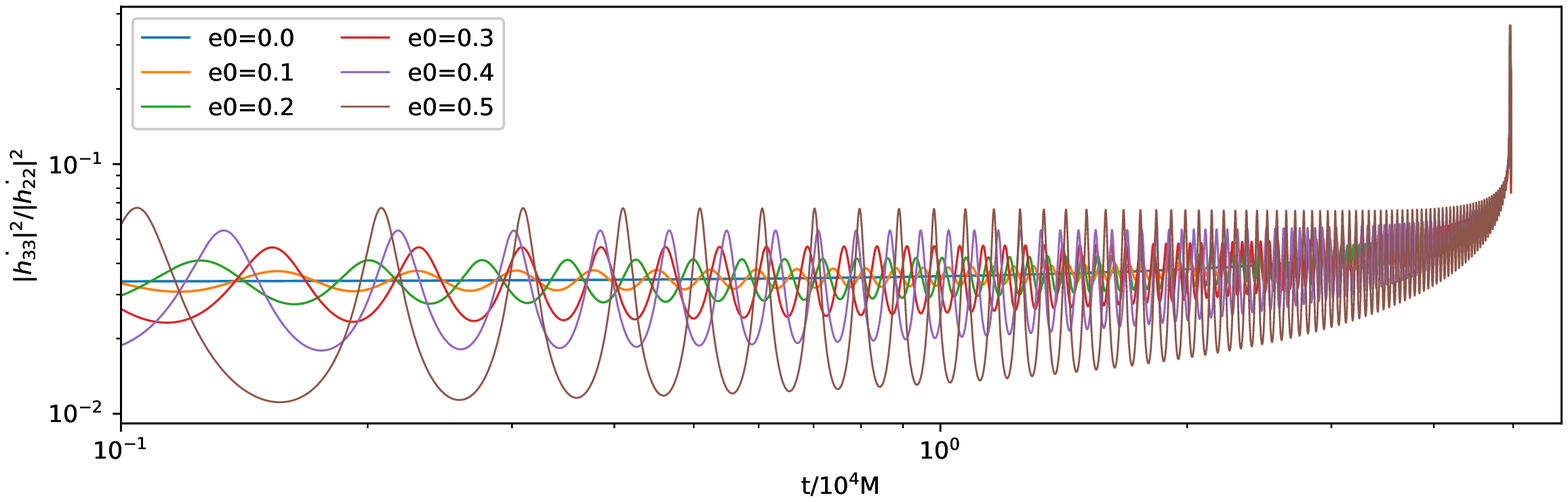}\\
\includegraphics[width=\textwidth]{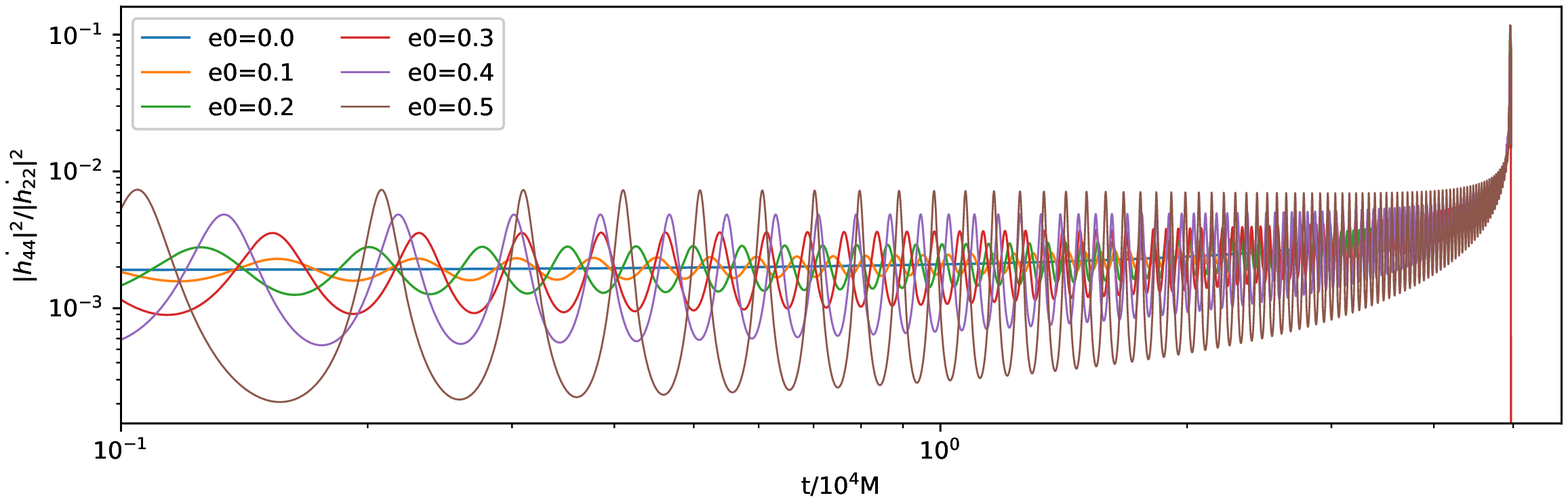}
\end{tabular}
\caption{The top panel: power ratio $\xi_{21}$ BBH mass ratio and spin parameters corresponding to SXS:BBH:1433 but different initial eccentricity shown in the legend at reference frequency $Mf_0=0.002$. The middle panel: similar to the top panel but for $\xi_{33}$ and SXS:BBH:1441. The bottom panel: similar to the top panel but for $\xi_{44}$ and SXS:BBH:1441.}\label{fig16}
\end{figure*}

Since the eccentricity does not change the power ratio $\eta_{lm}$, it is interesting to ask how about the following power ratio as a function of time
\begin{align}
\xi_{lm}(t)\equiv\frac{P_{lm}(t)}{P_{22}(t)}.
\end{align}
We pick the cases with largest $\eta_{lm}$ shown in the Fig.~\ref{fig15} as examples. For (2,1) mode, SXS:BBH:1433 is the one with largest $\eta_{21}\approx0.24$. SXS:BBH:1441 is the case with largest $\eta_{lm}$ for both (3,3) ($\eta_{33}\approx0.44$) and (4,4) mode ($\eta_{44}\approx0.18$). We plot these examples for $\xi(t)$ in the Fig.~\ref{fig16}. Compared to quasi-circular case, the eccentricity just introduces an oscillation for $\xi(t)$. There is no amplification of power ratio $\xi$ introduced by the eccentricity.

In the top panel of the Fig.~\ref{fig16}, there is a minor but interesting feature. Near the periastron point, $\xi_{21}$ shows two peaks. This is because eccentricity introduce a correction to the phase of the waveform and consequently a correction to the wave frequency of each mode. Such correction makes the frequency of $(l,m)$ mode a little different to the $m$ times of the orbital frequency. When the eccentricity is strong enough, such correction results in the mentioned two-peak behavior of $\xi_{21}$.

\begin{figure*}
\begin{tabular}{cccc}
\includegraphics[width=0.25\textwidth]{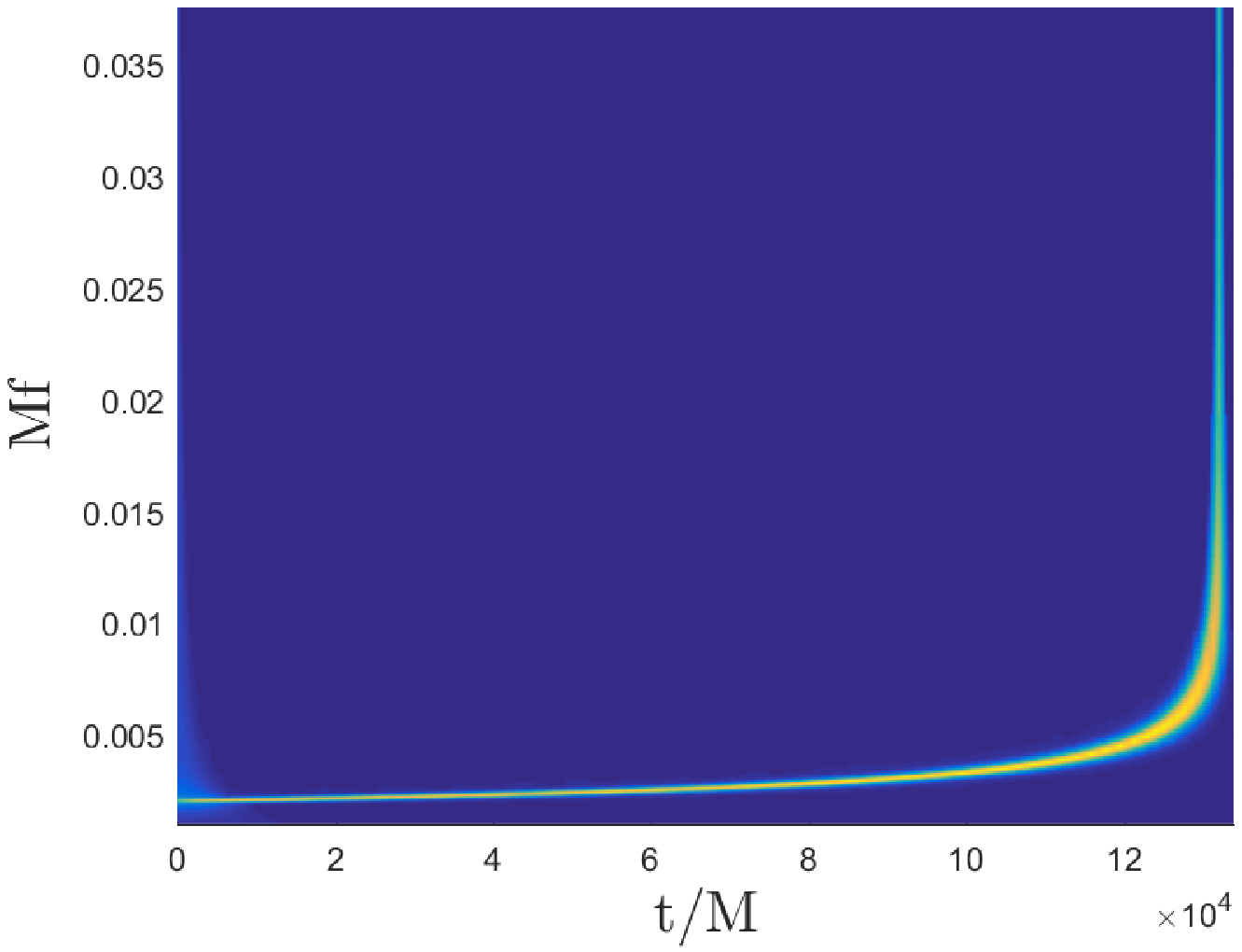}&
\includegraphics[width=0.25\textwidth]{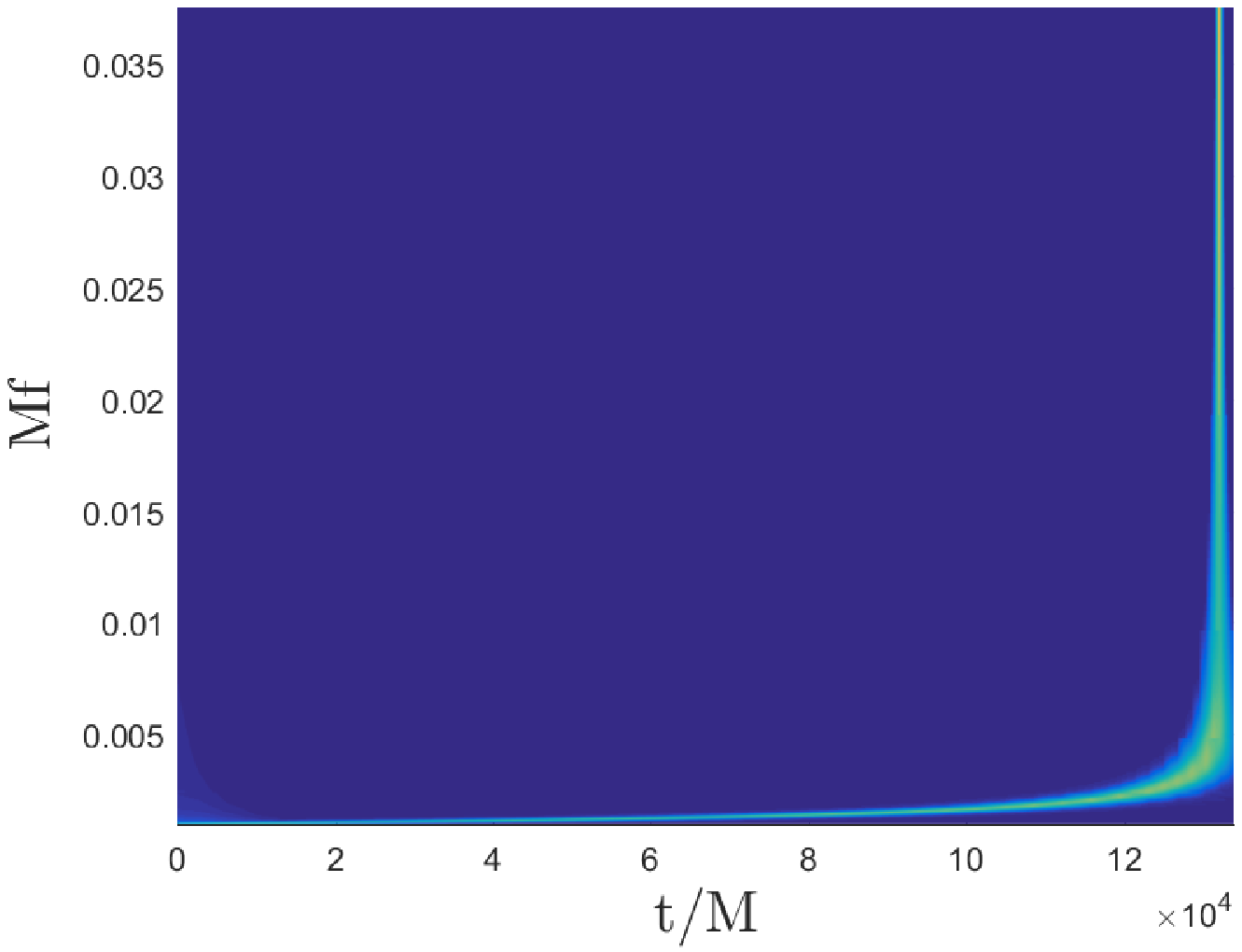}&
\includegraphics[width=0.25\textwidth]{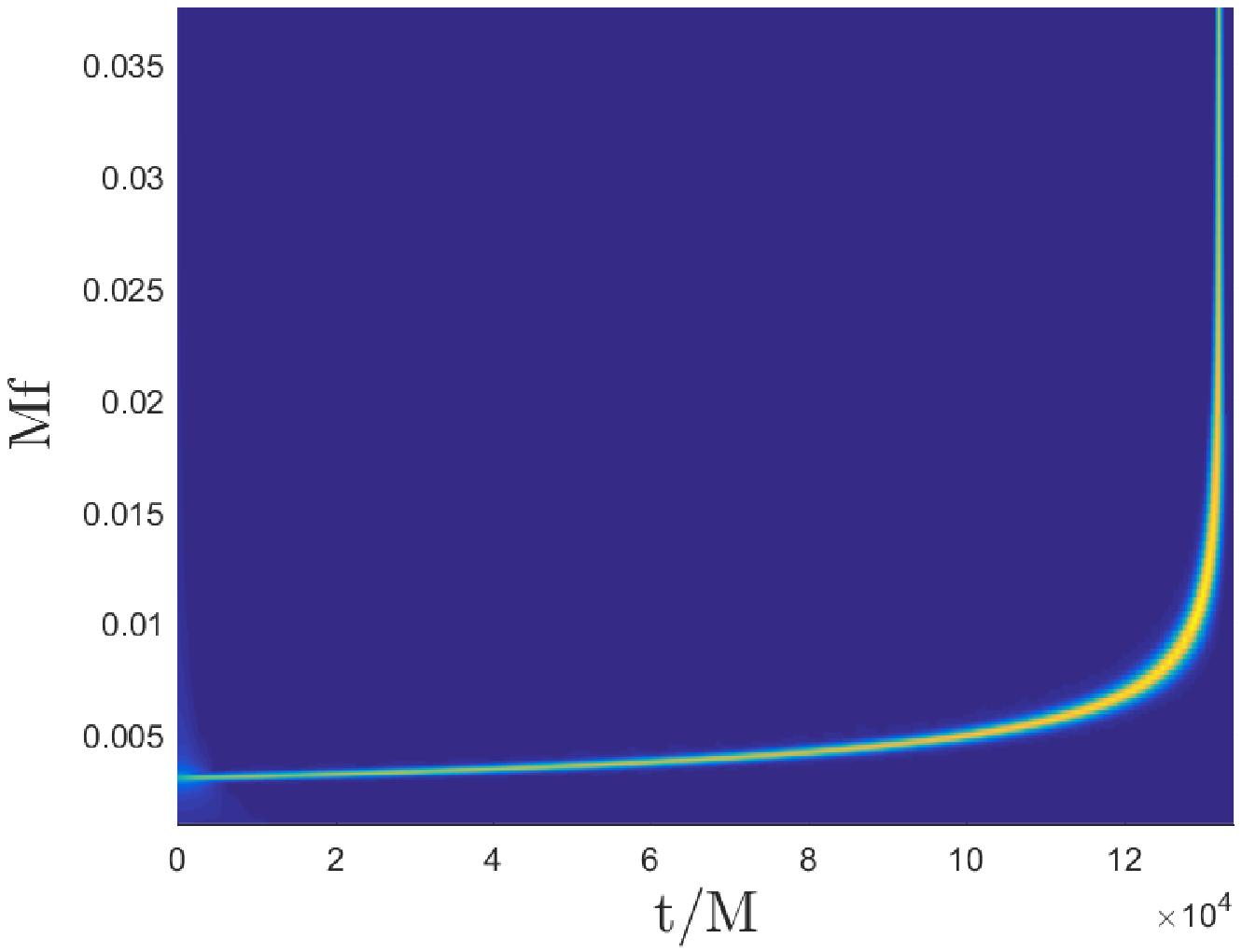}&
\includegraphics[width=0.25\textwidth]{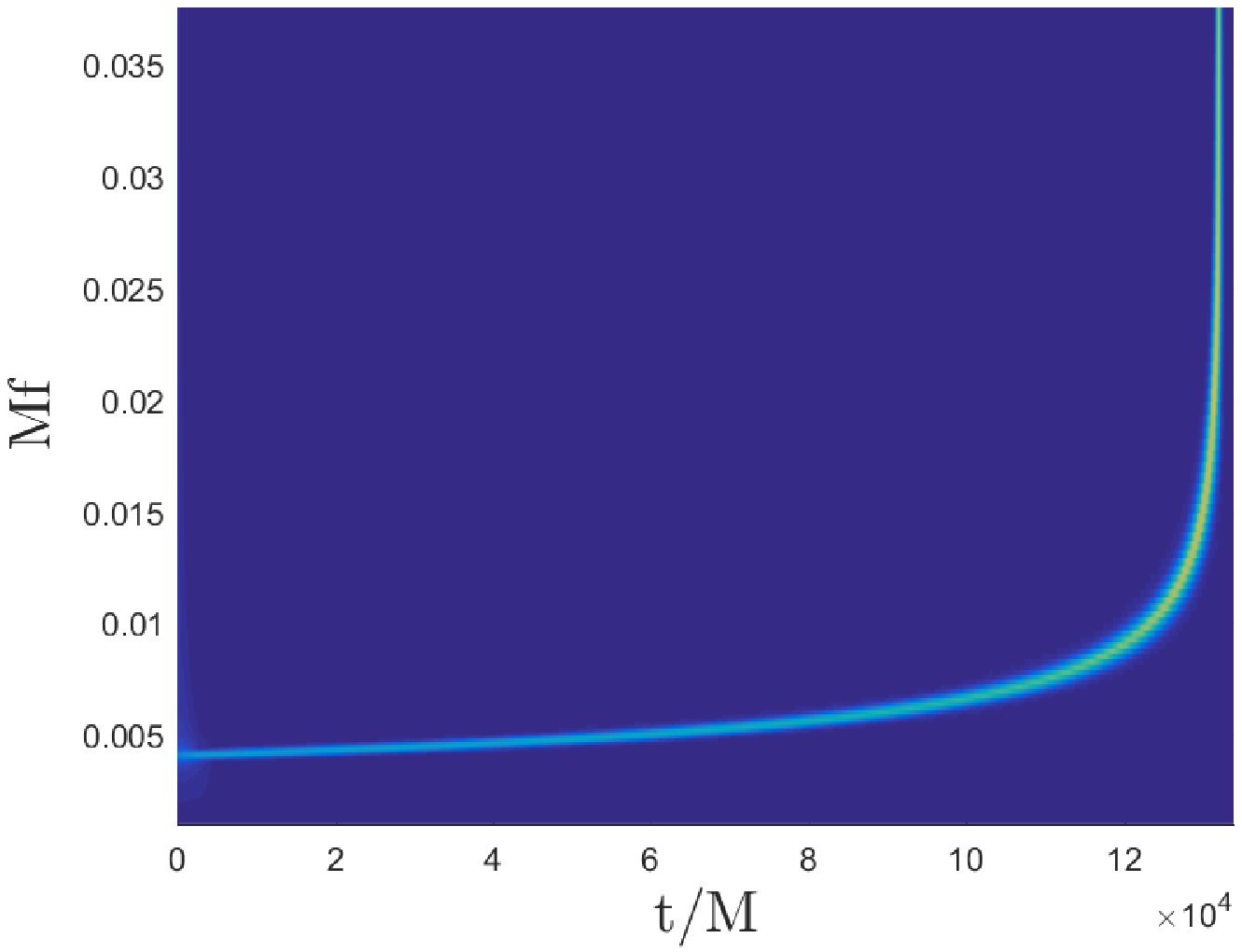}\\
\includegraphics[width=0.25\textwidth]{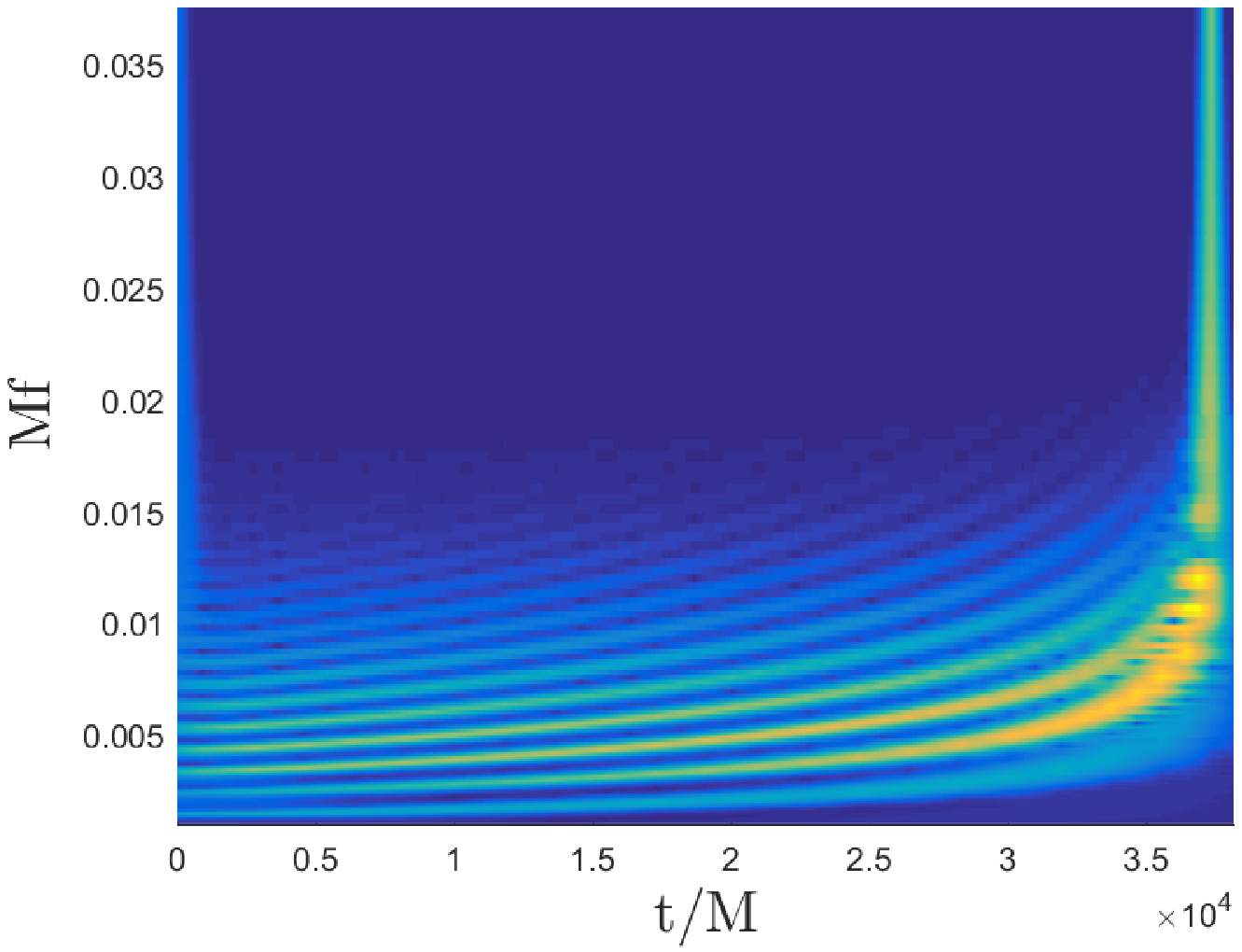}&
\includegraphics[width=0.25\textwidth]{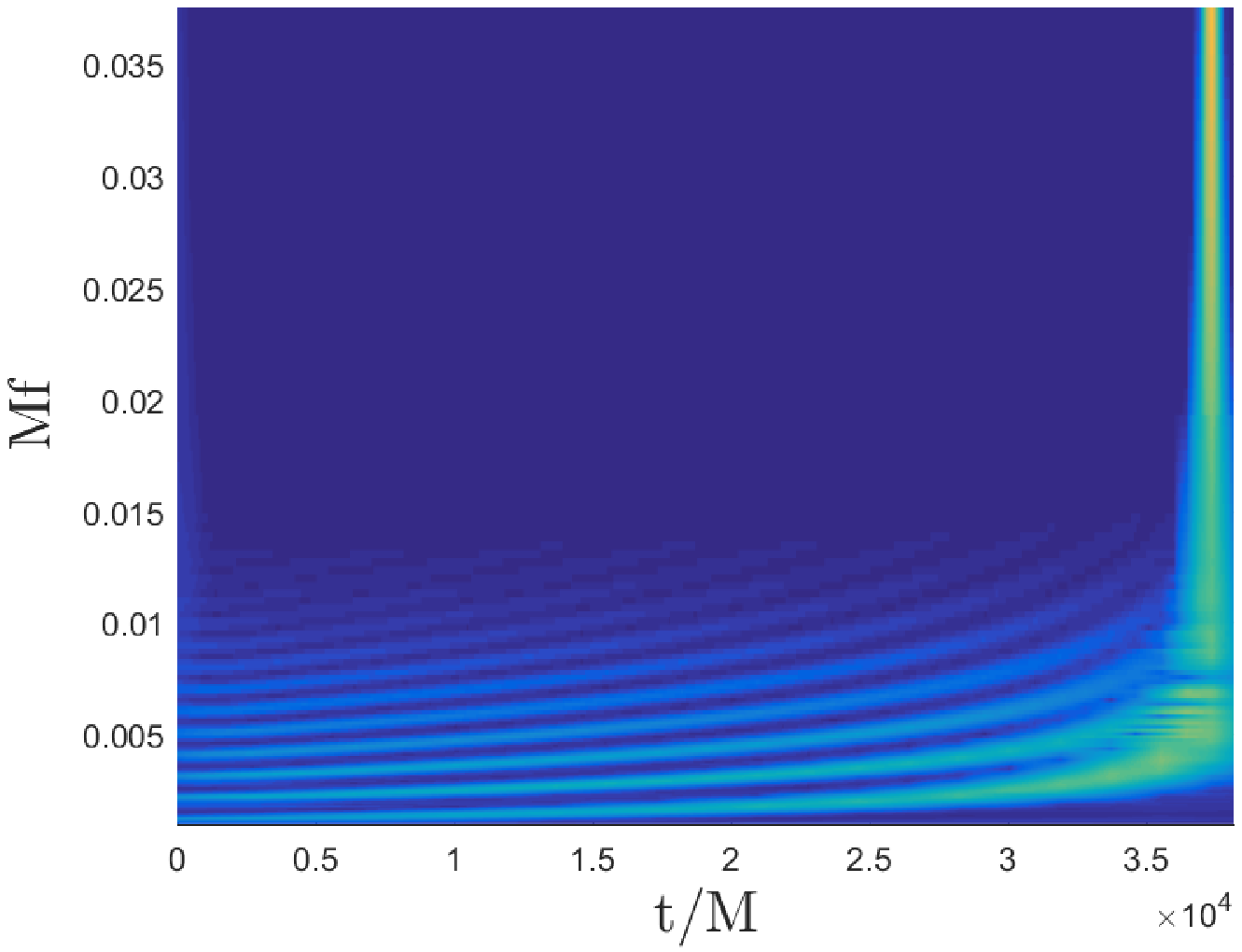}&
\includegraphics[width=0.25\textwidth]{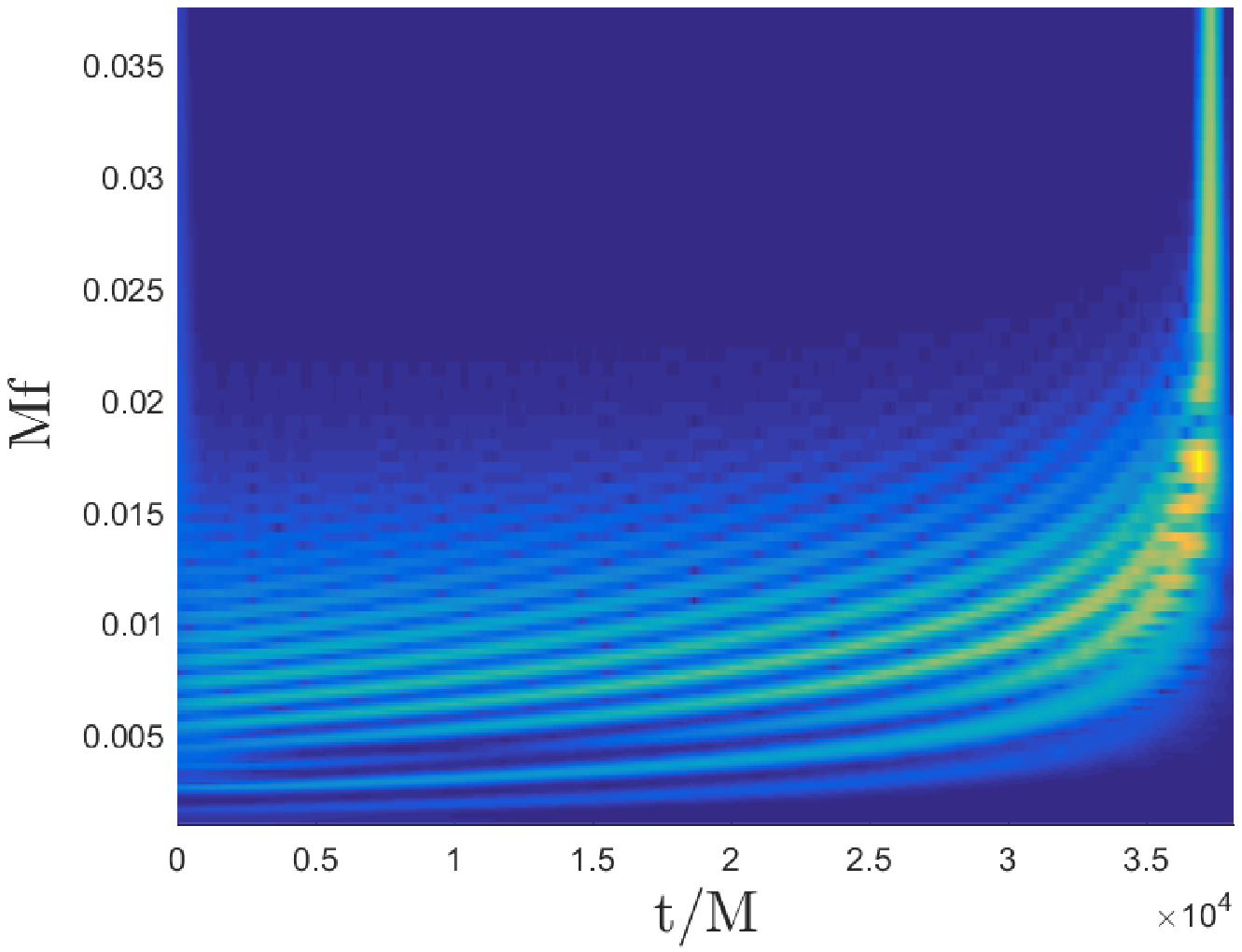}&
\includegraphics[width=0.25\textwidth]{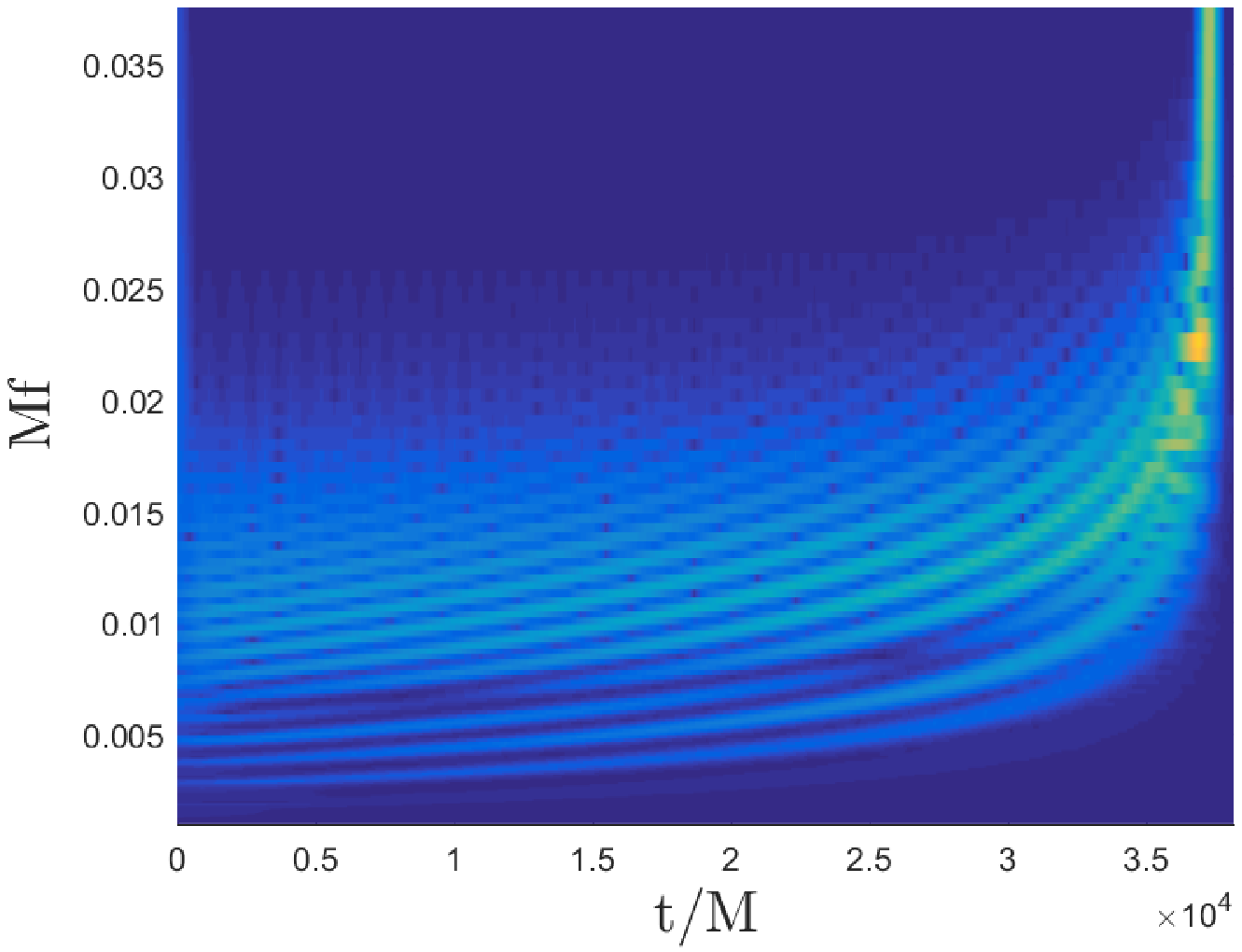}
\end{tabular}
\caption{The normalized time-frequency behavior through constant-Q-transformation for the gravitational wave modes. The BBH mass ratio and spin parameters corresponding to SXS:BBH:1433. The top row corresponds to the quasi-circular case. The bottom row corresponds to the eccentric case with initial eccentricity $e_0=0.5$ at reference frequency $Mf_0=0.002$. The columns from the left to the right correspond to (2,2), (2,1), (3,3) and (4,4) mode respectively.}\label{fig17}
\end{figure*}

\begin{figure*}
\begin{tabular}{cccc}
\includegraphics[width=0.25\textwidth]{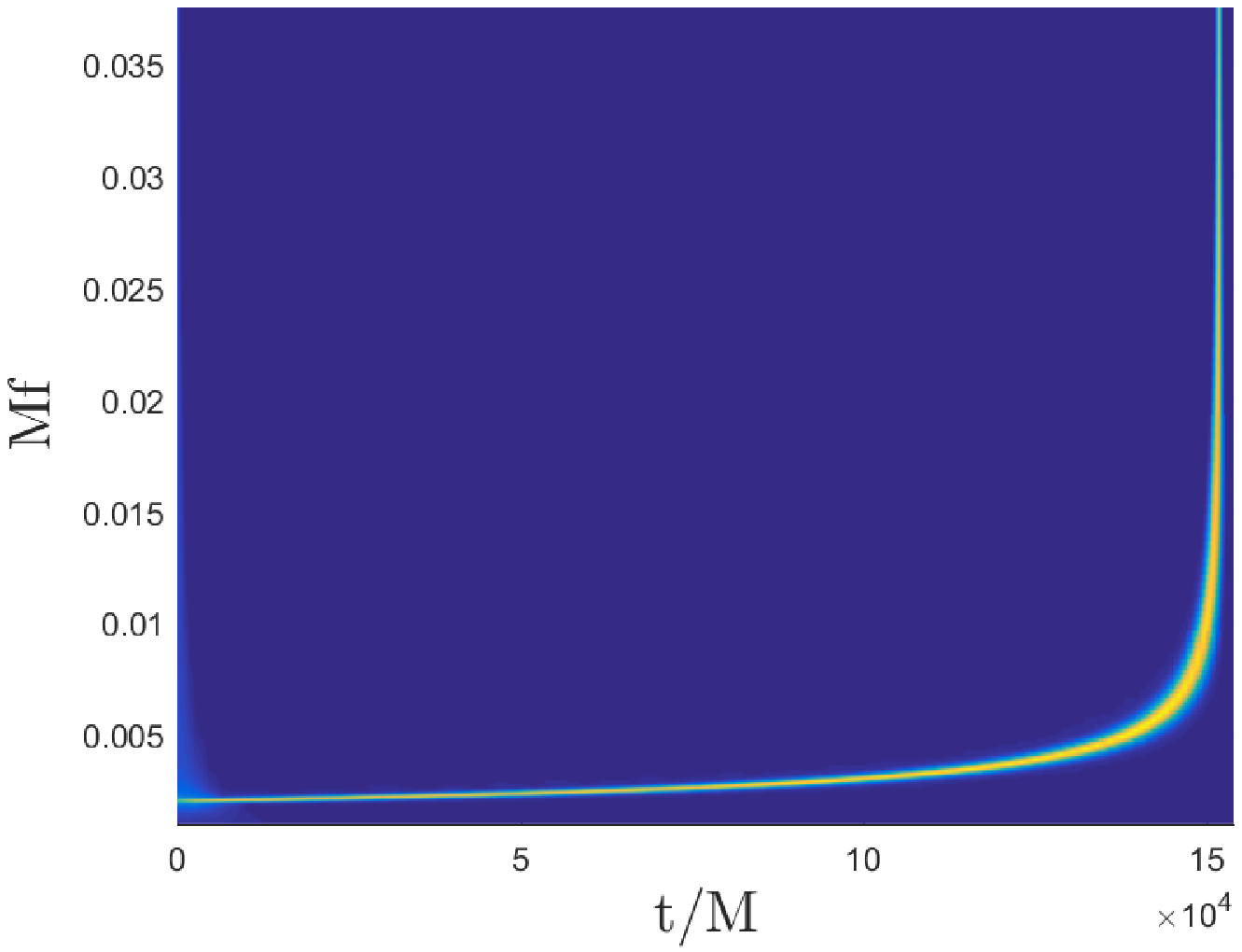}&
\includegraphics[width=0.25\textwidth]{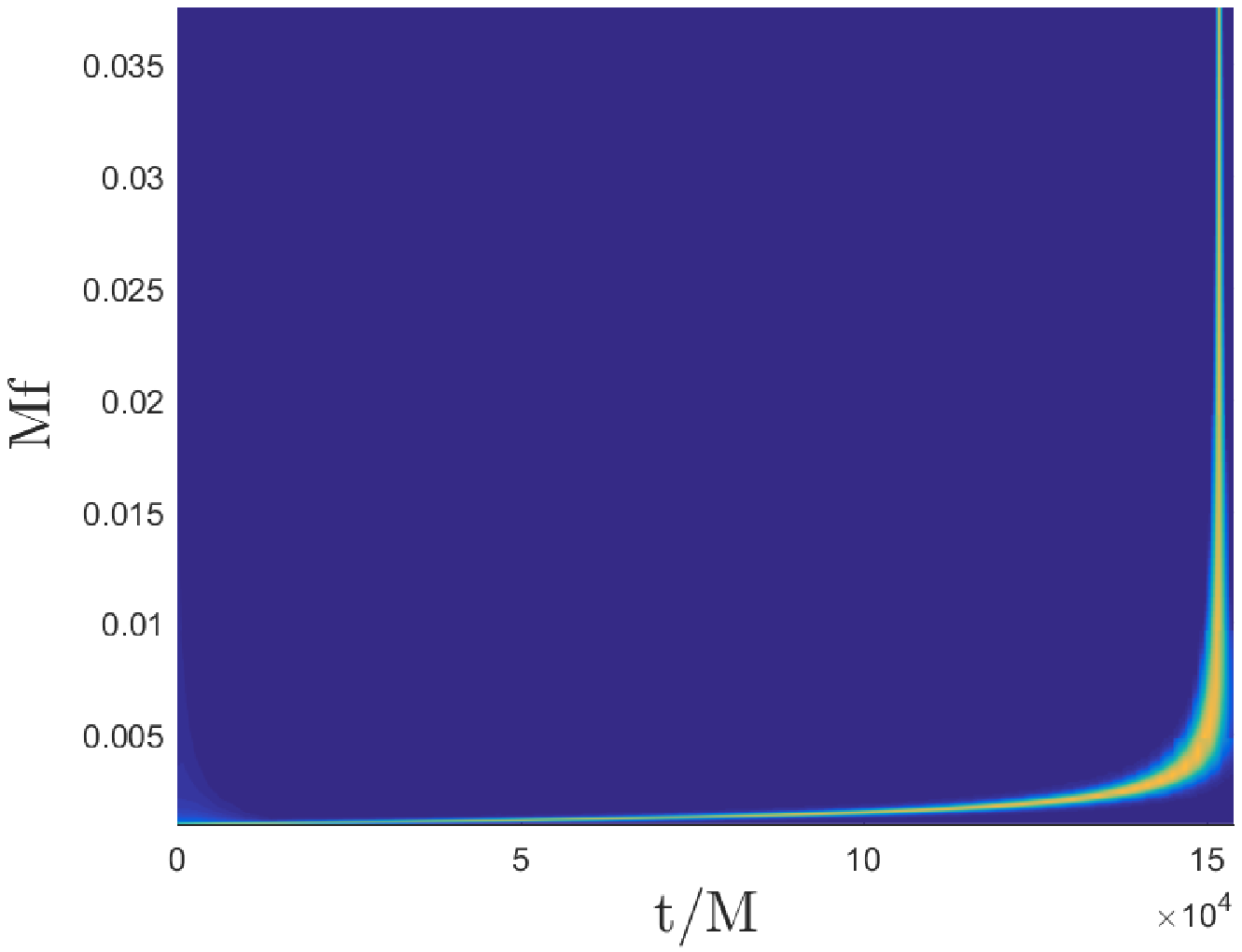}&
\includegraphics[width=0.25\textwidth]{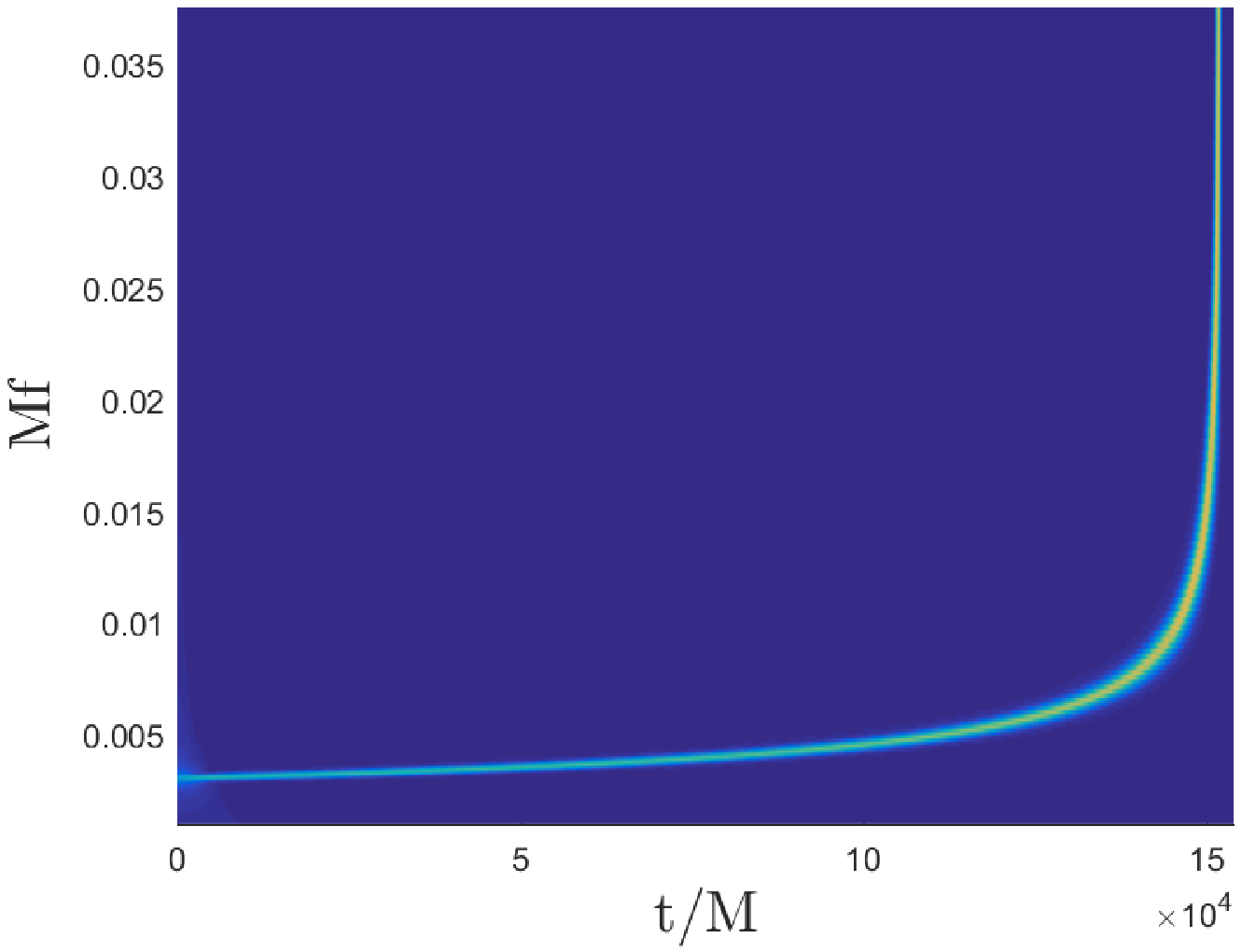}&
\includegraphics[width=0.25\textwidth]{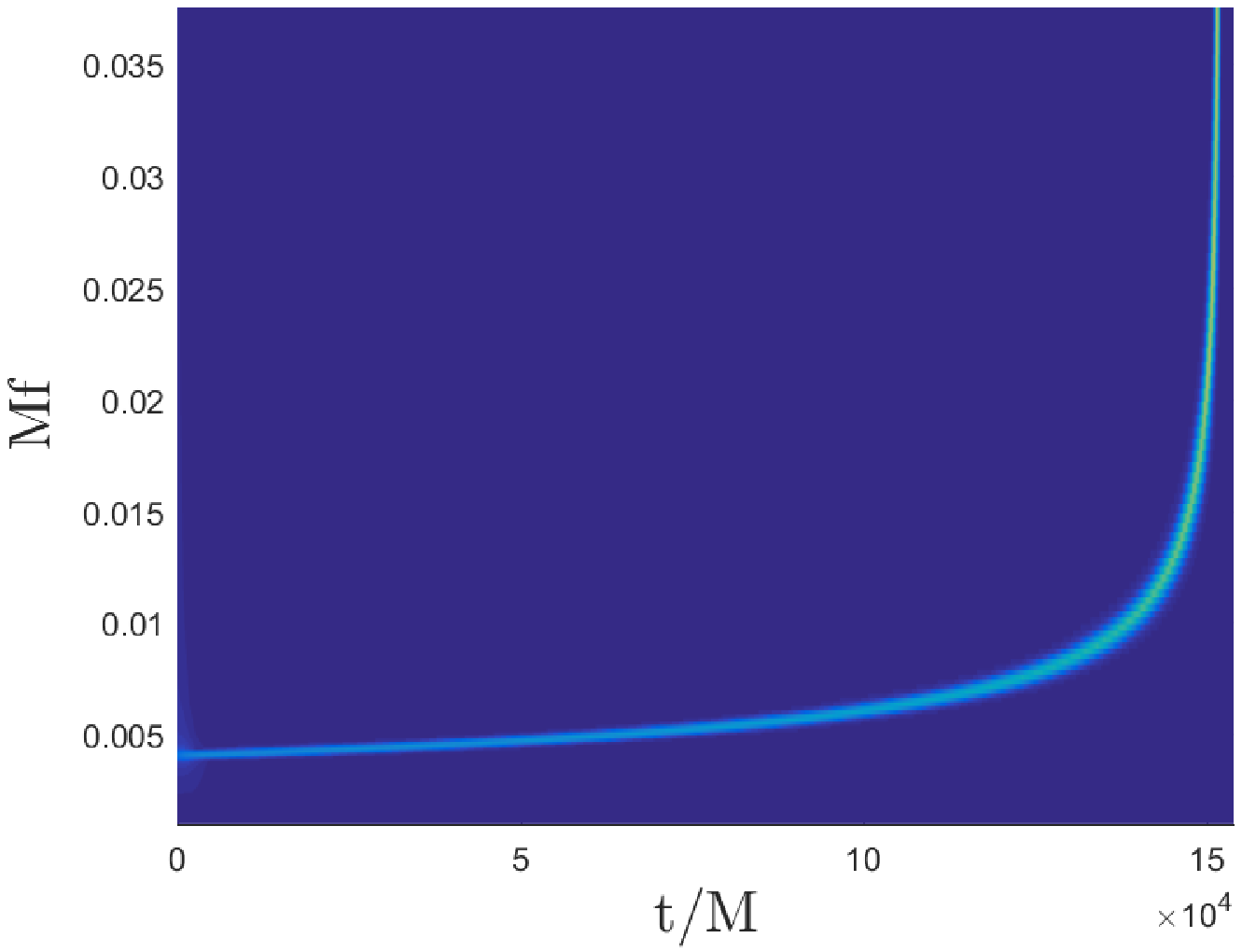}\\
\includegraphics[width=0.25\textwidth]{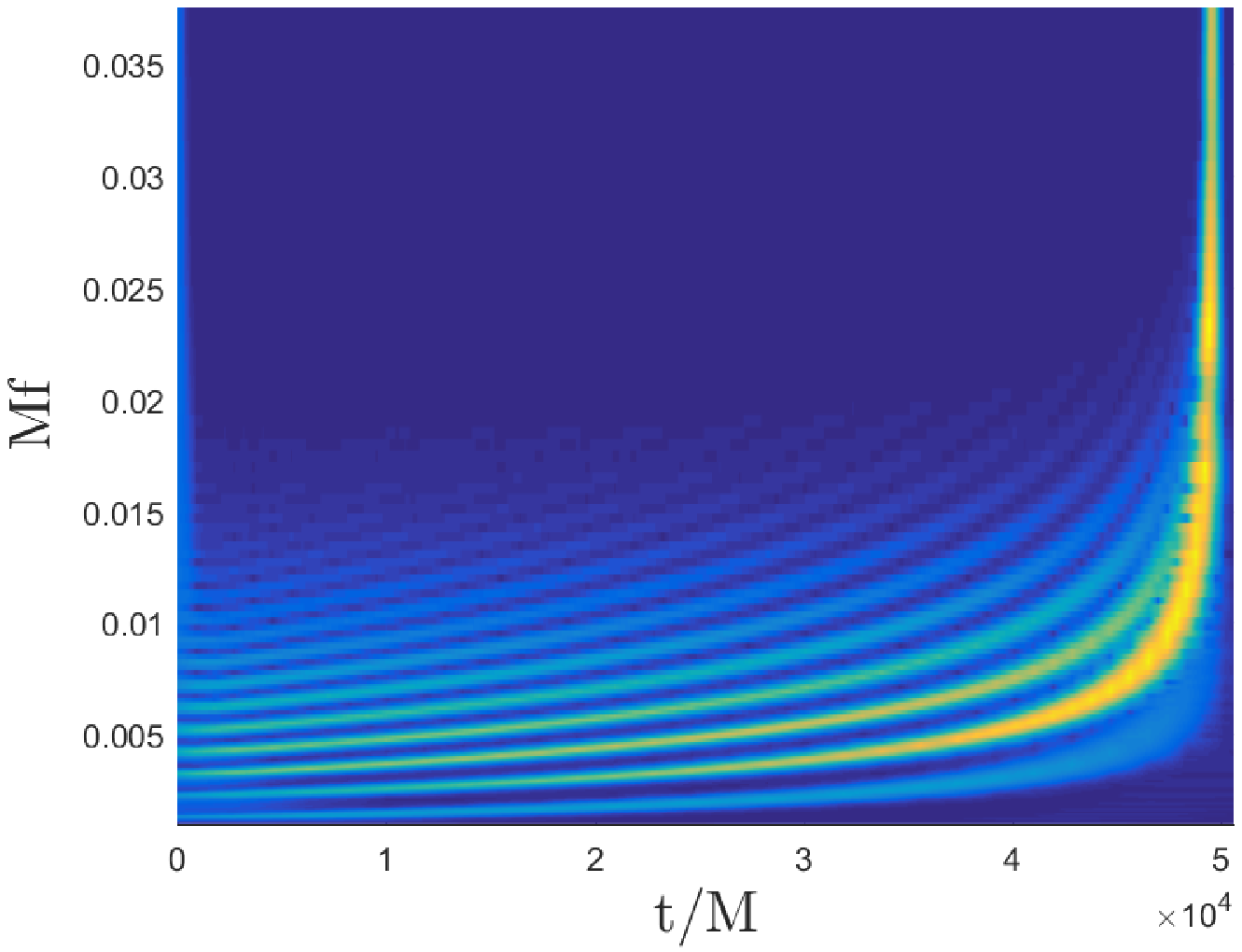}&
\includegraphics[width=0.25\textwidth]{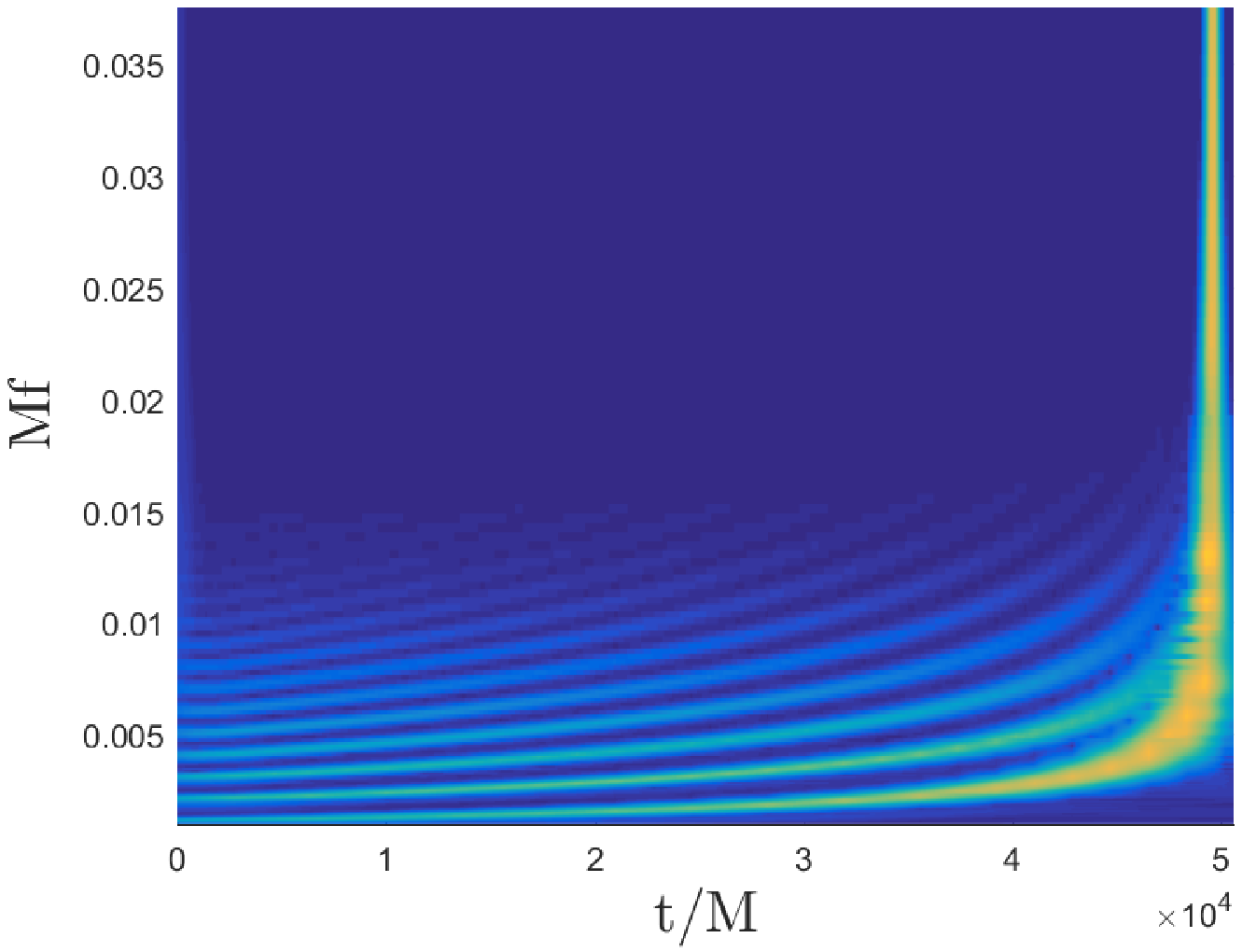}&
\includegraphics[width=0.25\textwidth]{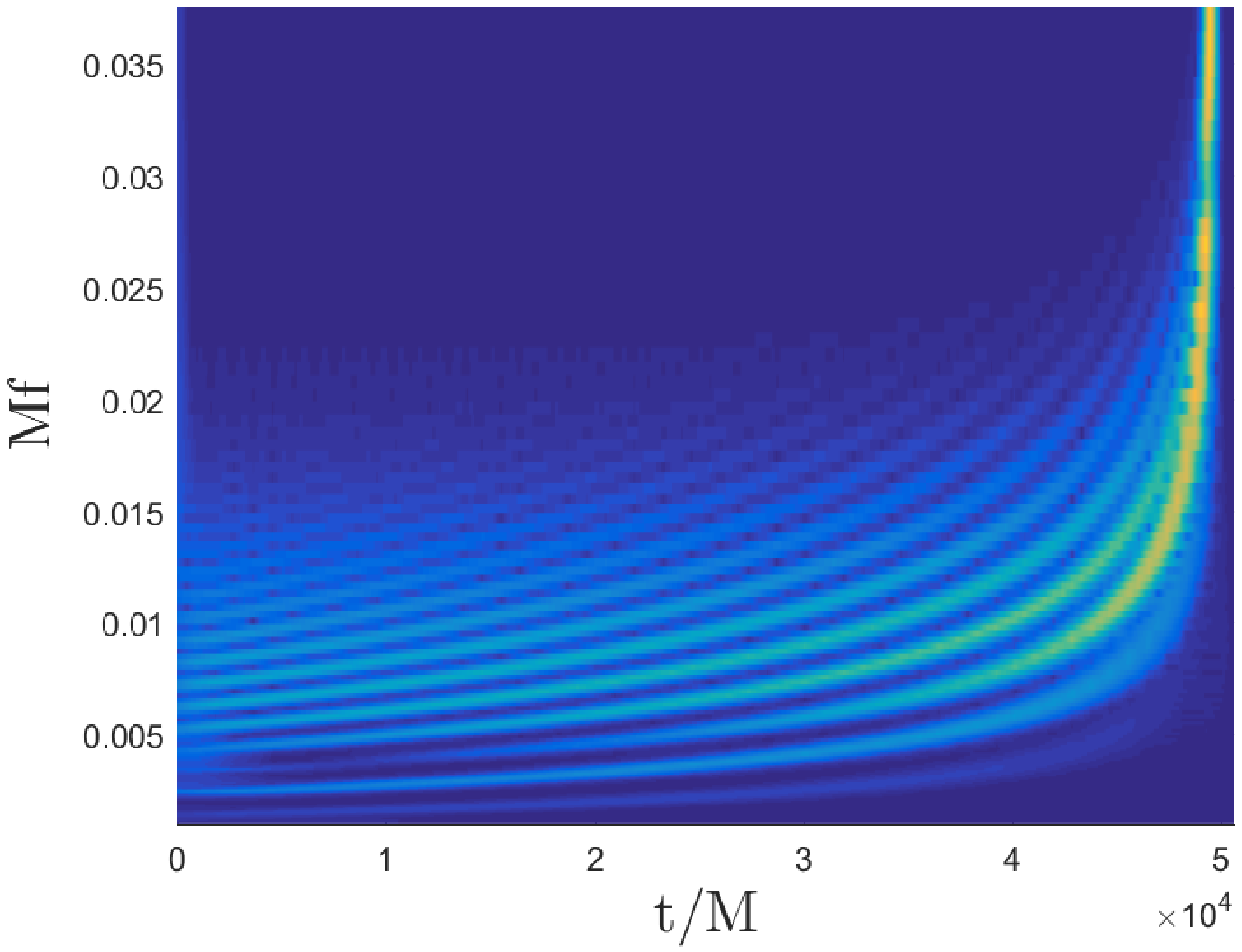}&
\includegraphics[width=0.25\textwidth]{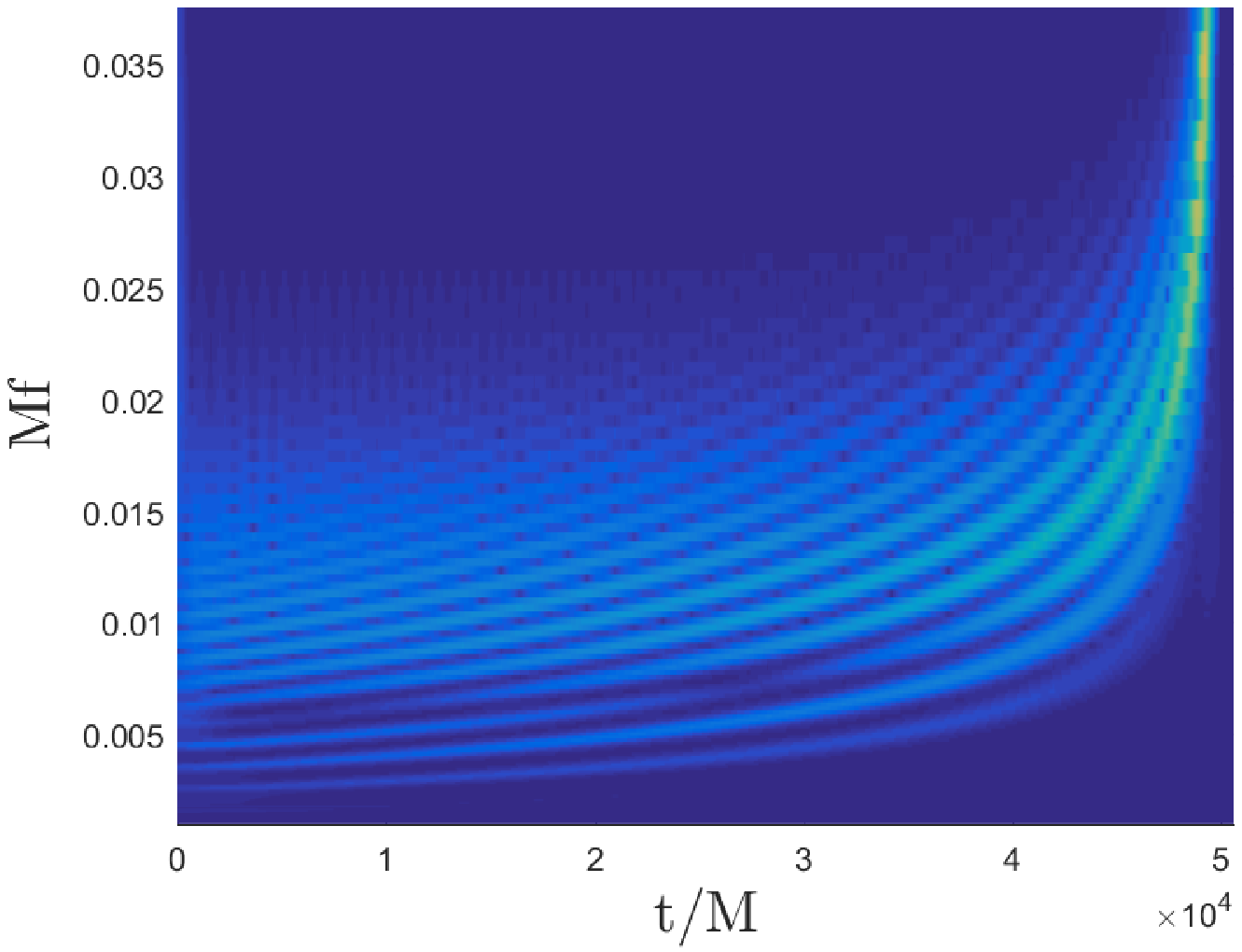}
\end{tabular}
\caption{The same to the Fig.~\ref{fig17} but for BBH mass ratio and spin parameters corresponding to SXS:BBH:1441.}\label{fig18}
\end{figure*}

So we conclude that eccentricity does not enhance the higher modes than (2,2) of the gravitational wave. Instead the eccentricity excite higher harmonic tones than the harmonic tone with frequency $\omega\approx |m| \omega_{\rm orb}$ for mode $(l,m)$, where $\omega_{\rm orb}$ means the orbital mean frequency \cite{Moore_2018,Moore_2019}. The Fig.~4 of \cite{Moore_2018} and the Fig.~5 of \cite{Moore_2019} have shown the higher harmonics excitation behavior by eccentricity based on post-Newtonian approximation. Here we correspondingly show the full general relativity results on such higher harmonics excitation in the Figs.~\ref{fig17} and \ref{fig18}. For the eccentric cases we can see the dot dash behavior of the power spectrum which is resulted from the burst near the periastron point from time to time. We once again see that the eccentricity does not amplify higher spherical-harmonic mode, instead excites the higher harmonics of each spherical-harmonic mode. The spherical-harmonic modes of quasci-circular BBHs admit solely the fundamental harmonic. In contrast, the spherical-harmonic modes of eccentric BBHs admit both the fundamental harmonic and other higher harmonics \cite{Han_2017}.
\section{Summary}\label{secV}
The recent found gravitational wave event GW190521 \cite{PhysRevLett.125.101102_GW190521,bustillo2021gw190521,bustillo2020confusing} inspired more interesting on the eccentric binary black holes \cite{2020arXiv200106492R,islam2021eccentric} for ground based GW detection. In order to properly treat the eccentricity during the gravitational wave data analysis, accurate waveform model for eccentric binary black hole is needed. Due to the GW190521 like events, the demand of waveform template for eccentric binary black holes increases. In addition, for the future gravitational wave detection projects in space, the residual eccentricity of BBH events in their frequency bands is still relatively large. For such GW sources, using gravitational wave template of circular orbit to detect gravitational wave signal will lead to the loss of signal-to-noise ratio \cite{PhysRevD.101.044049_validSEOBNRE}.

In the last few years, a handful of eccentric inspiral only \cite{PhysRevD.90.084016_EPCmodel,PhysRevD.98.104043,Moore_2018,PhysRevD.100.064006,PhysRevD.102.084042} and Inspiral-Merger-Ringdown (IMR) models \cite{PhysRevD.96.044028_SEOBNRE,PhysRevD.96.104048_eobecc,PhysRevD.97.024031_ENIGMA,PhysRevD.98.044015,PhysRevD.100.044016_ENIGMA_CalibNR,PhysRevD.101.101501_TEOBeccc} have been constructed. Among these waveform models, only the modified $\texttt{TEOBiResumS\_SM}$ \cite{PhysRevD.101.101501_TEOBeccc} is valid for higher modes than (2,2).

In this paper, we extended our previous $\texttt{SEOBNRE}$ waveform model \cite{PhysRevD.96.044028_SEOBNRE} to $\texttt{SEOBNREHM}$ to cover higher modes including (2,1), (3,3) and (4,4). At the mean time, the EOB factorized waveform is extended from quasi-circular cases to general equatorial cases, inspired by the work \cite{PhysRevD.79.064004_FWF01}. As the orbit of the binary system gradually circularizes along with the evolution, we just replace the low order post-Newtonian terms with the ones corrected by eccentricity.

Our new $\texttt{SEOBNREHM}$ waveform model can recover the $\texttt{SEOBNRv4HM}$ waveforms of quasi-circular BBHs very well. In the eccentric cases of BBH, the matching factor between the $\texttt{SEOBNREHM}$ waveform and the numerical relativistic waveform is greater than 98\%.

The Hamiltonian related to the $\texttt{SEOBNREHM}$ dynamics and the remnant black hole properties used in the ringdown part waveform are borrowed from the $\texttt{SEOBNRv4}$ model. After more numerical relativity simulations about eccentric BBHs are available \cite{islam2021eccentric}, it is straight forward to finetune the related parameters involved in the Hamiltonian and the remnant properties to improve the $\texttt{SEOBNREHM}$'s performance. Even without such finetuning, we have shown that $\texttt{SEOBNREHM}$ can match all SXS waveforms including highly eccentric ones as good as the $\texttt{SEOBNRv4}$ waveforms match the circular ones. Hopefully our $\texttt{SEOBNREHM}$ can facilitate the gravitational wave data analysis about eccentric BBHs \cite{stz2996_searchEcc02,2020arXiv200106492R}.

\section*{Acknowledgments}
We would like to thank Alessandra Buonanno, Lijing Shao and Antoni Ramos Buades for many helpful discussions. This work was supported by the NSFC (No.~11690023, No.~11633001, No.~11920101003 and No.~12021003) and by the Collaborative research program of the Institute for Cosmic Ray Research (ICRR), the University of Tokyo. Z. Cao and Z.-H. Zhu were supported by ``the Interdiscipline Research Funds of Beijing Normal University" and the Strategic Priority Research Program of the Chinese Academy of Sciences (No.~XDB23000000 and No.~XDB23040100).

\appendix
\begin{widetext}
\section{Gauge transformation between the EOB coordinates and the harmonic coordinates}\label{appendixA}
Ref.~\cite{PhysRevD.86.124012_GaugeTransform} has given the coordinate transformation among the harmonic coordinates, the ADM coordinates and the EOB coordinates, and the transformation among different canonical velocities and canonical momentums. According to such transformations we have
\begin{align}
y_0=&x_0\left[ {1-\frac{1}{4} \nu  x_2^2-\frac{1}{32} \nu  (3 \nu +14) x_1^4+\frac{1}{32} \left(-3 \nu ^2-6 \nu \right) x_2^4}\right. \label{y0tox} \\
&\left.{+x_1^2 \left(-\frac{3 \nu }{4}-\frac{1}{16} \nu  (13 \nu +10) x_2^2\right) + x_0^2 \left(\frac{1}{32} (8 \nu +16)+\frac{1}{2} \nu  x_2 \chi _S+\frac{1}{32} \left(30 \nu ^2-68 \nu \right) x_1^2+\frac{1}{32} \left(4 \nu ^2-26 \nu
   \right) x_2^2\right)}\right.\nonumber \\
&\left.{+\frac{1}{32} x_0^4 \left(32 \nu  \left(\sqrt{1-4 \nu } \chi _A \chi _S+\chi _A^2+\chi _S^2-2\right)-4 \left(4 \sqrt{1-4 \nu } \chi _A \chi _S+2 \chi
   _A^2+2 \chi _S^2-3\right)+\nu ^2 \left(7-32 \chi _S^2\right)\right)}\nonumber \right] \\
y_1=&x_1\left[ {1+\frac{1}{32} x_0^4 \left(32 \nu  \left(\sqrt{1-4 \nu } \chi _A \chi _S+\chi _A^2+\chi _S^2-2\right)-4 \left(4 \sqrt{1-4 \nu } \chi _A \chi _S+2 \chi
   _A^2+2 \chi _S^2-3\right)+\nu ^2 \left(7-32 \chi _S^2\right)\right)}\right.\label{y1tox} \\
&\left.{+x_0^2 \left(-3 \nu -\frac{1}{2} \nu  (9 \nu -11) x_1^2-\frac{1}{4} \nu  (29 \nu +17) x_2^2\right)}\right. \nonumber \\
&\left.{-\frac{1}{4} x_0^4 \left(-\nu  \left(8 \sqrt{1-4 \nu } \chi _A \chi _S+8 \chi _A^2+8 \chi _S^2-11\right)+2 \left(2 \sqrt{1-4 \nu } \chi_A \chi _S+\chi_A^2+\chi _S^2\right)+\nu ^2 \left(8 \chi _S^2-1\right)\right)} \right] \nonumber \\
y_2=&x_2\left[ {1+\frac{1}{8} \left(15 \nu ^2+7 \nu \right) x_2^4+\frac{3}{8} \nu  (\nu +1) x_1^4+x_1^2 \left(\frac{\nu }{2}+\frac{1}{4} \nu  (3 \nu +5)x_2^2\right)+\frac{3 \nu  x_2^2}{2}}\right. \label{y2tox}\\
&\left.{+x_0^2 \left(\frac{1}{8} (-12 \nu -8)+\frac{1}{8} \left(48 \nu -4 \nu ^2\right) x_1^2+\frac{1}{8} \left(-23 \nu ^2-19 \nu \right) x_2^2\right)}\right. \nonumber \\
&\left.{+\frac{1}{2} x_0^4 \left(-\nu  \left(4 \sqrt{1-4 \nu } \chi _A \chi _S+4 \chi _A^2+4 \chi _S^2-15\right)+2 \sqrt{1-4 \nu } \chi _A \chi _S+\chi _A^2+\nu^2 \left(4 \chi _S^2-1\right)+\chi _S^2\right)} \right]-\nu  x_0^4 \chi _S\nonumber \\
e^{-im\phi_h}=&e^{-im\phi}\left[ {1-\frac{1}{2} m^2 \nu ^2 x_2^2 x_1^2-\frac{1}{2} i m \nu  x_2 x_1^3-\frac{1}{4} i m (1-5 \nu ) \nu  x_0^2 x_2 x_1}\right. \label{phitrans}\\
&\left.{+ x_1 \left(\frac{1}{4} x_2^3 \left(-4 i m \nu ^2-2 i m \nu \right)-i m \nu  x_2\right)}\nonumber \right]
\end{align}
where $\chi_S=(\chi_1+\chi_2)/2, \chi_A=(\chi_1-\chi_2)/2$.

\section{2PN factorized waveform}\label{appendixB}
Here we introduce two auxiliary parameters
\begin{equation}
A_1=\frac{x_1}{x_0},\; A_2=\frac{x_2}{x_0}. \label{AppA12}
\end{equation}

We list the $\zeta_{lm}^e$ in the following. Firstly, $(2,2)$ mode:
\begin{align}
\zeta_{22}^e=&c_0^{h_{22}}+c_2^{h_{22}}x_0^2+c_3^{h_{22}}x_0^3+c_4^{h_{22}}x_0^4\\
c_0^{h_{22}} =& -\frac{A_1^2}{2}+i A_2 A_1+\frac{A_2^2}{2}+\frac{1}{2} \label{C0_22} \\
c_2^{h_{22}}=&-\frac{3}{2}+\frac{\nu }{2}+A_1^2 \left(\frac{31 \nu
   }{28}-\frac{9}{7}\right)+A_2 A_1 \left(\frac{25 i}{42}-\frac{16 i \nu }{7}\right)+A_2^4\left(\frac{15 \nu }{28}+\frac{1}{14}\right)\label{C2_22} \\&
   +A_2^2 \left(\frac{3 \nu }{28}-\frac{13}{21}\right)+A_1^4 \left(-\frac{15 \nu }{28}-\frac{1}{14}\right)+A_2 A_1^3 \left(\frac{15 i \nu }{14}+\frac{i}{7}\right)+A_2^3 A_1 \left(\frac{15 i \nu }{14}+\frac{i}{7}\right) \nonumber \\
c_3^{h_{22}}=&A_2 \left(-\sqrt{1-4 \nu } \chi _A+\frac{7 \nu  \chi _S}{6}-\chi _S\right)-\frac{1}{3} i A_1 \left(3 \sqrt{1-4 \nu } \chi _A+(3-5 \nu ) \chi _S\right)\label{C3_22} \\
c_4^{h_{22}}=&\frac{205 \nu ^2}{252}-\frac{31 \nu }{36}+\frac{65}{252} + A_1^4 \left(\frac{335 \nu ^2}{336}-\frac{923 \nu }{336}-\frac{11}{21}\right) + A_1^6 \left(-\frac{25 \nu ^2}{48}-\frac{179 \nu }{336}-\frac{1}{42}\right)\label{C4_22} \\
+&A_1 A_2 \left(-\frac{307 i \nu ^2}{378}-\frac{592 i \nu }{189}-\frac{359 i}{378}\right) + A_1^3 A_2 \left(-\frac{131 i \nu ^2}{63}+\frac{935 i \nu }{252}+\frac{71 i}{63}\right) + A_1^5 A_2 \left(\frac{25 i \nu ^2}{24}+\frac{179 i \nu }{168}+\frac{i}{21}\right)\nonumber\\
&+A_1^2 A_2^2 \left(-\frac{109 \nu ^2}{168}-\frac{45 \nu }{56}+\frac{23}{84}\right) + A_1^4 A_2^2 \left(-\frac{25 \nu ^2}{48}-\frac{179 \nu }{336}-\frac{1}{42}\right)+A_1 A_2^3 \left(-\frac{58 i \nu ^2}{21}+\frac{29 i \nu }{12}-\frac{i}{168}\right)\nonumber\\
&+A_1^3 A_2^3 \left(\frac{25 i \nu ^2}{12}+\frac{179 i \nu }{84}+\frac{2 i}{21}\right) + A_2^4 \left(-\frac{101 \nu ^2}{144}-\frac{1045 \nu }{1008}+\frac{317}{1008}\right)+A_1^2 A_2^4 \left(\frac{25 \nu ^2}{48}+\frac{179 \nu }{336}+\frac{1}{42}\right)\nonumber \\
&+A_2^6 \left(\frac{25 \nu ^2}{48}+\frac{179 \nu }{336}+\frac{1}{42}\right)+A_1 A_2^5 \left(\frac{25 i \nu ^2}{24}+\frac{179 i \nu}{168}+\frac{i}{21}\right)\nonumber \\
&+A_1^2 \left(-\frac{20 \nu ^2}{63}+\frac{1355 \nu }{252}-\frac{425}{504}\right) + A_2^2 \left(\frac{26 \nu ^2}{189}-\frac{635 \nu }{189}-\frac{10379}{3024}\right) +  \frac{A_1^2+A_2^2}{2}\left( \sqrt{1-4\nu}\chi_A + (1-2\nu)\chi_S \right)^2\nonumber \\
&+\frac{1}{2}\left[ \left(\sqrt{1-4\nu}\chi_A+(1+\nu)\chi_S\right)^2-3\nu^2\chi_S^2 \right].\nonumber
\end{align}

$(2,1)$ mode:
\begin{align}
\zeta_{21}^e=&c_0^{h_{21}}+c_1^{h_{21}}x_0+c_2^{h_{21}}x_0^2\\
A_2^{\frac{4}{3}}c_0^{h_{21}}=&A_2\label{C0_21} \\
A_2^{\frac{4}{3}}c_1^{h_{21}}=&-\frac{3 \chi _A}{2 \sqrt{1-4 \nu }}-\frac{3 \chi _S}{2} \label{C1_21} \\
A_2^{\frac{4}{3}}c_2^{h_{21}}=&A_2^3 \left(\frac{5}{28}-\frac{5 \nu }{14}\right)+A_1 A_2^2 \left(\frac{83 i}{14}-\frac{6 i \nu }{7}\right)+A_1^2 A_2 \left(-\frac{11 \nu
   }{14}-\frac{20}{7}\right)+A_2 \left(\frac{11 \nu }{7}-\frac{16}{7}\right) \label{C2_21} \\
 A_2^{\frac{4}{3}}c_3^{h_{21}}=&\frac{1}{14} \left(\frac{(84-95 \nu ) \chi _A}{\sqrt{1-4 \nu }}+(84-59 \nu ) \chi _S\right) + i \frac{A_1 A_2}{14} \left(\frac{(104 \nu -147)}{\sqrt{1-4 \nu }} \chi _A+(8 \nu -147) \chi _S\right)\label{C3_21}\\
 & + \frac{1}{28} A_1^2 \left(\frac{(95 \nu +126) \chi _A}{\sqrt{1-4 \nu }}+3 (29 \nu +42) \chi _S\right) + \frac{1}{28} A_2^2 \left(\frac{(327 \nu -63) \chi _A}{\sqrt{1-4 \nu }}+(107 \nu -63) \chi _S\right).\nonumber
 \end{align}

$(3,3)$ mode:
\begin{align}
\zeta_{33}^e=&c_0^{h_{33}}+c_2^{h_{33}}x_0^2+c_3^{h_{33}}x_0^3\\
c_0^{h_{33}}=&-\frac{2}{9} i A_1^3-\frac{2}{3} A_2 A_1^2+\frac{2}{3} i A_2^2 A_1+\frac{4 i A_1}{9}+\frac{2 A_2^3}{9}+\frac{7 A_2}{9} \label{C0_33} \\
c_2^{h_{33}}=&A_1^5 \left(-\frac{8 i \nu }{27}-\frac{2 i}{27}\right)+A_1^3 \left(\frac{20 i \nu }{27}-\frac{28 i}{27}\right)+A_1 \left(-\frac{4 i \nu }{81}-\frac{37i}{81}\right) \label{C2_33} \\
&+A_2 A_1^4 \left(-\frac{8 \nu }{9}-\frac{2}{9}\right)+A_2 A_1^2 \left(\frac{19 \nu }{9}-\frac{20}{9}\right)+A_2 \left(\frac{22 \nu}{27}-\frac{80}{27}\right) \nonumber\\
&+A_2^2 A_1^3 \left(\frac{16 i \nu }{27}+\frac{4 i}{27}\right)+A_2^2 A_1 \left(\frac{2 i}{3}-2 i \nu \right)+A_2^3 \left(\frac{2 \nu}{3}-\frac{11}{18}\right) \nonumber \\
&+A_2^5 \left(\frac{8 \nu }{27}+\frac{2}{27}\right)+A_1 A_2^4 \left(\frac{8 i \nu }{9}+\frac{2 i}{9}\right)+A_1^2 A_2^3 \left(-\frac{16 \nu}{27}-\frac{4}{27}\right) \nonumber \\
c_3^{h_{33}}=&\frac{1}{9} \left(\frac{2 (5 \nu -1) \chi _A}{\sqrt{1-4 \nu }}+(3 \nu -2) \chi _S\right) + \frac{1}{9} A_1^2 \left(3 (2-3 \nu ) \chi _S-\frac{(25 \nu -6) \chi _A}{\sqrt{1-4 \nu }}\right)\label{C3_33} \\
&+\frac{i A_1 A_2}{9} \left( \frac{(77 \nu -18)}{\sqrt{1-4 \nu }} \chi _A+(31 \nu -18) \chi _S\right) + \frac{1}{18} A_2^2 \left(\frac{(119 \nu -24) \chi _A}{\sqrt{1-4 \nu }}+(35 \nu -24) \chi _S\right). \nonumber
\end{align}

$(4,4)$ mode:
\begin{align}
\zeta_{44}^e=&c_0^{h_{44}}+c_2^{h_{44}}x_0^2 \\
c_0^{h_{44}}=&\frac{3 A_1^4}{32}-\frac{3}{8} i A_2 A_1^3-\frac{9}{16} A_2^2 A_1^2-\frac{9 A_1^2}{32}+\frac{3}{8} i A_2^3 A_1+\frac{27}{32} i A_2 A_1+\frac{3
   A_2^4}{32}+\frac{51 A_2^2}{64}+\frac{7}{64} \label{C0_44} \\
c_2^{h_{44}}=&\frac{1}{1-3\nu}\left[ {\frac{1}{704} \left(-621 \nu ^2+1358 \nu -397\right) +\frac{3 A_1^3 \left(A_1^3 \left(-525 \nu ^2+30 \nu +60\right)-i A_2 \left(6075 \nu ^2-9650 \nu +2348\right)\right)}{3520} }\right. \label{C2_44}\\
&\left.{-\frac{3 A_2 \left(A_2 \left(3 A_2^2 \left(4570 \nu ^2-5120 \nu +859\right)+12910 \nu ^2-57460 \nu +17971\right)-240 i A_1^5 \left(35 \nu ^2-2 \nu
   -4\right)\right)}{14080}}\right. \nonumber\\
   &\left.{+\frac{9}{704} \left(A_2^6 \left(-35 \nu ^2+2 \nu +4\right)+A_1^4 \left(5 A_2^2 \left(35 \nu ^2-2 \nu -4\right)+109 \nu ^2-186 \nu +46\right)\right)}\right.\nonumber\\
   &\left.{+\frac{A_1^2 \left(450 A_2^4 \left(35 \nu ^2-2 \nu -4\right)-3 A_2^2 \left(16185 \nu ^2-22730 \nu +5426\right)-5 \left(783 \nu ^2-2290 \nu
   +579\right)\right)}{7040}}\right. \nonumber\\
   &\left.{-\frac{i A_1 A_2 \left(180 A_2^4 \left(35 \nu ^2-2 \nu -4\right)-3 A_2^2 \left(4425 \nu ^2-2940 \nu +577\right)-4650 \nu ^2-5360 \nu
   +2832\right)}{3520}} \right].\nonumber
\end{align}

\end{widetext}

\section{Numerical-relativity waveforms for quasi-circular BBH used in the current work}\label{CIRCtable}
We have used 281 numerical relativity simulations in the current paper for quasi-circular BBH corresponding different parameters combination of symmetric mass ratio $\nu$ and the black holes' spin parameters $\chi_{1z}$ and $\chi_{2z}$. In the following we list the parameters for these numerical relativity simulations, the corresponding minimal matching factor FF${}_{\rm 22min}$ for (2,2) mode (check the Sec.~\ref{secIII} for detail explanation) and the averaged matching factor $\overline{\rm FF}(\boldsymbol{\theta})$, the minimal matching factor ${\rm FF}_{\rm min}(\boldsymbol{\theta})$ respect to the three angles $(\kappa, \iota, \varphi)$ (check the Sec.~\ref{secIV} for detail explanation).
\begingroup
\begin{longtable}{ccccccc}
\hline \hline
  \input{sxswaveformtablec_new}
\hline \hline
\end{longtable}
\endgroup

\bibliography{refs}

\end{document}

%% file: sxswaveformtablec_new.tex
ID & $q$ & $\chi_1$ & $\chi_2$ & FF${}_{\rm 22min}$ & $\overline{\rm FF}(\boldsymbol{\theta})$ & ${\rm FF}_{\rm min}(\boldsymbol{\theta})$ \\
\hline
0001 & $1.00$ & $0.00$ & $0.00$ & $99.58\%$ & $99.58\%$ & $99.56\%$ \\
0070 & $1.00$ & $0.00$ & $0.00$ & $99.58\%$ & $99.58\%$ & $99.49\%$ \\
0004 & $1.00$ & $-0.50$ & $0.00$ & $99.77\%$ & $99.68\%$ & $99.48\%$ \\
0005 & $1.00$ & $+0.50$ & $0.00$ & $99.88\%$ & $99.72\%$ & $99.45\%$ \\
0007 & $1.50$ & $0.00$ & $0.00$ & $99.49\%$ & $99.49\%$ & $99.47\%$ \\
0008 & $1.50$ & $0.00$ & $0.00$ & $99.57\%$ & $99.57\%$ & $99.53\%$ \\
0009 & $1.50$ & $+0.50$ & $0.00$ & $99.71\%$ & $99.62\%$ & $99.35\%$ \\
0013 & $1.50$ & $+0.50$ & $0.00$ & $99.86\%$ & $99.63\%$ & $99.35\%$ \\
0014 & $1.50$ & $-0.50$ & $0.00$ & $99.57\%$ & $99.31\%$ & $98.92\%$ \\
0016 & $1.50$ & $-0.50$ & $0.00$ & $99.56\%$ & $99.38\%$ & $99.03\%$ \\
0019 & $1.50$ & $-0.50$ & $+0.50$ & $99.70\%$ & $99.59\%$ & $99.30\%$ \\
0025 & $1.50$ & $+0.50$ & $-0.50$ & $99.47\%$ & $99.53\%$ & $99.00\%$ \\
0030 & $3.00$ & $0.00$ & $0.00$ & $99.66\%$ & $99.72\%$ & $99.63\%$ \\
0031 & $3.00$ & $+0.50$ & $0.00$ & $99.52\%$ & $99.56\%$ & $99.36\%$ \\
0036 & $3.00$ & $-0.50$ & $0.00$ & $99.95\%$ & $99.87\%$ & $99.78\%$ \\
0038 & $3.00$ & $-0.50$ & $0.00$ & $99.69\%$ & $99.79\%$ & $99.55\%$ \\
0039 & $3.00$ & $-0.50$ & $0.00$ & $99.92\%$ & $99.81\%$ & $99.53\%$ \\
0040 & $3.00$ & $-0.50$ & $0.00$ & $99.47\%$ & $99.65\%$ & $98.78\%$ \\
0041 & $3.00$ & $+0.50$ & $0.00$ & $99.10\%$ & $99.52\%$ & $99.30\%$ \\
0045 & $3.00$ & $+0.50$ & $-0.50$ & $99.40\%$ & $99.60\%$ & $99.33\%$ \\
0046 & $3.00$ & $-0.50$ & $-0.50$ & $99.61\%$ & $99.80\%$ & $99.58\%$ \\
0047 & $3.00$ & $+0.50$ & $+0.50$ & $99.49\%$ & $99.44\%$ & $99.26\%$ \\
0054 & $5.00$ & $0.00$ & $0.00$ & $99.69\%$ & $99.77\%$ & $99.45\%$ \\
0055 & $5.00$ & $0.00$ & $0.00$ & $99.84\%$ & $99.79\%$ & $99.56\%$ \\
0056 & $5.00$ & $0.00$ & $0.00$ & $99.89\%$ & $99.73\%$ & $99.41\%$ \\
0060 & $5.00$ & $-0.50$ & $0.00$ & $99.75\%$ & $99.66\%$ & $99.05\%$ \\
0061 & $5.00$ & $+0.50$ & $0.00$ & $98.99\%$ & $98.81\%$ & $98.50\%$ \\
0063 & $8.00$ & $0.00$ & $0.00$ & $99.80\%$ & $99.44\%$ & $98.55\%$ \\
0064 & $8.00$ & $-0.50$ & $0.00$ & $99.38\%$ & $99.37\%$ & $98.03\%$ \\
0065 & $8.00$ & $+0.50$ & $0.00$ & $98.96\%$ & $98.38\%$ & $97.69\%$ \\
0066 & $1.00$ & $0.00$ & $0.00$ & $99.58\%$ & $99.58\%$ & $99.55\%$ \\
0083 & $1.00$ & $+0.50$ & $0.00$ & $99.85\%$ & $99.78\%$ & $99.65\%$ \\
0084 & $1.00$ & $+0.50$ & $0.00$ & $99.91\%$ & $99.72\%$ & $99.46\%$ \\
0085 & $1.00$ & $+0.50$ & $0.00$ & $99.90\%$ & $99.75\%$ & $99.45\%$ \\
0086 & $1.00$ & $0.00$ & $0.00$ & $99.58\%$ & $99.57\%$ & $99.56\%$ \\
0087 & $1.00$ & $0.00$ & $0.00$ & $99.65\%$ & $99.64\%$ & $99.34\%$ \\
0089 & $1.00$ & $-0.50$ & $0.00$ & $98.95\%$ & $98.40\%$ & $98.36\%$ \\
0090 & $1.00$ & $0.00$ & $0.00$ & $99.60\%$ & $99.54\%$ & $99.30\%$ \\
0091 & $1.00$ & $0.00$ & $0.00$ & $99.61\%$ & $99.60\%$ & $99.44\%$ \\
0093 & $1.50$ & $0.00$ & $0.00$ & $99.49\%$ & $99.49\%$ & $99.44\%$ \\
0100 & $1.50$ & $0.00$ & $0.00$ & $99.51\%$ & $99.48\%$ & $99.44\%$ \\
0101 & $1.50$ & $-0.50$ & $0.00$ & $99.55\%$ & $99.43\%$ & $99.08\%$ \\
0105 & $3.00$ & $-0.50$ & $0.00$ & $99.88\%$ & $99.86\%$ & $99.77\%$ \\
0106 & $5.00$ & $0.00$ & $0.00$ & $99.56\%$ & $99.54\%$ & $99.24\%$ \\
0107 & $5.00$ & $0.00$ & $0.00$ & $99.91\%$ & $99.80\%$ & $99.54\%$ \\
0108 & $5.00$ & $-0.50$ & $0.00$ & $99.45\%$ & $99.45\%$ & $99.31\%$ \\
0109 & $5.00$ & $-0.50$ & $0.00$ & $99.58\%$ & $99.71\%$ & $99.36\%$ \\
0110 & $5.00$ & $+0.50$ & $0.00$ & $99.05\%$ & $98.55\%$ & $98.01\%$ \\
0111 & $5.00$ & $-0.50$ & $0.00$ & $99.59\%$ & $99.34\%$ & $98.08\%$ \\
0112 & $5.00$ & $0.00$ & $0.00$ & $99.85\%$ & $99.77\%$ & $99.48\%$ \\
0113 & $5.00$ & $0.00$ & $0.00$ & $99.80\%$ & $99.77\%$ & $99.45\%$ \\
0114 & $8.00$ & $-0.50$ & $0.00$ & $99.56\%$ & $99.57\%$ & $98.83\%$ \\
0148 & $1.00$ & $-0.44$ & $-0.44$ & $99.13\%$ & $99.05\%$ & $98.38\%$ \\
0149 & $1.00$ & $-0.20$ & $-0.20$ & $99.80\%$ & $99.78\%$ & $99.63\%$ \\
0150 & $1.00$ & $+0.20$ & $+0.20$ & $99.82\%$ & $99.89\%$ & $99.75\%$ \\
0151 & $1.00$ & $-0.60$ & $-0.60$ & $99.87\%$ & $99.83\%$ & $99.60\%$ \\
0152 & $1.00$ & $+0.60$ & $+0.60$ & $99.48\%$ & $99.44\%$ & $99.34\%$ \\
0153 & $1.00$ & $+0.85$ & $+0.85$ & $99.68\%$ & $99.46\%$ & $99.20\%$ \\
0154 & $1.00$ & $-0.80$ & $-0.80$ & $99.72\%$ & $99.66\%$ & $99.32\%$ \\
0155 & $1.00$ & $+0.80$ & $+0.80$ & $99.69\%$ & $99.54\%$ & $99.36\%$ \\
0157 & $1.00$ & $+0.95$ & $+0.95$ & $99.61\%$ & $99.25\%$ & $98.75\%$ \\
0158 & $1.00$ & $+0.97$ & $+0.97$ & $99.38\%$ & $98.77\%$ & $97.06\%$ \\
0159 & $1.00$ & $-0.90$ & $-0.90$ & $99.86\%$ & $99.80\%$ & $99.47\%$ \\
0160 & $1.00$ & $+0.90$ & $+0.90$ & $99.61\%$ & $99.36\%$ & $99.03\%$ \\
0162 & $2.00$ & $+0.60$ & $0.00$ & $99.83\%$ & $99.73\%$ & $99.63\%$ \\
0166 & $6.00$ & $0.00$ & $0.00$ & $99.70\%$ & $99.66\%$ & $99.16\%$ \\
0167 & $4.00$ & $0.00$ & $0.00$ & $99.66\%$ & $99.84\%$ & $99.68\%$ \\
0168 & $3.00$ & $0.00$ & $0.00$ & $99.61\%$ & $99.78\%$ & $99.73\%$ \\
0169 & $2.00$ & $0.00$ & $0.00$ & $99.48\%$ & $99.62\%$ & $98.98\%$ \\
0170 & $1.00$ & $+0.44$ & $+0.44$ & $99.19\%$ & $99.35\%$ & $99.18\%$ \\
0171 & $1.00$ & $-0.44$ & $-0.44$ & $98.90\%$ & $98.80\%$ & $98.49\%$ \\
0172 & $1.00$ & $+0.98$ & $+0.98$ & $99.33\%$ & $99.08\%$ & $98.47\%$ \\
0174 & $3.00$ & $+0.50$ & $0.00$ & $99.58\%$ & $99.52\%$ & $99.36\%$ \\
0175 & $1.00$ & $+0.75$ & $+0.75$ & $99.76\%$ & $99.66\%$ & $99.49\%$ \\
0176 & $1.00$ & $+0.96$ & $+0.96$ & $99.54\%$ & $99.21\%$ & $98.71\%$ \\
0177 & $1.00$ & $+0.99$ & $+0.99$ & $99.09\%$ & $98.85\%$ & $98.34\%$ \\
0178 & $1.00$ & $+0.99$ & $+0.99$ & $98.80\%$ & $98.66\%$ & $97.96\%$ \\
0180 & $1.00$ & $0.00$ & $0.00$ & $99.61\%$ & $99.59\%$ & $99.57\%$ \\
0181 & $6.00$ & $0.00$ & $0.00$ & $99.85\%$ & $99.61\%$ & $99.09\%$ \\
0182 & $4.00$ & $0.00$ & $0.00$ & $99.82\%$ & $99.83\%$ & $99.64\%$ \\
0183 & $3.00$ & $0.00$ & $0.00$ & $99.71\%$ & $99.76\%$ & $99.72\%$ \\
0184 & $2.00$ & $0.00$ & $0.00$ & $99.59\%$ & $99.64\%$ & $99.56\%$ \\
0185 & $9.99$ & $0.00$ & $0.00$ & $99.86\%$ & $99.29\%$ & $98.57\%$ \\
0186 & $8.27$ & $0.00$ & $0.00$ & $99.83\%$ & $99.35\%$ & $98.48\%$ \\
0187 & $5.04$ & $0.00$ & $0.00$ & $99.76\%$ & $99.70\%$ & $99.42\%$ \\
0188 & $7.19$ & $0.00$ & $0.00$ & $99.82\%$ & $99.45\%$ & $98.78\%$ \\
0189 & $9.17$ & $0.00$ & $0.00$ & $99.84\%$ & $98.83\%$ & $97.29\%$ \\
0190 & $4.50$ & $0.00$ & $0.00$ & $99.78\%$ & $99.70\%$ & $99.37\%$ \\
0191 & $2.51$ & $0.00$ & $0.00$ & $99.49\%$ & $99.55\%$ & $99.51\%$ \\
0192 & $6.58$ & $0.00$ & $0.00$ & $99.87\%$ & $99.52\%$ & $98.96\%$ \\
0193 & $3.50$ & $0.00$ & $0.00$ & $99.70\%$ & $99.74\%$ & $99.67\%$ \\
0194 & $1.52$ & $0.00$ & $0.00$ & $99.62\%$ & $99.62\%$ & $99.54\%$ \\
0195 & $7.76$ & $0.00$ & $0.00$ & $99.88\%$ & $99.48\%$ & $98.88\%$ \\
0196 & $9.66$ & $0.00$ & $0.00$ & $99.84\%$ & $99.22\%$ & $98.17\%$ \\
0197 & $5.52$ & $0.00$ & $0.00$ & $99.83\%$ & $99.66\%$ & $99.28\%$ \\
0198 & $1.20$ & $0.00$ & $0.00$ & $99.77\%$ & $99.64\%$ & $99.62\%$ \\
0199 & $8.73$ & $0.00$ & $0.00$ & $99.82\%$ & $99.26\%$ & $98.26\%$ \\
0200 & $3.27$ & $0.00$ & $0.00$ & $99.56\%$ & $99.69\%$ & $99.63\%$ \\
0201 & $2.32$ & $0.00$ & $0.00$ & $99.44\%$ & $99.55\%$ & $99.48\%$ \\
0202 & $7.00$ & $+0.60$ & $0.00$ & $99.04\%$ & $98.09\%$ & $97.44\%$ \\
0203 & $7.00$ & $+0.40$ & $0.00$ & $99.24\%$ & $98.86\%$ & $98.40\%$ \\
0204 & $7.00$ & $+0.40$ & $0.00$ & $99.23\%$ & $98.66\%$ & $98.19\%$ \\
0205 & $7.00$ & $-0.40$ & $0.00$ & $99.21\%$ & $99.55\%$ & $99.11\%$ \\
0206 & $7.00$ & $-0.40$ & $0.00$ & $99.54\%$ & $99.56\%$ & $98.83\%$ \\
0207 & $7.00$ & $-0.60$ & $0.00$ & $99.78\%$ & $98.84\%$ & $98.36\%$ \\
0208 & $5.00$ & $-0.90$ & $0.00$ & $99.26\%$ & $99.13\%$ & $98.10\%$ \\
0209 & $1.00$ & $-0.90$ & $-0.50$ & $99.25\%$ & $99.17\%$ & $98.81\%$ \\
0210 & $1.00$ & $-0.90$ & $0.00$ & $99.12\%$ & $98.92\%$ & $98.57\%$ \\
0211 & $1.00$ & $-0.90$ & $+0.90$ & $99.86\%$ & $99.75\%$ & $99.49\%$ \\
0212 & $1.00$ & $-0.80$ & $-0.80$ & $99.81\%$ & $99.70\%$ & $99.40\%$ \\
0213 & $1.00$ & $-0.80$ & $+0.80$ & $99.89\%$ & $99.80\%$ & $99.59\%$ \\
0214 & $1.00$ & $-0.62$ & $-0.25$ & $99.12\%$ & $99.01\%$ & $98.74\%$ \\
0215 & $1.00$ & $-0.60$ & $-0.60$ & $99.86\%$ & $99.84\%$ & $99.64\%$ \\
0216 & $1.00$ & $-0.60$ & $0.00$ & $99.64\%$ & $99.56\%$ & $99.36\%$ \\
0217 & $1.00$ & $-0.60$ & $+0.60$ & $99.80\%$ & $99.81\%$ & $99.61\%$ \\
0218 & $1.00$ & $-0.50$ & $+0.50$ & $99.76\%$ & $99.62\%$ & $99.21\%$ \\
0219 & $1.00$ & $-0.50$ & $+0.90$ & $99.72\%$ & $99.66\%$ & $99.51\%$ \\
0220 & $1.00$ & $-0.40$ & $-0.80$ & $99.87\%$ & $99.82\%$ & $99.55\%$ \\
0221 & $1.00$ & $-0.40$ & $+0.80$ & $99.81\%$ & $99.77\%$ & $99.56\%$ \\
0222 & $1.00$ & $-0.30$ & $0.00$ & $99.67\%$ & $99.55\%$ & $99.13\%$ \\
0223 & $1.00$ & $+0.30$ & $0.00$ & $99.86\%$ & $99.74\%$ & $99.45\%$ \\
0224 & $1.00$ & $+0.40$ & $-0.80$ & $99.83\%$ & $99.72\%$ & $99.49\%$ \\
0225 & $1.00$ & $+0.40$ & $+0.80$ & $99.33\%$ & $99.28\%$ & $99.18\%$ \\
0226 & $1.00$ & $+0.50$ & $-0.90$ & $99.79\%$ & $99.69\%$ & $99.16\%$ \\
0227 & $1.00$ & $+0.60$ & $0.00$ & $99.79\%$ & $99.74\%$ & $99.65\%$ \\
0228 & $1.00$ & $+0.60$ & $+0.60$ & $99.49\%$ & $99.43\%$ & $99.35\%$ \\
0229 & $1.00$ & $+0.65$ & $+0.25$ & $99.44\%$ & $99.39\%$ & $99.30\%$ \\
0230 & $1.00$ & $+0.80$ & $+0.80$ & $99.72\%$ & $99.57\%$ & $99.41\%$ \\
0231 & $1.00$ & $+0.90$ & $0.00$ & $98.77\%$ & $98.88\%$ & $98.77\%$ \\
0232 & $1.00$ & $+0.90$ & $+0.50$ & $99.55\%$ & $99.49\%$ & $99.39\%$ \\
0233 & $2.00$ & $-0.87$ & $+0.85$ & $99.15\%$ & $99.14\%$ & $99.01\%$ \\
0234 & $2.00$ & $-0.85$ & $-0.85$ & $99.23\%$ & $99.14\%$ & $98.59\%$ \\
0235 & $2.00$ & $-0.60$ & $-0.60$ & $99.94\%$ & $99.88\%$ & $99.74\%$ \\
0236 & $2.00$ & $-0.60$ & $0.00$ & $99.80\%$ & $99.74\%$ & $99.57\%$ \\
0237 & $2.00$ & $-0.60$ & $+0.60$ & $99.49\%$ & $99.44\%$ & $99.29\%$ \\
0238 & $2.00$ & $-0.50$ & $-0.50$ & $99.83\%$ & $99.69\%$ & $99.46\%$ \\
0239 & $2.00$ & $-0.37$ & $+0.85$ & $99.77\%$ & $99.57\%$ & $99.30\%$ \\
0240 & $2.00$ & $-0.30$ & $-0.30$ & $99.73\%$ & $99.61\%$ & $99.28\%$ \\
0241 & $2.00$ & $-0.30$ & $0.00$ & $99.81\%$ & $99.47\%$ & $99.23\%$ \\
0242 & $2.00$ & $-0.30$ & $+0.30$ & $99.79\%$ & $99.64\%$ & $99.43\%$ \\
0243 & $2.00$ & $-0.13$ & $-0.85$ & $99.12\%$ & $98.97\%$ & $98.93\%$ \\
0244 & $2.00$ & $0.00$ & $-0.60$ & $98.69\%$ & $98.96\%$ & $98.93\%$ \\
0245 & $2.00$ & $0.00$ & $-0.30$ & $99.30\%$ & $99.28\%$ & $99.21\%$ \\
0246 & $2.00$ & $0.00$ & $+0.30$ & $99.75\%$ & $99.53\%$ & $99.47\%$ \\
0247 & $2.00$ & $0.00$ & $+0.60$ & $99.85\%$ & $99.74\%$ & $99.57\%$ \\
0248 & $2.00$ & $+0.13$ & $+0.85$ & $99.82\%$ & $99.65\%$ & $99.45\%$ \\
0249 & $2.00$ & $+0.30$ & $-0.30$ & $99.58\%$ & $99.53\%$ & $99.47\%$ \\
0250 & $2.00$ & $+0.30$ & $0.00$ & $99.82\%$ & $99.74\%$ & $99.55\%$ \\
0251 & $2.00$ & $+0.30$ & $+0.30$ & $99.91\%$ & $99.74\%$ & $99.62\%$ \\
0252 & $2.00$ & $+0.37$ & $-0.85$ & $98.53\%$ & $98.80\%$ & $98.75\%$ \\
0253 & $2.00$ & $+0.50$ & $+0.50$ & $99.90\%$ & $99.83\%$ & $99.73\%$ \\
0254 & $2.00$ & $+0.60$ & $-0.60$ & $99.93\%$ & $99.70\%$ & $99.40\%$ \\
0255 & $2.00$ & $+0.60$ & $0.00$ & $99.86\%$ & $99.77\%$ & $99.65\%$ \\
0256 & $2.00$ & $+0.60$ & $+0.60$ & $99.89\%$ & $99.83\%$ & $99.75\%$ \\
0257 & $2.00$ & $+0.85$ & $+0.85$ & $99.76\%$ & $99.39\%$ & $99.05\%$ \\
0258 & $2.00$ & $+0.87$ & $-0.85$ & $99.33\%$ & $99.13\%$ & $98.93\%$ \\
0259 & $2.50$ & $0.00$ & $0.00$ & $99.57\%$ & $99.54\%$ & $99.48\%$ \\
0260 & $3.00$ & $-0.85$ & $-0.85$ & $98.40\%$ & $98.37\%$ & $98.05\%$ \\
0261 & $3.00$ & $-0.73$ & $+0.85$ & $99.79\%$ & $99.62\%$ & $99.01\%$ \\
0262 & $3.00$ & $-0.60$ & $0.00$ & $99.96\%$ & $99.79\%$ & $99.48\%$ \\
0263 & $3.00$ & $-0.60$ & $+0.60$ & $99.88\%$ & $99.75\%$ & $99.46\%$ \\
0264 & $3.00$ & $-0.60$ & $-0.60$ & $99.72\%$ & $99.74\%$ & $99.54\%$ \\
0265 & $3.00$ & $-0.60$ & $-0.40$ & $99.86\%$ & $99.79\%$ & $99.54\%$ \\
0266 & $3.00$ & $-0.60$ & $+0.40$ & $99.93\%$ & $99.81\%$ & $99.56\%$ \\
0267 & $3.00$ & $-0.50$ & $-0.50$ & $99.79\%$ & $99.66\%$ & $99.51\%$ \\
0268 & $3.00$ & $-0.40$ & $-0.60$ & $99.58\%$ & $99.63\%$ & $99.56\%$ \\
0269 & $3.00$ & $-0.40$ & $+0.60$ & $99.82\%$ & $99.62\%$ & $99.39\%$ \\
0270 & $3.00$ & $-0.30$ & $-0.30$ & $99.80\%$ & $99.75\%$ & $99.71\%$ \\
0271 & $3.00$ & $-0.30$ & $0.00$ & $99.92\%$ & $99.81\%$ & $99.62\%$ \\
0272 & $3.00$ & $-0.30$ & $+0.30$ & $99.93\%$ & $99.85\%$ & $99.72\%$ \\
0273 & $3.00$ & $-0.27$ & $-0.85$ & $99.09\%$ & $99.15\%$ & $99.04\%$ \\
0274 & $3.00$ & $-0.23$ & $+0.85$ & $99.82\%$ & $99.71\%$ & $99.00\%$ \\
0275 & $3.00$ & $0.00$ & $-0.60$ & $98.79\%$ & $99.04\%$ & $98.95\%$ \\
0276 & $3.00$ & $0.00$ & $-0.30$ & $99.49\%$ & $99.41\%$ & $99.33\%$ \\
0277 & $3.00$ & $0.00$ & $+0.30$ & $99.84\%$ & $99.77\%$ & $99.59\%$ \\
0278 & $3.00$ & $0.00$ & $+0.60$ & $99.87\%$ & $99.81\%$ & $99.66\%$ \\
0279 & $3.00$ & $+0.23$ & $-0.85$ & $98.53\%$ & $98.61\%$ & $98.24\%$ \\
0280 & $3.00$ & $+0.27$ & $+0.85$ & $99.54\%$ & $99.39\%$ & $99.18\%$ \\
0281 & $3.00$ & $+0.30$ & $-0.30$ & $99.74\%$ & $99.71\%$ & $99.55\%$ \\
0282 & $3.00$ & $+0.30$ & $0.00$ & $99.79\%$ & $99.58\%$ & $99.40\%$ \\
0283 & $3.00$ & $+0.30$ & $+0.30$ & $99.68\%$ & $99.50\%$ & $99.28\%$ \\
0284 & $3.00$ & $+0.40$ & $-0.60$ & $99.60\%$ & $99.41\%$ & $98.78\%$ \\
0285 & $3.00$ & $+0.40$ & $+0.60$ & $99.57\%$ & $99.41\%$ & $99.26\%$ \\
0286 & $3.00$ & $+0.50$ & $+0.50$ & $99.63\%$ & $99.46\%$ & $99.29\%$ \\
0287 & $3.00$ & $+0.60$ & $-0.60$ & $99.79\%$ & $99.59\%$ & $99.41\%$ \\
0288 & $3.00$ & $+0.60$ & $-0.40$ & $99.75\%$ & $99.58\%$ & $99.40\%$ \\
0289 & $3.00$ & $+0.60$ & $0.00$ & $99.71\%$ & $99.53\%$ & $99.38\%$ \\
0290 & $3.00$ & $+0.60$ & $+0.40$ & $99.72\%$ & $99.62\%$ & $99.47\%$ \\
0291 & $3.00$ & $+0.60$ & $+0.60$ & $99.81\%$ & $99.69\%$ & $99.59\%$ \\
0292 & $3.00$ & $+0.73$ & $-0.85$ & $99.85\%$ & $99.68\%$ & $99.42\%$ \\
0293 & $3.00$ & $+0.85$ & $+0.85$ & $99.46\%$ & $98.95\%$ & $98.33\%$ \\
0294 & $3.50$ & $0.00$ & $0.00$ & $99.77\%$ & $99.68\%$ & $99.61\%$ \\
0295 & $4.50$ & $0.00$ & $0.00$ & $99.88\%$ & $99.80\%$ & $99.66\%$ \\
0296 & $5.50$ & $0.00$ & $0.00$ & $99.91\%$ & $99.61\%$ & $99.24\%$ \\
0297 & $6.50$ & $0.00$ & $0.00$ & $99.90\%$ & $99.57\%$ & $99.04\%$ \\
0298 & $7.00$ & $0.00$ & $0.00$ & $99.90\%$ & $99.58\%$ & $98.96\%$ \\
0299 & $7.50$ & $0.00$ & $0.00$ & $99.91\%$ & $99.38\%$ & $98.67\%$ \\
0300 & $8.50$ & $0.00$ & $0.00$ & $99.79\%$ & $99.43\%$ & $98.48\%$ \\
0301 & $9.00$ & $0.00$ & $0.00$ & $99.79\%$ & $99.43\%$ & $98.61\%$ \\
0302 & $9.50$ & $0.00$ & $0.00$ & $99.77\%$ & $99.19\%$ & $98.21\%$ \\
0303 & $10.00$ & $0.00$ & $0.00$ & $99.76\%$ & $99.04\%$ & $97.75\%$ \\
0304 & $1.00$ & $+0.50$ & $-0.50$ & $99.89\%$ & $99.72\%$ & $99.67\%$ \\
0305 & $1.22$ & $+0.33$ & $-0.44$ & $99.60\%$ & $99.40\%$ & $98.86\%$ \\
1420 & $8.00$ & $-0.80$ & $+0.80$ & $99.32\%$ & $98.06\%$ & $95.38\%$ \\
1422 & $7.95$ & $-0.80$ & $-0.46$ & $99.27\%$ & $98.75\%$ & $96.97\%$ \\
1423 & $8.00$ & $-0.60$ & $-0.75$ & $99.47\%$ & $99.35\%$ & $98.01\%$ \\
1424 & $6.46$ & $-0.66$ & $-0.80$ & $99.08\%$ & $98.87\%$ & $96.98\%$ \\
1425 & $6.12$ & $-0.80$ & $+0.67$ & $99.35\%$ & $98.98\%$ & $97.06\%$ \\
1427 & $7.41$ & $-0.61$ & $-0.73$ & $99.18\%$ & $99.27\%$ & $98.26\%$ \\
1428 & $5.52$ & $-0.80$ & $-0.70$ & $99.13\%$ & $98.69\%$ & $96.91\%$ \\
1429 & $7.75$ & $-0.20$ & $-0.78$ & $99.26\%$ & $99.23\%$ & $98.31\%$ \\
1430 & $8.00$ & $+0.28$ & $-0.75$ & $99.53\%$ & $99.41\%$ & $98.73\%$ \\
1431 & $8.00$ & $+0.08$ & $-0.78$ & $99.40\%$ & $99.41\%$ & $98.65\%$ \\
1433 & $8.00$ & $-0.74$ & $+0.21$ & $99.00\%$ & $98.86\%$ & $97.06\%$ \\
1434 & $4.37$ & $+0.80$ & $+0.80$ & $99.13\%$ & $98.34\%$ & $98.03\%$ \\
1435 & $6.59$ & $-0.79$ & $+0.07$ & $98.90\%$ & $99.00\%$ & $98.15\%$ \\
1436 & $6.28$ & $+0.01$ & $-0.80$ & $99.09\%$ & $99.47\%$ & $98.98\%$ \\
1437 & $6.04$ & $+0.80$ & $+0.15$ & $99.03\%$ & $98.06\%$ & $97.66\%$ \\
1438 & $5.87$ & $+0.13$ & $+0.80$ & $99.65\%$ & $99.07\%$ & $98.11\%$ \\
1439 & $6.48$ & $+0.72$ & $-0.32$ & $99.40\%$ & $99.02\%$ & $98.51\%$ \\
1440 & $5.64$ & $+0.77$ & $+0.31$ & $99.07\%$ & $98.60\%$ & $98.18\%$ \\
1441 & $8.00$ & $+0.60$ & $-0.48$ & $98.98\%$ & $98.22\%$ & $97.62\%$ \\
1442 & $6.58$ & $-0.71$ & $-0.18$ & $99.21\%$ & $98.87\%$ & $97.22\%$ \\
1443 & $5.68$ & $+0.41$ & $-0.74$ & $99.18\%$ & $98.89\%$ & $98.33\%$ \\
1444 & $5.94$ & $-0.06$ & $-0.76$ & $99.16\%$ & $99.53\%$ & $99.28\%$ \\
1445 & $4.67$ & $-0.50$ & $+0.80$ & $99.48\%$ & $99.72\%$ & $99.35\%$ \\
1446 & $3.15$ & $-0.80$ & $+0.78$ & $99.80\%$ & $99.62\%$ & $99.15\%$ \\
1447 & $3.16$ & $+0.74$ & $+0.80$ & $99.88\%$ & $99.02\%$ & $98.77\%$ \\
1448 & $6.94$ & $-0.48$ & $+0.52$ & $99.86\%$ & $99.69\%$ & $99.08\%$ \\
1449 & $4.19$ & $-0.80$ & $-0.34$ & $99.24\%$ & $99.57\%$ & $98.99\%$ \\
1450 & $4.07$ & $-0.28$ & $-0.80$ & $98.99\%$ & $99.26\%$ & $98.78\%$ \\
1451 & $4.06$ & $+0.31$ & $-0.80$ & $99.17\%$ & $99.52\%$ & $99.41\%$ \\
1452 & $3.64$ & $+0.80$ & $-0.43$ & $99.64\%$ & $99.44\%$ & $99.17\%$ \\
1453 & $2.35$ & $+0.80$ & $-0.78$ & $99.77\%$ & $99.55\%$ & $99.15\%$ \\
1454 & $2.45$ & $-0.80$ & $-0.73$ & $99.26\%$ & $99.63\%$ & $99.37\%$ \\
1455 & $8.00$ & $-0.40$ & $0.00$ & $99.31\%$ & $98.94\%$ & $97.73\%$ \\
1456 & $3.00$ & $+0.74$ & $+0.70$ & $99.83\%$ & $99.68\%$ & $99.51\%$ \\
1457 & $3.25$ & $+0.54$ & $+0.80$ & $99.62\%$ & $99.48\%$ & $99.37\%$ \\
1458 & $3.80$ & $-0.06$ & $+0.80$ & $99.51\%$ & $99.81\%$ & $99.55\%$ \\
1459 & $2.26$ & $+0.76$ & $+0.80$ & $99.45\%$ & $99.20\%$ & $98.98\%$ \\
1460 & $8.00$ & $+0.12$ & $+0.11$ & $99.40\%$ & $99.03\%$ & $98.16\%$ \\
1461 & $2.88$ & $-0.45$ & $-0.80$ & $99.15\%$ & $99.43\%$ & $99.34\%$ \\
1462 & $2.63$ & $-0.80$ & $+0.51$ & $99.89\%$ & $99.69\%$ & $99.34\%$ \\
1463 & $4.98$ & $+0.61$ & $+0.24$ & $98.68\%$ & $98.47\%$ & $98.03\%$ \\
1464 & $6.54$ & $-0.05$ & $-0.32$ & $99.31\%$ & $99.04\%$ & $98.35\%$ \\
1466 & $1.90$ & $+0.70$ & $-0.80$ & $99.93\%$ & $99.57\%$ & $99.18\%$ \\
1467 & $2.23$ & $-0.56$ & $+0.80$ & $99.64\%$ & $99.42\%$ & $99.19\%$ \\
1468 & $2.27$ & $+0.51$ & $+0.80$ & $99.40\%$ & $99.81\%$ & $99.71\%$ \\
1469 & $1.85$ & $+0.80$ & $+0.67$ & $99.74\%$ & $99.38\%$ & $99.18\%$ \\
1470 & $1.52$ & $-0.73$ & $-0.79$ & $99.93\%$ & $99.72\%$ & $99.43\%$ \\
1471 & $1.33$ & $-0.78$ & $-0.80$ & $99.95\%$ & $99.69\%$ & $99.22\%$ \\
1472 & $2.37$ & $-0.80$ & $-0.12$ & $99.91\%$ & $99.75\%$ & $99.52\%$ \\
1473 & $1.45$ & $+0.70$ & $+0.79$ & $99.75\%$ & $99.13\%$ & $99.04\%$ \\
1474 & $1.28$ & $+0.72$ & $-0.80$ & $99.60\%$ & $98.48\%$ & $97.47\%$ \\
1475 & $1.00$ & $-0.80$ & $-0.80$ & $99.78\%$ & $99.64\%$ & $99.33\%$ \\
1476 & $1.00$ & $-0.80$ & $+0.80$ & $99.81\%$ & $99.76\%$ & $99.51\%$ \\
1477 & $1.00$ & $+0.80$ & $+0.80$ & $99.70\%$ & $99.56\%$ & $99.37\%$ \\
1478 & $1.97$ & $+0.80$ & $+0.13$ & $99.31\%$ & $99.35\%$ & $99.23\%$ \\
1479 & $1.55$ & $-0.56$ & $-0.80$ & $99.97\%$ & $99.87\%$ & $99.68\%$ \\
1480 & $1.55$ & $-0.80$ & $-0.31$ & $99.74\%$ & $99.64\%$ & $99.38\%$ \\
1481 & $1.00$ & $+0.73$ & $+0.79$ & $99.75\%$ & $99.63\%$ & $99.46\%$ \\
1482 & $1.39$ & $-0.58$ & $+0.80$ & $99.77\%$ & $99.62\%$ & $99.43\%$ \\
1483 & $3.17$ & $+0.56$ & $-0.19$ & $99.30\%$ & $99.38\%$ & $99.04\%$ \\
1484 & $2.90$ & $-0.56$ & $+0.30$ & $99.84\%$ & $99.73\%$ & $99.51\%$ \\
1485 & $3.09$ & $+0.35$ & $-0.40$ & $99.32\%$ & $99.67\%$ & $99.45\%$ \\
1486 & $3.72$ & $+0.43$ & $-0.03$ & $99.24\%$ & $99.26\%$ & $98.93\%$ \\
1487 & $1.25$ & $-0.80$ & $+0.51$ & $99.49\%$ & $99.40\%$ & $99.16\%$ \\
1488 & $1.59$ & $-0.33$ & $+0.75$ & $99.86\%$ & $99.65\%$ & $99.39\%$ \\
1489 & $3.46$ & $+0.30$ & $-0.17$ & $99.73\%$ & $99.48\%$ & $99.19\%$ \\
1490 & $1.25$ & $+0.41$ & $+0.76$ & $99.73\%$ & $99.24\%$ & $99.13\%$ \\
1491 & $1.66$ & $+0.20$ & $-0.70$ & $99.13\%$ & $98.92\%$ & $98.87\%$ \\
1492 & $1.00$ & $-0.47$ & $-0.79$ & $99.67\%$ & $99.27\%$ & $98.77\%$ \\
1493 & $1.28$ & $+0.01$ & $+0.80$ & $99.80\%$ & $99.64\%$ & $99.46\%$ \\
1494 & $2.21$ & $-0.47$ & $-0.39$ & $99.72\%$ & $99.79\%$ & $99.22\%$ \\
1495 & $1.00$ & $+0.78$ & $+0.53$ & $99.59\%$ & $99.22\%$ & $99.10\%$ \\
1496 & $1.16$ & $+0.80$ & $+0.03$ & $99.08\%$ & $98.73\%$ & $98.59\%$ \\
1497 & $1.00$ & $+0.68$ & $+0.67$ & $99.67\%$ & $99.59\%$ & $99.45\%$ \\
1498 & $1.03$ & $+0.22$ & $-0.78$ & $99.73\%$ & $99.55\%$ & $99.24\%$ \\
1499 & $1.00$ & $-0.75$ & $+0.34$ & $99.73\%$ & $99.65\%$ & $99.40\%$ \\
1500 & $1.00$ & $-0.77$ & $-0.20$ & $99.13\%$ & $98.99\%$ & $98.72\%$ \\